\documentclass[structabstract]{aa}
\usepackage{graphicx}
\usepackage{aalongtable}
\usepackage{lscape}
\usepackage{txfonts}
\usepackage{natbib}
\bibpunct{(}{)}{;}{a}{}{,}
\begin{document}
   \title{A kinematic study and membership analysis of the Lupus star-forming region\thanks{Based partly on observations collected at the European Southern Observatory, Chile
(ESO Programme 087.C-0315).}}

   \author{P.A.B. Galli
          \inst{1}
          \and
          C. Bertout\inst{2}
          \and
          R. Teixeira\inst{1}
          \and
          C. Ducourant\inst{3}
          }

   \institute{Instituto de Astronomia, Geof\'isica e Ci\^encias Atmosf\'ericas, Universidade de S\~ao Paulo, Rua do Mat\~ao, 1226 - Cidade Universit\'aria, 05508-900, S\~ao Paulo - SP, Brazil\\
              \email{galli@astro.iag.usp.br}
           \and
          Institut d'Astrophysique, 98bis, Bd. Arago, 75014 Paris, France
         \and
             Observatoire Aquitain des Sciences de l 'Univers, CNRS-UMR 5804, BP 89, Floirac, France
             }

\date{Received / Accepted}

 \abstract{}{A precise determination of the distance to individual stars is required to reliably determine the fundamental parameters (mass and age) of young stellar objects.
 This paper is dedicated to investigating the kinematic properties of the Lupus moving group of young stars with the primary objective of deriving individual parallaxes for each group member.}
 {We identify those stars in the Lupus star-forming region that define the comoving association of young stars by utilizing our new and improved convergent point search method that allows us to derive the precise position of the convergent point of the comoving association from the stars' proper motions. We used published proper motion catalogs and searched the literature for radial velocities, which are needed to compute individual parallaxes. We supplemented the radial velocity data with new measurements from spectroscopic observations performed with the FEROS spectrograph mounted on the MPG/ESO 2.2m telescope at La Silla. }
 {We identify a comoving group with 109 pre-main sequence stars and candidates that define the kinematic properties of the Lupus low-mass star-forming region.  We derive individual parallaxes for stars with known radial velocity and tentative parallaxes for the remaining group members by assuming that all stars share the same space motion. The convergent point method, combined with the k-NN algorithm, makes it possible to distinguish the Lupus and Upper Centaurus Lupus stars from the adjacent Scorpius-Centaurus association. We find significant depth effects in this region and show that the classical T~Tauri stars, located in the close vicinity of the Lupus molecular clouds, form a background population, while the weak-emission line T~Tauri stars are dispersed not only in angular extent but also in depth.}
 {The newly derived individual parallaxes will be used in a forthcoming paper to refine the masses and ages of Lupus T~Tauri stars, with the aim of better constraining the lifetimes of their circumstellar, protoplanetary disks.}

\keywords{methods: data analysis - technique: radial velocities - astrometry - parallaxes - proper motions - stars: distances - stars: kinematics and dynamics - open clusters and associations: individual: Lupus}

\maketitle

%----------------------------------------------------------------------------------------------------------------										1. INTRODUCTION
%----------------------------------------------------------------------------------------------------------------
\section{Introduction}

The inferred fundamental parameters of T Tauri stars (TTSs), the young solar-type pre-main sequence stars first discovered by \cite{Joy(1945)}, are sensitive to their assumed distance. TTSs are usually associated with molecular clouds, the distances of which can be estimated, say, from the photometry of a few bright stars enshrouded in reflection nebulosity \citep[see, for example,][]{Racine(1968)} or from 2MASS extinction maps combined with known stellar parallaxes \citep{Lombardi(2008)}. The distances to the nearby star-forming regions (SFRs) are thus known to relatively good accuracy, providing a first estimate of the distances to the members of the stellar population associated with them. Although these average parallaxes provide valuable information, we must know the distances to the individual members of the young associations more precisely to better constrain their ages and masses by comparing observed stellar properties to evolutionary models.

TTSs are late-type objects and usually faint in the visible, suffering typically from 1 to 2 magnitudes of extinction in that range, while the nearby SFRs to which they belong are typically in the distance range 0.1 -- 1 kpc. Parallax determination from the ground is usually impossible for these objects, and they even presented a challenge for the \textsc{Hipparcos} mission \citep{HIP97}, which observed only a few of them. \cite{Bertout(1999)} computed new astrometric solutions using the \textsc{Hipparcos} data for groups of TTSs in various SFRs, thus providing post-\textsc{Hipparcos} average distances to these groups.

Some progress in determining the distance of individual TTSs has been made in recent years, using two different methods. Loinard and collaborators used the Very Long Baseline Array (VLBA) to determine parallaxes of a few selected objects, mainly in the Taurus-Auriga SFR \citep{Loinard(2007),Torres(2007),Torres(2009),Dzib(2010),Dzib(2011),Torres(2012)}. The VLBA data allow for a very precise parallax determination, but the method is observationally intensive, limiting its application to a few remarkable stars. Another approach rests on the fact that the members of young associations share the same spatial motion and uses their proper motions to determine individual parallaxes for members of some nearby associations. This was done, for example, for the TW Hydrae association by \cite{Mamajek(2005)} and for the Taurus-Auriga SFR by \cite{Bertout(2006)}. The derived parallaxes using this method are not as precise as those obtained by VLBA observations, but it yields usable parallaxes for all members of the moving group that have measured radial velocities. 

Thus, while we have progressed, we are still far from knowing the distances to the large sample of pre-main sequence stars (PMSs), which is required to study such timely topics as the effect of differing protostellar environments on disk lifetimes and the timescales of planet formation. The situation will change considerably with the launch of the Gaia mission, because its instruments will measure the parallaxes and proper motions of millions of faint stars. Although the satellite's launch date is 2013, its catalog will be published several years later, so one would appreciate making some progress in determining the distances of TTSs in the meantime. Also, the astrometric methods that are developed and tested for this purpose will certainly be useful in the Gaia era.

In this paper, we study the largest southern SFR, located in Lupus, using the recently developed new version of the convergent point (CP) method presented by \cite{Galli(2012)}. Section~2 is devoted to a brief presentation of the Lupus SFR and its previous distance determinations, while Sect.~3 discusses the sample of Lupus candidate stars found in recent catalogs and proper motion information needed for the CP analysis. Section~4 discusses our search for radial velocity data, including observations carried out for a number of stars in our sample with FEROS mounted on the 2.2m MPG/ESO telescope at La Silla, Chile. The CP and membership analysis are discussed in Sect.~5, while Sect.~6 presents the parallax computations for the moving group members. We discuss the results of this investigation in Sect.~7. Finally, our conclusions are given in Sect.~8.

%----------------------------------------------------------------------------------------------------------------							2. THE LUPUS ASSOCIATION OF YOUNG STARS
%----------------------------------------------------------------------------------------------------------------
\section{The Lupus association of young stars}

The Lupus dark cloud complex is a low-mass star-forming region that contains four main star-forming clouds (Lupus~1~to~4) and constitutes one of the richest nearby associations of TTSs.  Many of the Lupus TTSs have been identified from ROSAT X-ray observations and from the spectroscopic surveys conducted by \citet{Krautter(1997)} and \citet{Wichmann(1997a), Wichmann(1997b)}. These surveys showed that the Lupus molecular clouds are surrounded by an extended halo of late-type, X-ray active stars, which bear resemblance to the so-called weak emission-line TTSs (WTTSs), the X-ray active TTSs without circumstellar accretion disks \citep{Walter(1986)}. If the ROSAT-detected objects were all WTTSs belonging to the Lupus association, then they would greatly exceed the number of so-called classical TTSs (CTTSs), the TTS subgroup showing evidence of circumstellar accretion disks \citep{Bertout(1987),Bertout(1988)}. Lupus CTTSs were identified by \citet{Schwartz(1977)} on the basis of their association with the molecular clouds and their H$\alpha$ emission. This early census of Lupus young stars has recently been expanded by infrared observations with the \emph{Spitzer} Space Observatory \citep{Merin(2008)}, and a proper motion study of these objects was performed by \citet{Lopez-Marti(2011)}.

The focus of most kinematic studies over the past century was not the Lupus clouds, but the adjacent Scorpius-Centaurus (Sco-Cen) association that contains the nearest OB association. It is divided into the three subgroups Upper Scorpius (US), Upper Centaurus-Lupus (UCL) and Lower Centaurus-Crux (LCC).  The Lupus molecular clouds, which are probably signposts of a more recent episode of star formation in that region \citep[see][]{Comeron(2008)}, occupy a gap between US and UCL. The distance of the Sco-Cen subgroups was investigated by \citet{deZeeuw(1999)}, who assessed association membership again using the \textsc{Hipparcos} data and derived a mean distance of 140 pc for the UCL subgroup. The Lupus SFR was first estimated to be at the same distance as UCL \citep{Hughes(1993)}. The close proximity of the UCL subgroup makes it difficult to distinguish between members of the Lupus association and of the UCL association. \citet{Mamajek(2002)} conducted a spectroscopic survey of UCL candidate members and selected 56 stars that had the same proper motion as UCL association members. A more complete list containing 81 UCL candidate stars is given in the recent review on the Sco-Cen association by \citet{Preibisch(2008)}. As it turned out, many of these stars were previously reported as WTTS members of the Lupus association by \citet{Krautter(1997)} and \citet{Wichmann(1997a),Wichmann(1997b)}. Since the membership status of these objects remains unclear, a more detailed study is clearly necessary.

In the past decade, several works have cast doubt on the distance to the Lupus star-forming region. From the angular extent of the molecular clouds (about $15\deg$), and by assuming that the depth of the molecular region is comparable, the distances of individual association members are expected to range from about 110~pc to 190~pc. Two bright TTSs observed by \textsc{Hipparcos}, RY~Lup, and V856~Sco  are apparently associated with clouds of the Lupus star-forming region, and their parallaxes are compatible with the above range, but the cloud distances reported in the literature range from 100 pc \citep{Knude(1998)} to 360 pc for Lupus~2 \citep{Knude(2001)}.

\cite{Bertout(1999)} used five stars connected with the Lupus complex that had been observed by \textsc{Hipparcos} to compute the average parallax of the Lupus SFR. They found a distance equal to 206$^{+34}_{-20}$ pc, which is a high value compared to the previous estimates, but noticed that the group parallax determination was dominated by the brightest stars HIP 79080 and 79081, two components of a Herbig Ae/Be (HAeBe) system located in Lupus 3. Considering only the three fainter stars  HIP 77157, 78094, and 78317, located in Lupus 1, 2, and 4, respectively, they obtained a distance of 147$^{+42}_{-27}$ pc, in agreement with previous determinations of the Lupus association distance, but with a large uncertainty, and a value of 228$^{+42}_{-30}$ pc for HIP 79080 and 79081. \cite{Bertout(1999)} concluded that either HIP 79080 and 79081 are not members of Lupus~3, which appears unlikely, or that this cloud is farther away than the other subgroups. \citet{Lombardi(2008)} conclude from their detailed study of 2MASS extinction maps that Lupus has a depth of 51$^{+61}_{-35}$ pc and that this \emph{``might be the result of different Lupus subclouds being at different distances"}. The kinematic study presented hereafter aims to shed light on the structure of the Lupus clouds, as well as on the membership of the Lupus association of PMS stars.

%----------------------------------------------------------------------------------------------------------------								3. THE SAMPLE OF PMS STARS IN LUPUS
%----------------------------------------------------------------------------------------------------------------
\section{Sample of Lupus candidate members with known proper motions} 

\subsection{Proper motions properties}

The \citet[][hereafter D05]{Ducourant(2005)} proper motion catalog for PMS stars contains 197 stars in the general area of the Lupus SFR, which approximately ranges from  $325^{\circ}\leq l\leq 342^{\circ}$ in Galactic longitude and $0^{\circ}\leq b\leq 25^{\circ}$ in Galactic latitude. The stars listed in this catalog were mostly identified by \citet{Herbig(1988)}, \citet{Krautter(1997)}, and \citet{Wichmann(1997a),Wichmann(1997b)}. We then included in the sample 37 stars of the comprehensive review performed by \citet{Comeron(2008)} that were not considered in D05, and another 24 stars from the \textit{c2d Spitzer} Legacy Program detected in the recent paper of \citet[][only group A sources of their study]{Lopez-Marti(2011)}. Our first list of presumed members of the Lupus association therefore consists of 258 stars.

To access the more recent measurements, we searched for proper motion data in the \rm{\textsc{PPMXL} }\citep{PPMXL}, \rm{\textsc{SPM4}} \citep{SPM4}, and \rm{\textsc{{UCAC4}} \citep{UCAC4} catalogs for all the 258 stars in our sample. The main source of proper motions is the \rm{\textsc{SPM4}} catalog, which represents the best present-day compromise between proper motion precision and target coverage for the stars in our sample. For the brightest stars $(V\leq12)$ in the sample we also searched the \rm{\textsc{Tycho2}} \citep{TYCHO2} catalog and used these values when the proper motions given in \rm{\textsc{SPM4}} were of lower precision.  Doing so, we found proper motion information for 241 stars in our initial sample, and for the remaining ones we kept the proper motions given in D05.  We used the proper motions of the \rm{\textsc{PPMXL}} and \rm{\textsc{UCAC4}} catalogs for two stars, since these are the only values available in the literature.

Whenever dealing with the kinematics of SFRs, the high fraction of binaries and multiple systems plays an important role. The overall binary fraction in Lupus is expected to reach about 30\% to 40\% \citep{Ghez(1997),Merin(2008)}. An advantage of the D05 catalog is that these systems are clearly identified with a mention AB indicating that the given proper motion is representative of the binary system. In these cases we decided to use the proper motion given in D05 for both resolved and unresolved binaries rather than taking the value provided by \rm\textsc{SPM4}}, where the existence of close companions is not mentioned.

The CP search method selects cluster members and defines the CP of a moving group based on proper motion data. Therefore, before starting our analysis, it is necessary to reject all stars with proper motion that carry poor information because of measurement errors. From the sample we reject 33 stars whose proper motion is dominated by errors (i.e., $\sigma_{\mu_{\alpha,\delta}}\geq\mu_{\alpha,\delta}$) in both components. After a $3\sigma$ elimination in both proper motion components we reject another eight stars that are in obvious disagreement with the common streaming motion of the Lupus moving group. These stars are given in Table~\ref{tab1}. The remaining 217 PMS stars will be used in this paper to investigate the kinematic properties of this SFR and discuss their membership status based on the CP analysis. Their average proper motion is $(\mu_{\alpha}\cos\delta,\mu_{\delta})=(-16,-21)$~mas/yr with an average precision of about 2~mas/yr in each component. 

\begin{figure}[!btp]
\begin{center}
\includegraphics[width=0.49\textwidth]{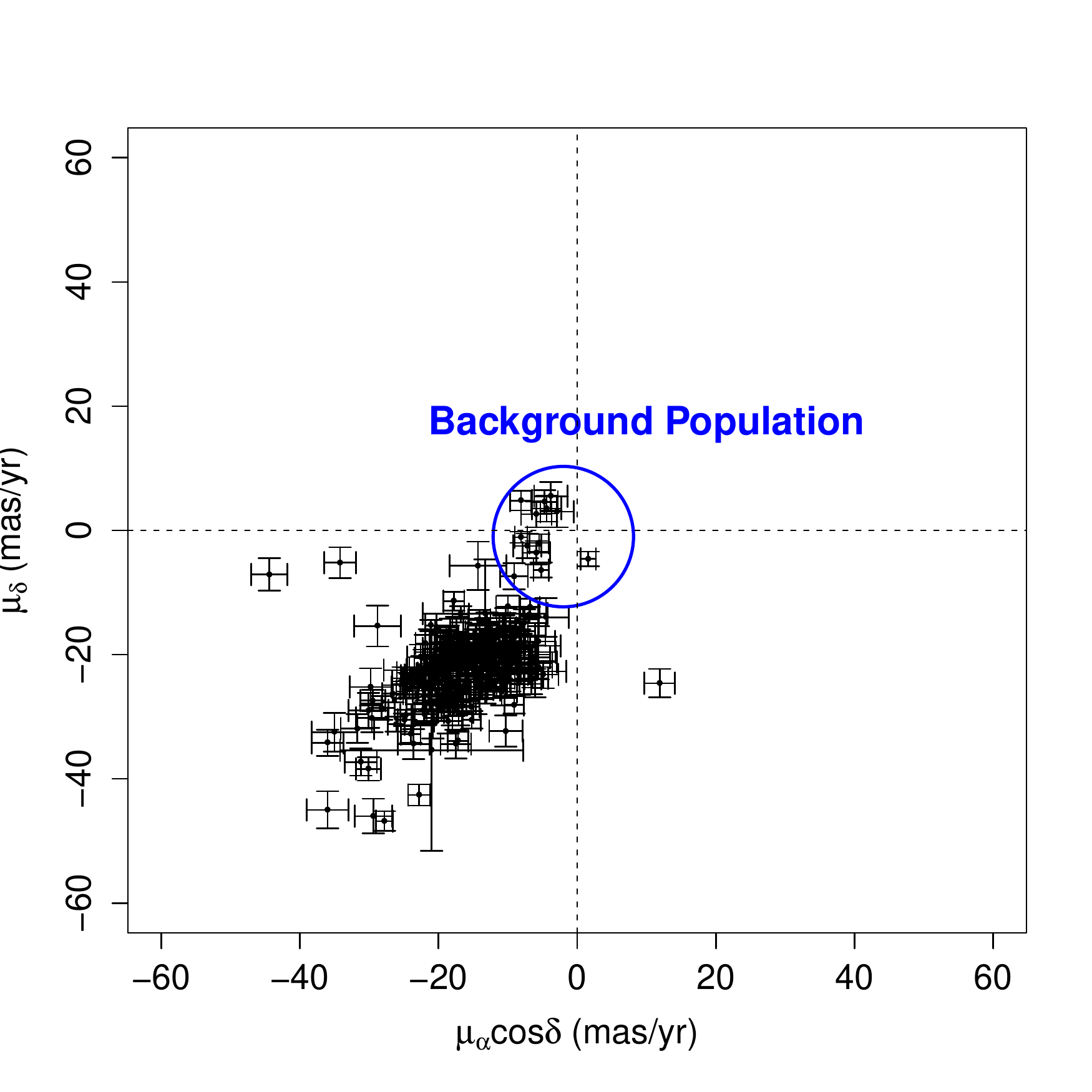}
\caption{\label{fig1}
Proper motion and associated errors for the 217 stars considered in this work. }
\end{center}
\end{figure}

\begin{figure}[!btp]
\begin{center}
\includegraphics[width=0.5\textwidth]{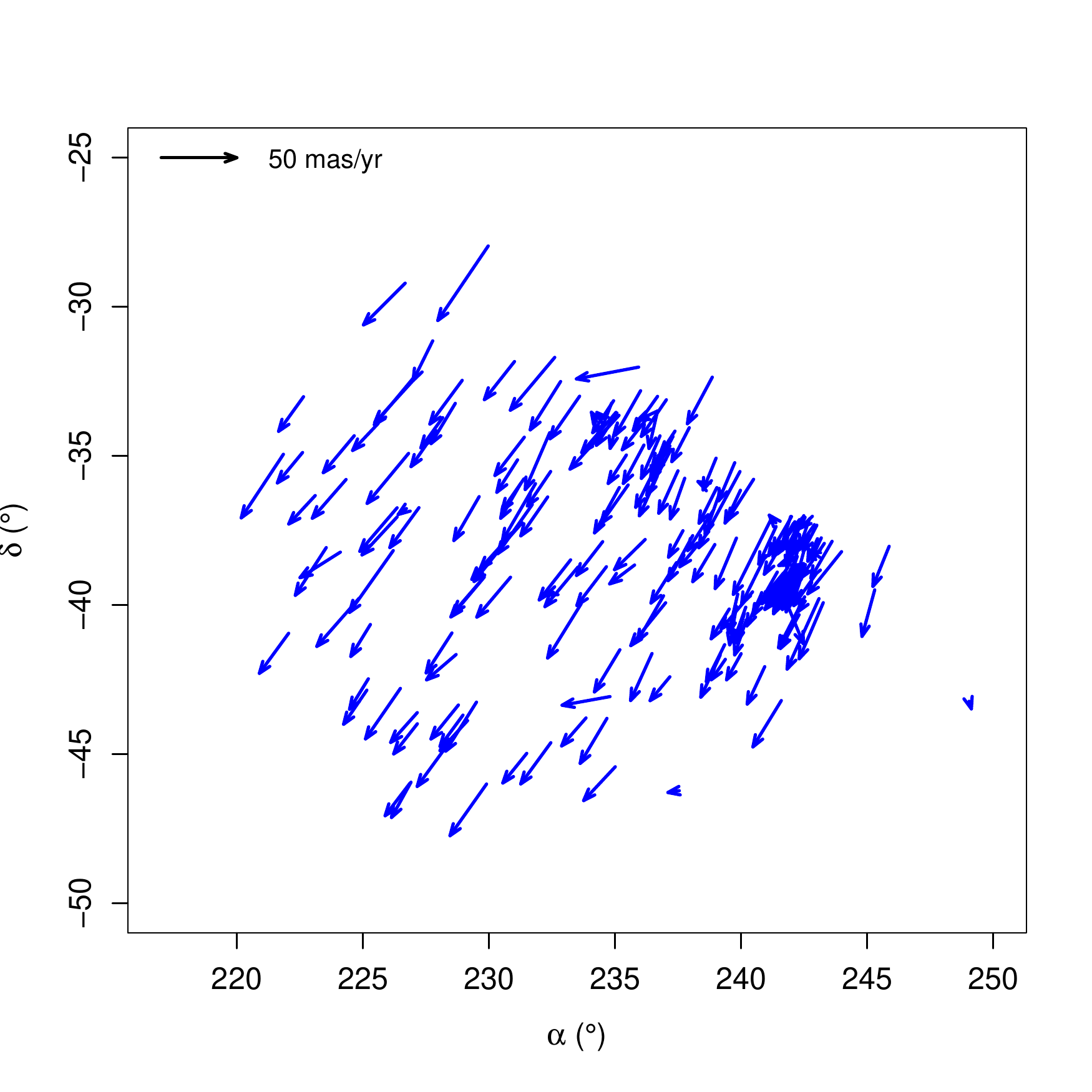}
\caption{\label{fig2}
Proper motion vectors of the 217 Lupus candidate stars.  }
\end{center}
\end{figure}

Figure~\ref{fig1} displays all proper motion values and their associated uncertainties. We note that a small group of these stars displays small proper motions. Because we considered only stars related to the Lupus SFR whose PMS status had already been established in previous works, it appears unlikely that field stars pollute our sample. We therefore assume for the moment that these stars form a background population belonging to the Lupus complex, and we come back to this point in Sect.~7. Figure~\ref{fig2} shows the proper motion vectors for our sample of Lupus stars. Although the position of the CP is not clearly apparent in this plot, one notices that most proper motion vectors point in a common direction. The precise coordinates of the CP will be derived in Sect.~5.

%FIGURE 1

%FIGURE 2

%TABLE 1
\begin{table*}[!htp]
\centering
\caption{Rejected stars from our sample of Lupus candidate members. 
\label{tab1}}
\begin{tabular}{lcccccc}
\hline
Star&$\alpha$&$\delta$&$\mu_{\alpha}\cos\delta$&$\mu_{\delta}$&Source&2MASSJ\\
&(h:m:s) &($^{\circ}$ $^\prime$ $^\prime$$^\prime$)&(mas/yr)&(mas/yr)&&\\
\hline
	RXJ1508.4-3337	&	15	08	26.2	&	-33	37	52	&$	+21.0	\pm	14.0	$&$	+16.0	\pm	14.0	$&	D05	&	15082621-3337517	\\
	RXJ1514.0-4629B	&	15	13	59.8	&	-46	29	54	&$	+29.0	\pm	13.0	$&$	+18.0	\pm	13.0	$&	D05	&	15135984-4629540	\\
	HD137727	&	15	28	44.0	&	-31	17	38	&$	+19.9	\pm	2.0	$&$	+48.0	\pm	1.9	$&	TYCHO2	&	15284402-3117387	\\
	Sz66	&	15	39	28.3	&	-34	46	18	&$	+160.3	\pm	5.4	$&$	-40.8	\pm	5.4	$&	UCAC4	&	15392828-3446180	\\
	HD140637	&	15	45	47.6	&	-30	20	55	&$	-70.7	\pm	2.4	$&$	-96.4	\pm	2.8	$&	SPM4	&	15454761-3020555	\\
	Sz101	&	16	08	28.4	&	-39	05	32	&$	+88.8	\pm	3.9	$&$	+29.1	\pm	4.0	$&	SPM4	&	16082843-3905324	\\
	RXJ1609.3-3855AB	&	16	09	23.2	&	-38	55	55	&$	+93.0	\pm	14.0	$&$	-119.0	\pm	14.0	$&	D05	&	16092320-3855547	\\
	V346Nor	&	16	32	32.1	&	-44	55	31	&$	+63.0	\pm	17.0	$&$	+130.0	\pm	17.0	$&	D05	&	16323219-4455306	\\
\hline
\end{tabular}
\tablefoot{We provide the most usual identifier, position (epoch 2000), proper motion, source of proper motion, and the 2MASS identifier for each star.}
\end{table*}

%----------------------------------------------------------------------------------------------------------------
\subsection{Refining the Lupus pre-main sequence star sample}

\begin{figure*}[!btp]
\begin{center}
\includegraphics[angle=-90,width=1\textwidth]{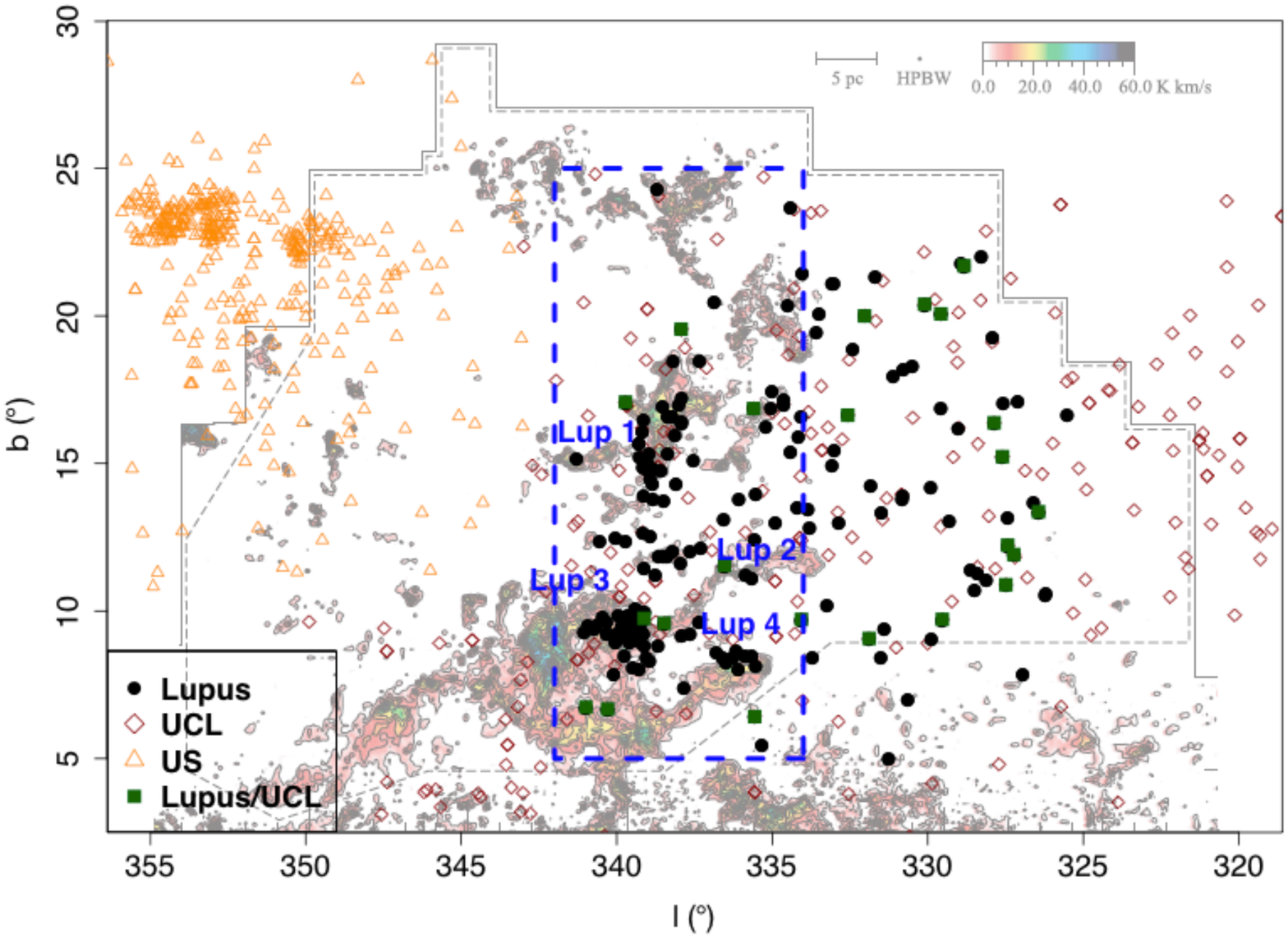}
\caption{Location of the 217 Lupus PMS stars overlaid on the $^{12}$CO intensity map from \citet{Tachihara(2001)}. Stars from US and UCL subgroups of the Sco-Cen association as given in \citet{deZeeuw(1999)} and \citet{Preibisch(2008)} are also included in this figure. Black filled circles denote the Lupus candidate stars and open symbols indicate US and UCL members. Stars marked with green filled squares were classified as both Lupus and UCL members in the literature. The blue dashed box encloses the Lupus star-forming clouds.}   
\label{fig3}
\vspace{1cm}
\end{center}
\end{figure*}

While the majority of the stars in our sample lie very close to the Lupus clouds, a fraction of the stars are scattered over a wider region (see Fig.~\ref{fig3}). Most of these stars were identified and classified as WTTSs by \citet{Krautter(1997)} and \citet{Wichmann(1997a),Wichmann(1997b)} from \textit{ROSAT} X-ray pointed observations and the ROSAT All Sky Survey (RASS). However, we mentioned that the Lupus SFR is located near the UCL subgroup of the Sco-Cen association, so one suspects that some of these stars might not be related to the Lupus SFR but might instead be part of the older Sco-Cen association. Indeed, a list of 81 low-mass candidate members of the UCL subgroup is given in the review by \citet{Preibisch(2008)}, and our sample of Lupus association candidate members includes 25 stars of their list. \cite{Preibisch(2008)} also claim that other stars previously identified as Lupus PMS stars by \cite{Krautter(1997)} and \cite{Wichmann(1997b)} appear to be UCL members because of their proper motions and positions in the HR-diagram.

It appears \citep{Preibisch(2008)} that the UCL stars are spread over a large extent on the sky, while the spatial distribution of Lupus stars exhibits two components: (i) an \textit{on-cloud} population concentrated in the immediate vicinity of the molecular clouds and (ii) a more dispersed \textit{off-cloud} population surrounding the clouds. The corresponding region spans the range of Galactic coordinates $334^{\circ}\leq l\leq342^{\circ}$ and $5^{\circ}\leq b\leq25^{\circ}$ and contains 160 stars, while the off-cloud population in our sample is located in the region with  $l<334^{\circ}$, which contains 57 stars. There is some arbitrariness involved in defining these regions, but the idea of separating our sample into two groups turned out to be necessary to our analysis, as is seen below, and has also been suggested by \citet{Preibisch(2008)}, who wrote \textit{``... it is probably wise for astrophysical studies to separate the on- or near-cloud Lupus members from the off-cloud UCL/US members."}. We show that the off-cloud population is indeed a mix of Lupus and UCL stars in Sect.~5.

In Table~\ref{tab2} we present the median positions, proper motions, and radial velocities for the Lupus subgroups. We find good agreement between the proper motions of the various star-forming clouds (Lupus 1 - 4). When comparing the proper motions of on-cloud and off-cloud stars, we note a difference of about 7~mas/yr in right ascension, although the median values for both populations are perfectly compatible in declination. One possibility to explain this difference is the existence of field stars or UCL members in the off-cloud sample as mentioned before. However, the reported difference of 7~mas/yr is still consistent e.g. with the observed extreme proper motion values for the various subgroups of the Taurus complex where a single UVW velocity value can be adopted assuming a velocity dispersion of about 1~km/s among the  different subgroups \citep[see][]{Luhman(2009)}. In the case of radial velocities, we consider both radial velocities from the literature and additional measurements derived in this paper (to be discussed in Sect.~4). Binaries and stars with insignificant measurements (i.e., $\sigma_{V_{r}}\geq V_{r}$) are excluded from the analysis presented in Table~\ref{tab2}. We note that stars in the off-cloud region exhibit radial velocities that are slightly higher than the values observed for the on-cloud population, which can be explained by geometrical effects, because both populations are at different angular separations from the CP. In the following, we assume that the Lupus stars spread over the various subgroups are comoving and use the CP search method to identify a moving group of PMS stars in this SFR.

%TABLE 2
\begin{table*}[!btp]
\centering
\caption{Median positions, proper motions, radial velocities, and the number of stars for the various subgroups in Lupus.}
\label{tab2}
\begin{tabular}{lcccccccc}
\hline
Sample&$\alpha$&$\delta$&$l$&$b$&$\mu_{\alpha}\cos\delta$&$\mu_{\delta}$&$V_{r}$&Stars\\
&(h:m:s) &($^{\circ}$ $^\prime$ $^\prime$$^\prime$)&($^{\circ}$)&($^{\circ}$)&(mas/yr)&(mas/yr)&(km/s)&\\
\hline
	Lupus 1	&	15	46	42.1	&	-35	00	46	&	338.8	&	15.3	&	-14.0	&	-21.1	&	$2.5\pm1.6$	&	30	\\
	Lupus 2	&	15	56	02.1	&	-37	56	06	&	338.6	&	12.0	&	-15.0	&	-23.1	&	$2.2\pm0.9$	&	15	\\
	Lupus 3	&	16	08	53.2	&	-39	05	34	&	339.5	&	9.4	&	-12.0	&	-20.4	&	$1.0\pm0.7$	&	73	\\
	Lupus 4	&	15	59	16.5	&	-41	57	10	&	336.2	&	8.5	&	-10.0	&	-20.8	&	$0.3\pm3.8$\tablefootmark{a}	&	9	\\
\hline
	Lupus (on-cloud)	&	16	00	47.0	&	-38	48	54	&	339.1	&	9.7	&	-13.0	&	-21.6	&	$2.5\pm0.4$	&	160\tablefootmark{b}	\\
	Lupus (off-cloud)	&	15	12	39.8	&	-40	50	52	&	329.9	&	14.0	&	-20.0	&	-21.8	&	$4.6\pm0.4$	&	57	\\
\hline
	Lupus (full sample)	&	15	49	30.7	&	-38	59	48	&	338.2	&	11.0	&	-16.0	&	-21.7	&	$3.7\pm0.4$	&	217	\\
\hline
\end{tabular}
\tablefoot{
The uncertainties in the median proper motion values for each group are about 1-2~mas/yr.\tablefoottext{a}{The radial velocity value presented for Lupus~4 is based on only two stars and should be regarded with caution.}
\tablefoottext{b}{The number of stars in the Lupus star-forming clouds (Lupus 1 - 4) do not add to the total number of stars (160 stars) in the on-cloud population (see definition in Sect.~3.2), because those stars that are spread beyond the limits of these clouds  (see $^{12}$CO intensity map in Fig.~\ref{fig3}) were not assigned to any cloud. \vspace{1cm}  }
}
\end{table*}

%FIGURE 3

%----------------------------------------------------------------------------------------------------------------									4. RADIAL VELOCITIES
%----------------------------------------------------------------------------------------------------------------

\section{Radial velocities} 

We mentioned that stellar radial velocities (RVs) are needed to determine individual parallaxes. We summarize here our search for RVs in the literature and some additional observations for stars with unknown RVs.

\subsection{Radial velocities from the literature}

We searched the CDS databases to access RV information for the stars in our sample. The search made use of the data mining tools available on the CDS site. We also looked for published RV data that are not available via the web-based CDS service. Our search for RVs, made as exhaustive as possible, is based on
\citet{Herbig(1988)},
\citet{Gregorio-Hetem(1992)},
\citet{Barbier-Brossat(1994)},
\citet{Duflot(1995)},
\citet{Dubath(1996)},
\citet{Grenier(1999)},
\citet{Wichmann(1999)},
\citet{Barbier-Brossat(2000)},
\citet{Madsen(2002)},
\citet{Melo(2003)},
\citet{Nordstrom(2004)},
\citet{Bobylev(2006)},
\citet{Gontcharov(2006)},
\citet{James(2006)},
\citet{Malaroda(2006)},
\citet{Torres(2006)},
\citet{Guenther(2007)},
\citet{Kharchenko(2007)}, and
\citet{White(2007)}.

\begin{figure*}[!btp]
\begin{center}
\includegraphics[width=0.33\textwidth]{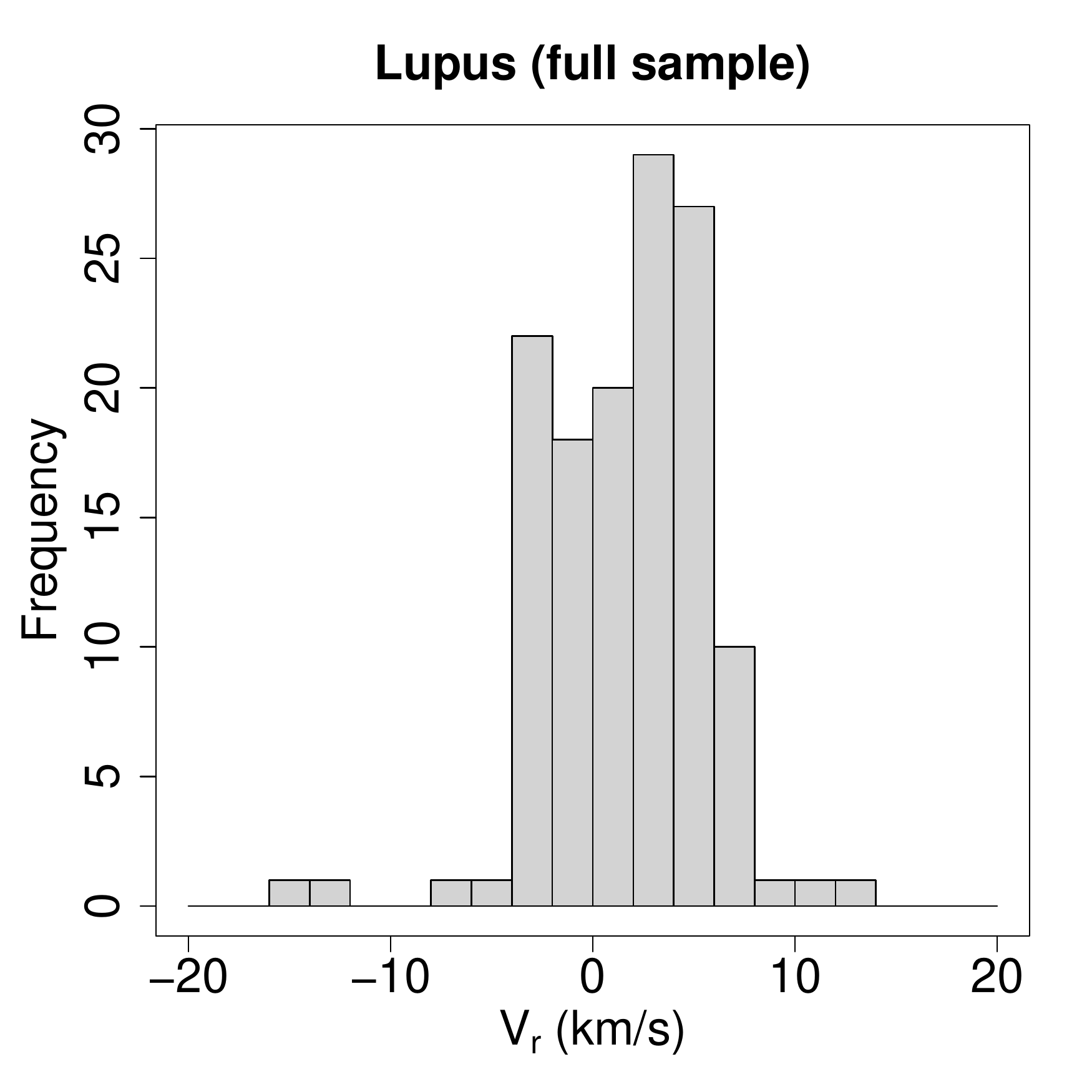}
\includegraphics[width=0.33\textwidth]{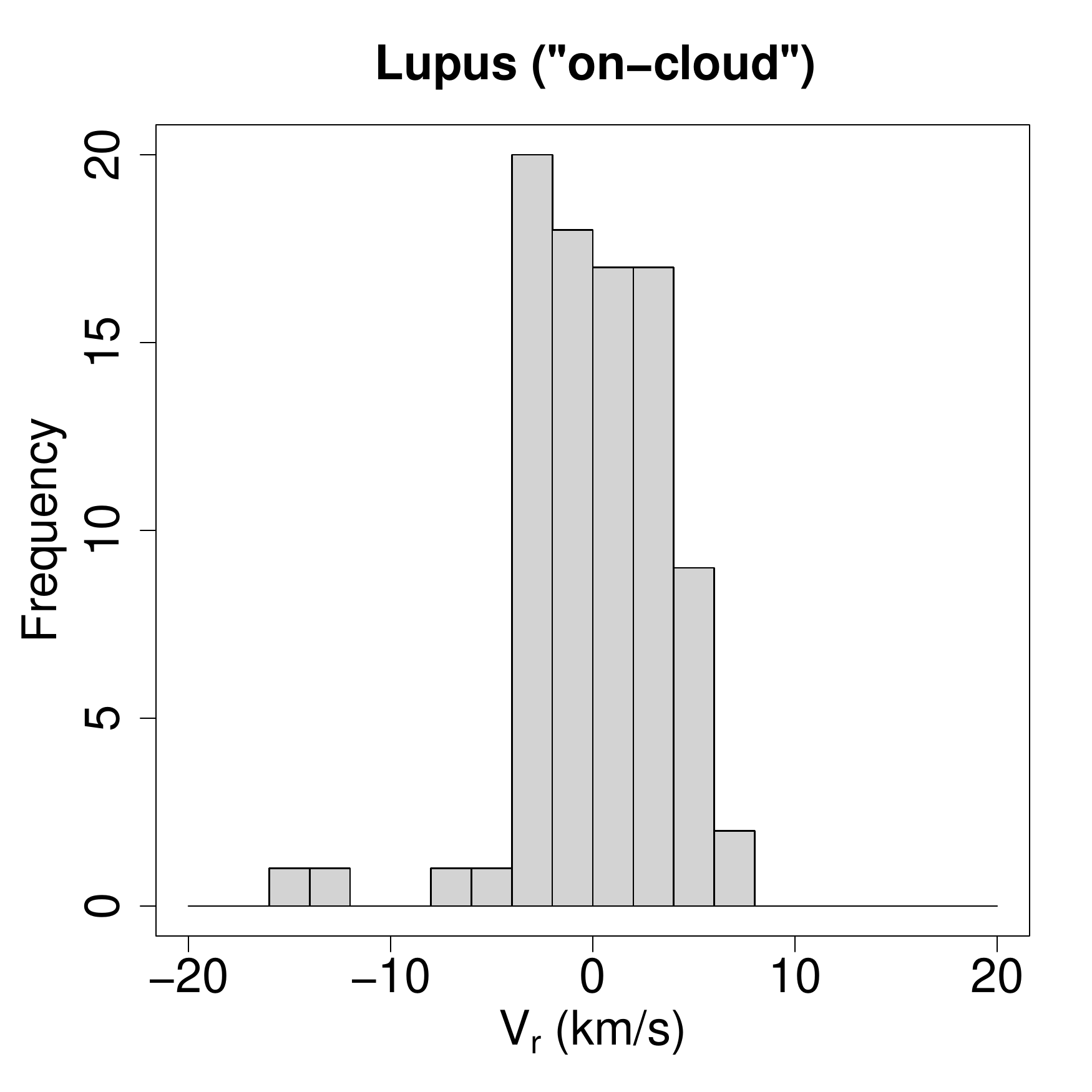}
\includegraphics[width=0.33\textwidth]{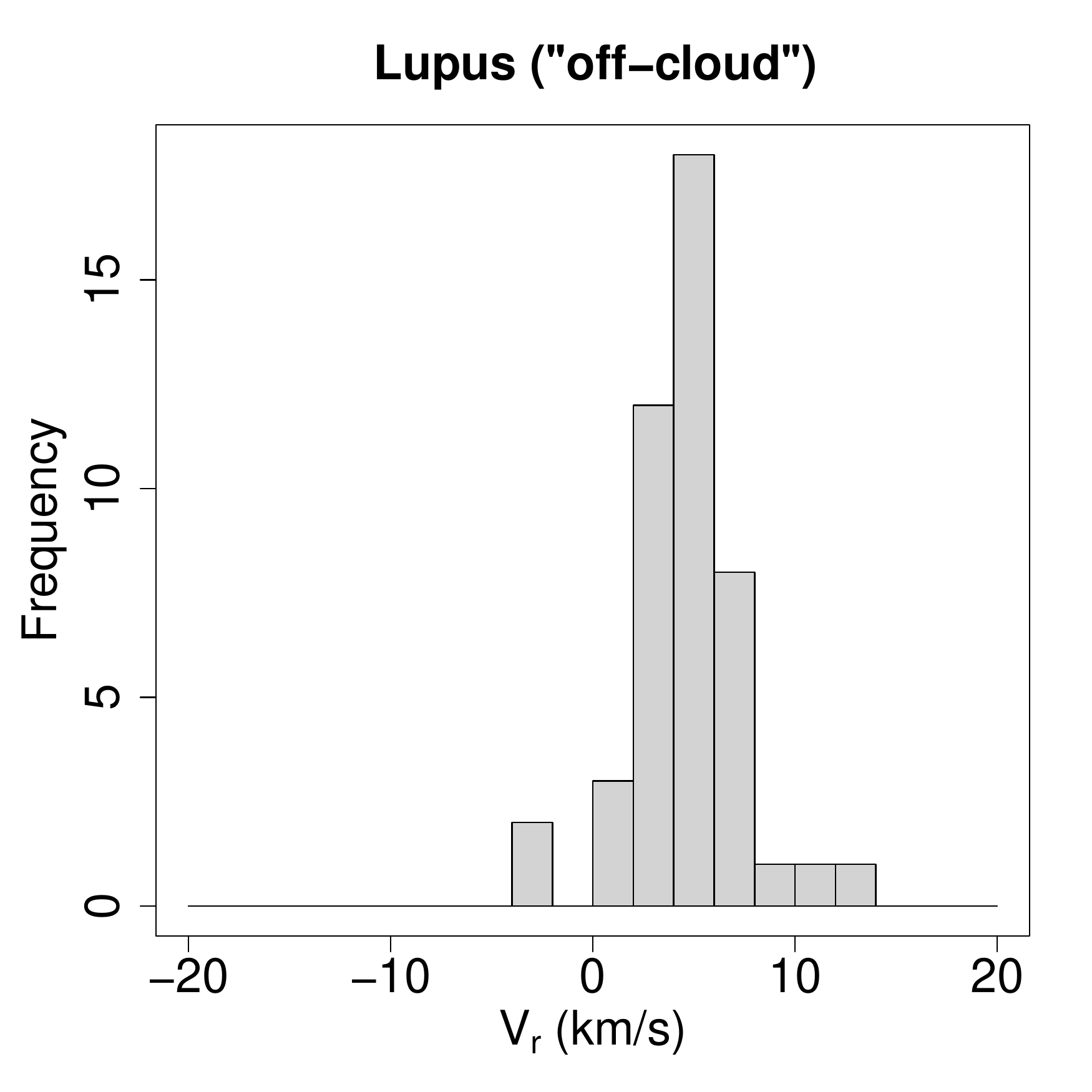}
\caption{\label{fig4}
Histogram of RVs found in the literature for Lupus stars.
}
\end{center}
\end{figure*}

We found RVs for only 108 stars of the full sample (on-cloud and off-cloud populations), which reflects the scarcity of this measurement in the literature. This is because previous spectroscopic investigations in Lupus have often focused on the few bright stars of this region. Figure~\ref{fig4} displays the RV distribution in our sample. 

%FIGURE 4

%----------------------------------------------------------------------------------------------------------------
\subsection{Additional radial velocity observations}

The scarcity of measured RVs is the main limitation in deriving individual distances in this work. To increase the number of Lupus stars with known RV information, we thus performed spectroscopic observations with the high-resolution $(R=48000)$ FEROS \citep{Kaufer(1999)} \'echelle spectrograph mounted at the ESO/MPG 2.2m telescope operated at La Silla (Chile). In addition to its high performance, FEROS provides a full wavelength coverage in the optical region (between 3500{\AA} and 9200{\AA}) over 39 spectral orders. The observations were taken in object calibration mode, which allows acquiring simultaneous spectra of the object and of the ThAr cell. Exposure times ranged from 5~min to 60~min, so that a S/N of about 20-30 was achieved. We observed 52 stars spread over the Lupus and Ophiuchus\footnote{The kinematic properties of the Ophiuchus SFR will be presented in a companion paper (Galli et al., in preparation). However, the results of our observations are presented in this paper since only a few stars of that region were observed in our program.  } SFRs during the nights of April 29 to May 05, 2011. Binaries and stars belonging to multiple systems as given in the literature were not included in our list of targets. The observed spectra were reduced with the standard FEROS data reduction pipeline, which performs bias subtraction, flat-fielding, scattered-light removal, \'echelle-order extraction, barycentric velocity correction, and wavelength calibration of the spectra. The extracted wavelength-calibrated spectra for each spectral order were not merged into a single spectrum, but used as 39 separate spectra (one per order) to determine the RV of the target. This procedure enabled us to eliminate the largest noise contributions that come from the orders in the red (orders $\leq 7$) and blue (orders $\geq30$) regions of the spectra \citep[see][]{Setiawan(2003)}.

We derived RVs by cross-correlating the reduced spectra of program stars with template spectra. To maximize the quality of our RV measurements we used both a standard-star spectrum and a numerical mask as template spectrum.

The cross-correlation with a stellar template utilized the standard star HD~82106 ($V=7.2$~mag, K3V), which was observed every night and whose RV is known to a high accuracy \citep[$V_{r}=+29.75\pm 0.05$~km/s,][]{Udry(1999)}. We computed the RVs of program stars by cross-correlating their spectrograms order-by-order with one reference spectrogram of HD~82106 (taken on May 4, 2011) using the IRAF task \rm\textit{{fxcor}}. We obtained the RV for each order separately and then averaged these values. The errors of our RV measurements were determined from the variance of all orders considered. Orders with discrepant values (due to a lower S/N) were obviously not considered. To investigate the accuracy of our results we calculated the RV of HD~82106 as derived from our observations. We cross-correlated the reference spectrogram with all spectrograms of HD~82106 spread over the whole observing campaign. The absolute RV of the reference spectrum was determined by cross-correlating it with the solar spectrum. The mean RV derived from our observations is $V_{r}=+29.71\pm 0.05$~km/s, in good accordance with \citet[][]{Udry(1999)}.

The other alternative to computing RVs consisted of a cross-correlation with a box-shaped binary template. The procedure follows the method outlined in \citet{Baranne(1996)}, which fits a Gaussian to the cross-correlation function of each order. The center of the Gaussian then gives the RV of the target \citep[see][for more details]{Weise(2010)}\nocite{Weise(2010)_thesis}. To better compare the RVs derived for HD~82106 using both techniques, we used a binary template of a G2V star. Our choice of the template was motivated by the use of the solar spectrum to derive the absolute RV of HD~82106 in our first approach (see above). For program stars, we used a K0V star template that is more representative of our targets' spectral types. The mean RV derived with this alternative strategy is $V_{r}=+29.74\pm 0.04$~km/s, which is perfectly consistent with the values mentioned above.

Using both a standard-star spectrum and a numerical template is valuable for gaining confidence in the derived results. To spot possible errors, the RVs of program stars were computed using only the common (not rejected) orders in both procedures. The final RV of our targets is the average of two independent values (one for each method). The uncertainties are calculated by propagating the individual errors and they are of a few hundred m/s. When one of the methods failed to return a RV value (due to low S/N in some orders) we only considered the result derived by the other method. A more realistic idea of our external precision is given in Table~\ref{tab3} by comparing the RVs derived in this work with published results for a control group of 3 stars in our sample with previously known RVs. We conclude that our results are fully compatible with the values found in the literature, with a rms uncertainty of about 500~m/s. We present the RVs of 52 stars belonging to the Lupus and Ophiuchus SFRs in Table~\ref{tab4}, together with the Li~I and H$\alpha$ equivalent widths (EWs) derived from our observations.

%TABLE 3
\begin{table}[!h]
\centering
\caption{Comparison of RVs derived in this paper with those published in the literature. 
\label{tab3}}
\resizebox{9cm}{!} {
\begin{tabular}{cccc}
%\hline
&This Work&Literature&\\
\hline\hline
Star&$V_{r}$&$V_{r}$&Reference\\
&(km/s)&(km/s)&\\
\hline
\hline
RXJ1524.5-3652 & $3.75\pm0.33$ &$4.2\pm1.0$ &\citet{Wichmann(1999)}\\
&&$4.1\pm0.4$&\citet{Torres(2006)}\\	
&&$4.4\pm0.4$&\citet{James(2006)}\\
&&$4.6\pm0.4$&\citet{James(2006)}\\
\hline
RXJ1547.6-4018 & $2.85\pm0.37$ &$2.8\pm1.0$ &\citet{Wichmann(1999)}\\
&&$3.2\pm0.1$&\citet{Torres(2006)}\\	
&&$3.1\pm0.4$&\citet{James(2006)}\\
&&$3.2\pm0.4$&\citet{James(2006)}\\
\hline
RXJ1538.0-3807 & $2.46\pm0.28$ &$3.4\pm1.0$ &\citet{Wichmann(1999)}\\
&&$3.0\pm0.2$&\citet{Guenther(2007)}\\	
\hline
\end{tabular}
}
\end{table}

%TABLE 4
\begin{table*}[!btp]
\centering
\caption{Results from our observations.
\label{tab4}}
\begin{tabular}{lccccccc}
\hline
Star & $\alpha$& $\delta$& $V_{r}$ & $EW(Li)$ & $EW(H\alpha)$ &Remarks\\ 
& (h:m:s) &($^{\circ}$ $^\prime$ $^\prime$$^\prime$)  & (km/s) & ({\AA}) & ({\AA})&   \\

\hline
&&&\textit{Lupus}&&\\
\hline

RXJ1448.2-4103&14 48 13.3&-41 02 58&$+5.95\pm0.17$&$0.336\pm0.017$&$\lesssim 0.1$&WTTS\\

RXJ1452.4-3740&14 52 26.2&-37 40 08&$+5.00\pm0.40$&$0.402\pm0.003$&$\lesssim 0.1$&WTTS\\

RXJ1454.2-3955&14 54 11.3&-39 55 23&$+3.83\pm0.38$&$0.382\pm0.002$&$-0.68\pm0.04$&WTTS\\

RXJ1502.4-3405&15 02 26.0&-34 05 13&$+3.88\pm0.17$&$0.437\pm0.001$&$-0.27\pm0.08$&WTTS\\

RXJ1505.4-3857&15 05 25.9&-38 57 03&$+4.92\pm0.33$&$0.416\pm0.004$&$-0.92\pm0.08$&WTTS\\

RXJ1505.9-4311&15 05 56.9&-43 12 02&$+3.99\pm0.27$&$0.465\pm0.001$&$-0.99\pm0.05$&WTTS\\

RXJ1506.7-3047&15 06 42.6&-30 47 33&$+0.81\pm0.24$&$0.480\pm0.002$&$-1.82\pm0.13$&WTTS\\

RXJ1508.0-3338&15 08 05.1&-33 37 55&$+1.65\pm0.23$&$0.447\pm0.010$&$-0.55\pm0.02$&WTTS\\

RXJ1508.4-3338&15 08 25.0&-33 37 55&$+1.85\pm0.43$&$0.438\pm0.011$&$-1.29\pm0.11$&WTTS\\

RXJ1508.8-3715&15 08 53.8&-37 15 46&$+4.51\pm0.28$&$0.421\pm0.012$&$-0.50\pm0.06$&WTTS\\

RXJ1515.1-4438&15 15 09.3&-44 38 36&$+6.34\pm0.34$&$0.367\pm0.012$&$-1.11\pm0.05$&WTTS\\

RXJ1518.0-4445&15 18 01.3&-44 44 26&$+5.97\pm0.15$&$0.099\pm0.007$&$0.81\pm0.05$&WTTS\\

RXJ1524.5-3652&15 24 32.4&-36 52 02&$+3.75\pm0.33$&$0.344\pm0.001$&$\lesssim 0.1$&WTTS\\

RXJ1526.8-3721&15 26 52.6&-37 22 06&$+1.92\pm0.21$&$0.505\pm0.002$&$-2.25\pm0.15$&WTTS\\

RXJ1529.8-4523&15 29 48.9&-45 22 45&$+4.64\pm0.41$&$0.442\pm0.008$&$-0.75\pm0.03$&WTTS\\

RXJ1534.3-3300&15 34 23.2&-33 00 09&$+1.40\pm0.20$&$0.525\pm0.008$&$-1.11\pm0.08$&WTTS\\

RXJ1538.0-3807&15 38 02.7&-38 07 23&$+2.46\pm0.28$&$0.423\pm0.021$&$-1.93\pm0.44$&WTTS\\

RXJ1539.7-3450&15 39 46.4& -34 51 02&$+5.39\pm0.29$&$0.284\pm0.013$&$-0.32\pm0.02$&WTTS\\

RXJ1542.0-3601&15 42 05.2&-36 01 32&$-0.05\pm0.26$&$0.492\pm0.011$&$-0.81\pm0.03$&WTTS\\

RXJ1544.5-3521&15 44 35.3&-35 21 49&$+2.47\pm0.23$&$0.491\pm0.008$&$-1.37\pm0.07$&WTTS\\

RXJ1547.1-3540&15 47 08.4&-35 40 19&$+0.82\pm 4.90$:&$0.472\pm0.016$&$-1.09\pm0.39$&WTTS\\

RXJ1547.6-4018&15 47 41.8&-40 18 26&$+2.85\pm0.37$&$0.387\pm0.012$&$0.17\pm0.01$&WTTS\\

RXJ1548.0-4004&15 48 02.1&-40 04 28&$+2.07\pm0.35$&$0.433\pm0.014$&$-2.01\pm0.06$&WTTS\\

RXJ1548.9-3513&15 48 54.1&-35 13 18&$+0.62\pm0.16$&$0.372\pm0.002$&$-0.41\pm0.03$&WTTS\\

RXJ1601.8-4026&16 01 49.5&-40 26 19&$+2.87\pm0.46$&$0.371\pm0.011$&$-1.39\pm0.08$&WTTS\\

RXJ1606.3-4447&16 06 23.4&-44 47 35&$+4.68\pm0.54$&$0.455\pm0.007$&$-0.48\pm0.06$&WTTS\\

RXJ1608.0-3857&16 08 00.0&-38 57 51&$-2.42\pm 0.84$:&$0.638\pm0.002$&$-1.91\pm0.14$&WTTS\\

F304&16 08 11.0&-39 10 46&$+2.76\pm0.11$&$0.449\pm0.001$&$-0.66\pm0.02$&WTTS\\

V908Sco&16 09 01.9&-39 05 12&$-0.77\pm 0.82$:&$0.570\pm0.004$&$-51.24\pm4.20$&CTTS\\

RXJ1609.9-3923&16 09 54.0&-39 23 27&$-0.21\pm0.46$&$0.550\pm0.003$&$-22.33\pm0.93$&CTTS\\

RXJ1611.6-3841&16 11 38.0&-38 41 35&$+2.52\pm0.39$&$0.438\pm0.007$&$-3.93\pm0.31$&WTTS\\

RXJ1615.9-3947&16 15 56.7&-39 47 16&$+0.09\pm0.27$&$0.456\pm0.011$&$-2.37\pm0.23$&WTTS\\

RXJ1615.9-3241&16 15 57.0&-32 41 24&$-0.45\pm0.26$&$0.458\pm0.010$&$-0.68\pm0.06$&WTTS\\

HD147454&16 23 32.3&-34 39 50&$-0.10\pm0.41$&$0.112\pm0.004$&$2.32\pm0.15$&WTTS\\

SAO207620&16 23 37.7&-34 40 21&$-0.41\pm0.14$&$0.195\pm0.005$&$0.72\pm0.09$&WTTS\\

\hline
&&&\textit{Ophiuchus}&&\\
\hline

GSC6780-1061&16 06 54.4&-24 16 11&$-5.43\pm0.19$&$0.561\pm0.001$&$-1.55\pm0.09$&WTTS\\

GSC6793-994&16 14 02.1&-23 01 02&$-2.28\pm0.51$&$0.356\pm0.013$&$\lesssim 0.1$&WTTS\\

PDS145&16 14 20.9&-19 06 05&$-7.67\pm 10.85$:&$0.261\pm0.007$&$-65.94\pm2.30$&CTTS\\

RXJ1620.7-2348&16 20 46.0&-23 48 21&$-3.21\pm0.21$&$0.465\pm0.002$&$-0.37\pm0.07$&WTTS\\

RXJ1621.4-2312&16 21 28.5&-23 12 11&$-8.37\pm 2.24$&$0.588\pm0.011$&$-1.27\pm0.09$&WTTS\\

Haro1-1&16 21 34.7&-26 12 27&$-4.29\pm0.50$&$0.455\pm0.024$&$-153.60\pm2.50$&CTTS\\

GSC6794-537&16 23 07.8&-23 01 00&$-10.05\pm 1.41$:&$0.491\pm0.011$&$-0.52\pm0.10$&WTTS\\

GSC6794-156&16 24 51.4&-22 39 32&$-5.27\pm 1.72$:&$0.338\pm0.001$&$-0.90\pm0.18$&WTTS\\

RXJ1625.4-2346&16 25 28.6&-23 46 27&$-10.38\pm0.64$&$0.381\pm0.001$&$0.24\pm0.01$&WTTS\\

DoAr25&16 26 23.7&-24 43 14&$-8.25\pm0.59$&$0.547\pm0.002$&$-8.28\pm0.50$&CTTS\\

RNO90&16 34 09.2&-15 48 17&$-12.92\pm 7.92$:&$0.360\pm0.006$&$-78.00\pm2.40$&CTTS\\

He3-1254&16 46 44.3&-15 14 38&$-8.98\pm0.29$&$0.444\pm0.008$&$-97.14\pm4.40$&CTTS\\

WaOph6&16 48 45.6&-14 16 36&$-10.09\pm 0.54$:&$0.520\pm0.009$&$-24.19\pm0.95$&CTTS\\

WaOph5&16 49 00.8&-14 17 11&$-9.54\pm 3.04$:&$0.668\pm0.001$&$-54.07\pm7.31$&CTTS\\

V1725Oph&17 16 13.9&-20 57 46&$-11.77\pm 0.84$:&$0.636\pm0.007$&$-22.51\pm1.60$&CTTS\\
\hline
GSC6213-194&16 09 41.0&-22 17 59&&&&SB2, WTTS\\
RXJ1613.1-3804&16 13 12.7&-38 03 51&&&&SB2, WTTS\\

\hline
\end{tabular}
\tablefoot{The \textit{upper panels} present the RVs derived in this paper, together with EWs for the Li and $H\alpha$ lines. We also provide the TTS subclass based on $EW(H\alpha)$. The symbol ``:" indicates uncertain values of RVs where one of the cross-correlation techniques failed to return a result. Negative and positive values of $EW(H\alpha)$ denote that the line is in emission and absorption, respectively. The \textit{lower panel} presents those stars that show evidence of spectroscopic binarity (SB2).}
\end{table*}

Equivalent widths of the Li~I and H$\alpha$ lines were measured from our spectra using IRAF \textit{splot} routine. The presence of Li absorption in late-type stars is one of the primary criteria for stellar youth \citep[see][]{Basri(1991)}. The Li~I $\lambda$6708{\AA} resonance doublet is blended by Fe lines and the lithium isotope $^{6}$Li. It is not possible to separate individual lines at our spectral resolution ($\sim$~48000), so we see no evidence of these features. The lithium EWs were measured using both a Gaussian fit and direct integration. The difference between these values is smaller than 25~m{\AA}, which we consider to be the upper limit of our measurement errors. The contribution from the neighboring blending lines is expected to be smaller than the uncertainty of our results, which we estimated by varying the location of the continuum adjacent to the line.

The H$\alpha$ emission line is one the most prominent spectroscopic features in the visible spectra of CTTSs \citep{Joy(1945),Herbig(1962)}. The line profile is often complex and takes different shapes, which were studied and classified by \cite{Reipurth(1996)}. We  measured the H$\alpha$ EWs by direct integration and by fitting a Voigt profile (when the target exhibited a H$\alpha$ profile with a single peak). Our measurement errors are mainly caused by the uncertainty on the continuum level in the vicinity of the line. We use the standard limit of 10{\AA} \citep[see, e.g.,][]{Appenzeller(1989)} to distinguish between CTTSs (EW(H$\alpha)\geq$10{\AA})  and WTTSs (EW(H$\alpha)<$10{\AA}) and find that our sample contains 10~CTTSs and 42~WTTSs. A more detailed inspection of the spectra in our sample revealed two PMS spectroscopic binaries (SB2), for which no evidence of binarity could be found in the literature prior to our observations. We present the results of our observations in Table~\ref{tab4}.

%----------------------------------------------------------------------------------------------------------------									5. ANALYSIS  AND  RESULTS
%----------------------------------------------------------------------------------------------------------------
\section{Analysis and results} \label{Analysis}

In the following we derive the CP of the comoving group of Lupus stars considered in this paper. We first use the k-nearest neighbor \citep[k-NN,][]{Fix(1951),Venables(2002)} algorithm to distinguish between Lupus and UCL stars based on their position, then we apply the CP search method to the sample of Lupus candidate members and perform a membership analysis. The technique that we use to find the CP position of the Lupus moving group is the new CP search method that was recently developed by our team. We refer the reader to the original paper \citep{Galli(2012)} for more details on the implementation of this method.

%----------------------------------------------------------------------------------------------------------------
\subsection{Preliminary analysis} 

The velocity dispersion in Lupus is a first input parameter that we must determine for performing the CP analysis. The velocity dispersion of young moving groups is expected to be low, only a few km/s \citep{Mathieu(1986)}. Typical values of velocity dispersion in nearby SFRs are 1-2~km/s \citep{Jones(1979),Dubath(1996),Makarov(2007),Luhman(2009)}. We adopt $\sigma_{v}=1$~km/s  as the one-dimensional velocity dispersion of the Lupus moving group, and we come back to this point in Sect.~6. Another input parameter in the CP method is the mean distance to the moving group. As discussed in Sect.~2, the distance to the Lupus SFR has undergone substantial revision in the literature over the past decade. Here we use the value of $d=150$~pc that seems to be adequate for most of the clouds in the complex \citep{Comeron(2008)}. The CP search method is rather insensitive to small variations in both parameters.

When we apply the CP search method to the 217 stars in the sample, we end up with a CP position given by
\begin{center}
$(\alpha_{cp},\delta_{cp})=(93.3^{\circ},-25.7^{\circ})\pm(1.8^{\circ},2.6^{\circ})$
\end{center}
\noindent with 125 moving group members. In a recent paper, \citet{Makarov(2007)} investigated the kinematics of the Lupus sample of 93 PMS stars identified by \citet{Krautter(1997)} using the UCAC2 catalog \citep{UCAC2} and derived a velocity dispersion of $1.3$~km/s. The CP position presented in his work is 
$(\alpha_{cp},\delta_{cp})=(92.8^{\circ},-28.1^{\circ})\pm(3.1^{\circ},5.0^{\circ})$. 

We note that there is good agreement between both solutions, and it is probable that the small differences are caused by the different samples of Lupus stars, the different proper motion sources, and the different CP search methods. However, whether the above result is the most appropriate for the sample of TTSs associated with the Lupus molecular clouds is questionable since the derived CP solution might be affected by UCL stars misidentified as Lupus WTTSs. We investigate this question below.

%----------------------------------------------------------------------------------------------------------------
\subsection{The k-NN algorithm applied to the sample of Lupus stars} 

The k-NN method is a nonparametric machine learning algorithm used in pattern recognition for classifying objects. Although the k-NN method is not yet widely used in astronomy, it has already proven to be effective, such as for the photometric search of brown dwarfs \citep{Marengo(2009)} and estimations of photometric redshifts for quasars \citep{Ball(2007)}.  It is used in this work to segregate potential Lupus and UCL members in our \textit{test sample} of 217 stars based on their position with respect to both the star-forming clouds and confirmed members of each association.

In the classic version of the k-NN method, a test element is classified by the majority vote of its $k$ nearest neighbors from the \textit{training set}. To begin with, we define the training set $T=\{(\mathbf{x}_{1},y_{1}),(\mathbf{x}_{2},y_{2}), ..., (\mathbf{x}_{N},y_{N})\}$, where $\mathbf{x}_{i}$ is the vector of attributes (stellar positions), $\mathbf{x}_{i}=(\alpha_{i},\delta_{i})$, and $y_{i}\in \{c_{1}, c_{2}\}$ denote the class membership (Lupus or UCL). The training set used in this work includes both Lupus and UCL stars, and it is constructed as follows. Since most members of the Lupus SFR are already included in our test sample  (and many of them lack membership confirmation) we use the position of the 105 molecular clouds detected in $^{12}$CO by \citet{Tachihara(2001)} for this purpose, and the 81 stars given by \citet{Preibisch(2008)} as UCL members.\footnote{Although a larger sample of UCL members from the \textsc{Hipparcos} catalog exists \citep[see e.g.][]{deZeeuw(1999)}, we prefer at this stage to use only those provided by \citet{Preibisch(2008)} in order to keep approximately the same fraction of Lupus and UCL data points in the training set to avoid a classification bias. The remaining UCL stars will be used later in this section to investigate the accuracy of our procedure and determine the optimum value of $k$ (see text). } Stars in our test sample are classified as potential Lupus or UCL members based on the majority vote of their $k$ nearest neighbors where the value of $k$ is optimized to our specific problem (see below). In this context, the probability $p\,(y\mid x)$ for a given star with attributes $(\mathbf{x},y)$ is given by

\begin{equation}
p\,(y=c\mid x)=\frac{k_{C}}{k}\, 
\end{equation}
where $k_{C}$ denotes the number of nearest neighbors in the training set labeled with class $c$ (Lupus or UCL). 

One important point to be considered in the k-NN algorithm is the best choice of $k$. While higher values of $k$ would make the boundaries between Lupus and UCL less clear, a low value, on the other hand, can lead to noisy classification. An odd number for $k$ is preferable and avoids tied votes since we only have to distinguish between two classes. To gain confidence in the derived results and determine the optimum value for $k$, we construct a total of 1000 synthetic test samples to be classified by the k-NN method using the procedure described above. To do so, we use the well-known UCL members from the \textsc{Hipparcos} catalog given by \citet{deZeeuw(1999)}, and simulate synthetic Lupus training stars by generating stellar positions from a Gaussian distribution using the position of the $^{12}$CO-peak-integrated intensity and the radius of the molecular clouds as given in Table~1 of \citet{Tachihara(2001)}. At each run our routine randomly chooses Lupus and UCL test stars to construct samples with 200 stars and classify them based on our training set. The fraction of Lupus and UCL stars in the synthetic test samples is not fixed and varies at each iteration to avoid any bias when evaluating the accuracy of the procedure. 

The first results of our simulations, however, revealed that many Lupus synthetic stars located in the core of the star-forming clouds were misclassified. This is because some data points in our training set refer to single stars, while others in principle represent the center position of molecular clouds. Neither the size of the molecular clouds nor the distance to nearer neighbors are taken into account, showing that the majority vote criteria is not suited to our specific case. One way to solve this problem is to not give the same weight (vote) to all the neighbors. In this context, we implement a \textit{weighted k-NN algorithm} that uses a distance weighted function. We distinguish whether the positions $(\alpha_{i},\delta_{i})$ in the training set refer to molecular clouds (Lupus) or stars (UCL) by assigning a different weight. The weight function $\omega_{i}$ is given by 

\begin{equation*}
    \omega_{i} = \begin{cases}
               1/(d-r)\,,               & \textit{molecular cloud}\\
               1/d\, ,              & \textit{star}
           \end{cases}
\end{equation*}
where $r$ is the radius of the molecular cloud and $d$ is calculated from the position of the $^{12}$CO-peak-integrated intensity of the Lupus molecular clouds \citep[see][]{Tachihara(2001)} or the position of the UCL star in the training set. The angular distance $d$ between a test star with given position $(\alpha_{j},\delta_{j})$ and a training star $(\alpha_{i},\delta_{i})$ is given by 

\begin{equation}
d=\cos^{-1}[\sin\delta_{i}\sin\delta_{j}+\cos\delta_{i}\cos\delta_{j}\cos(\alpha_{i}-\alpha_{j})]\, .
\end{equation}
In this case the posterior probability for a given star in the test sample takes the weight function into account and it is defined as
\begin{equation}
p\, (y=c\mid x)=\frac{\sum^{k_{C}}_{i=1}\, \omega_{i}}{\sum^{k}_{i=1}\, \omega_{i}}\, , 
\end{equation}
where the sum in the numerator runs only over the nearest neighbors in the training set with class $c$ (Lupus or UCL). We use these probabilities to classify the stars in our test sample as potential Lupus or UCL candidate members.

\begin{figure}[!h]
\begin{center}
\includegraphics[width=0.5\textwidth]{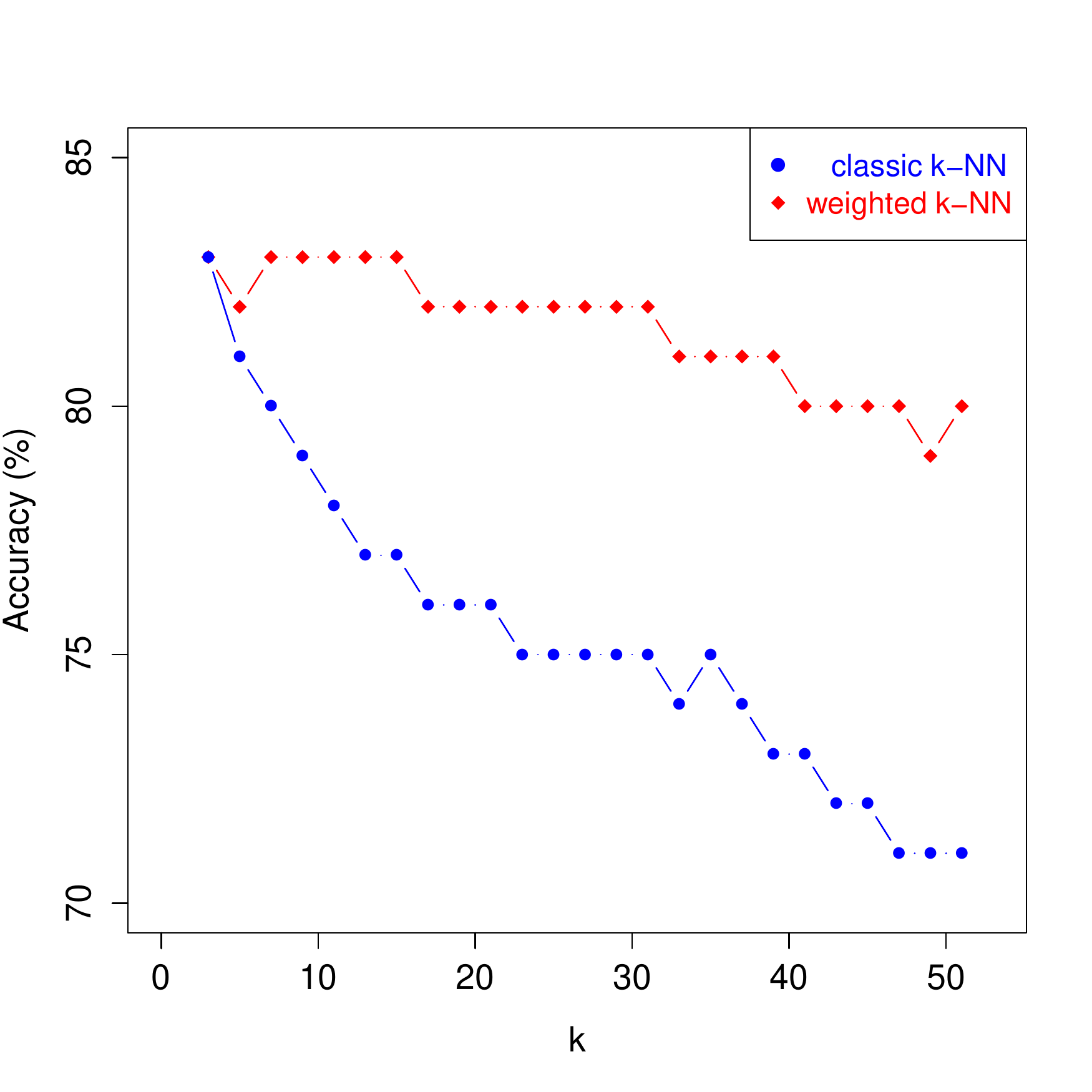}
\caption{\label{fig5}
Comparison of performance between the classic and weighted k-NN methods as a function of $k$ derived from our simulations. Each point represents an average value of 1000 iterations.   
 }
\end{center}
\end{figure}

\begin{figure*}[!btp]
\begin{center}
\includegraphics[width=0.59\textwidth,angle=-90]{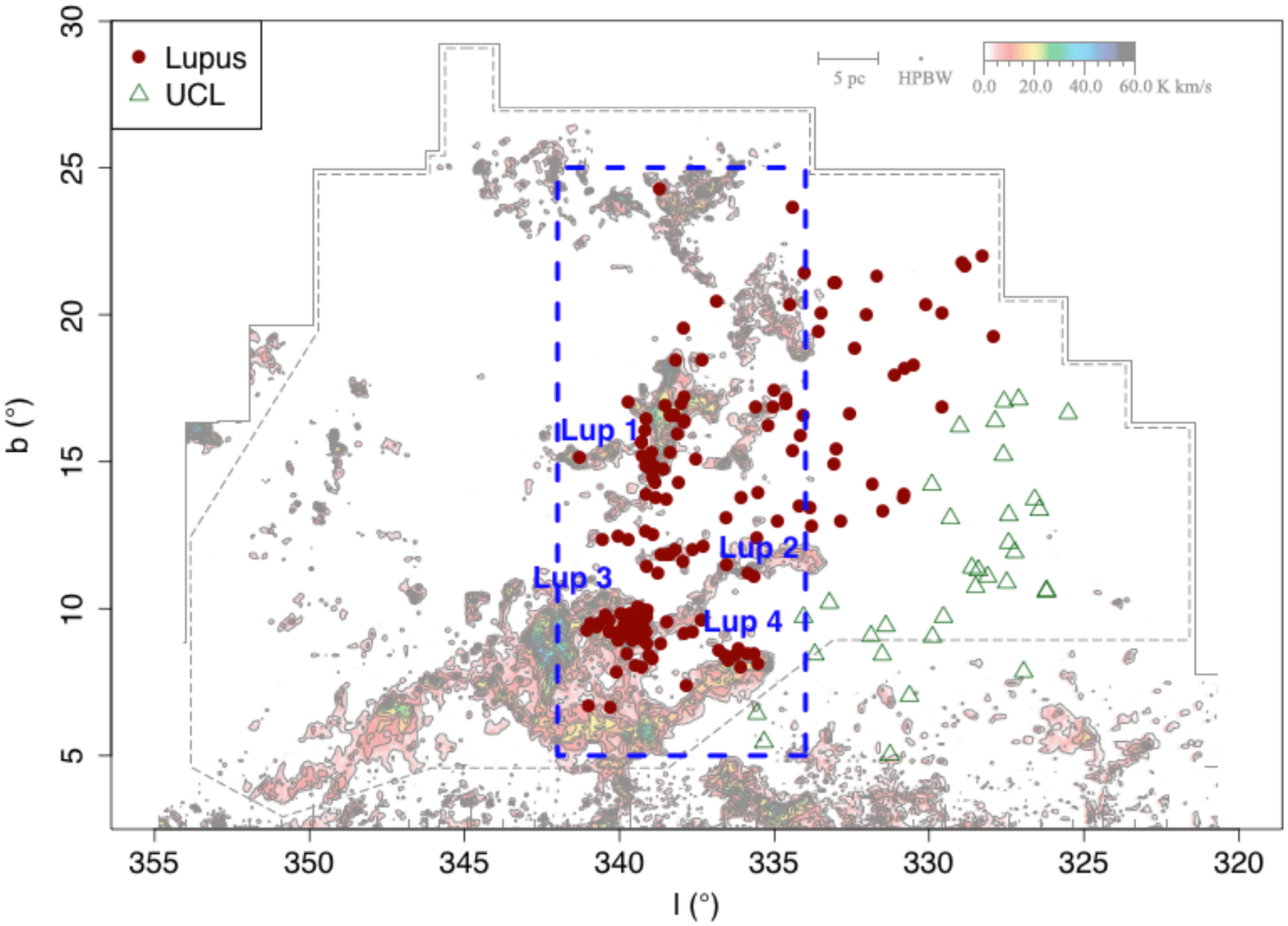}
\caption{
\label{fig6}
Location of the 217 PMS stars overlaid on the $^{12}$CO intensity map of \citet{Tachihara(2001)}. Different symbols and colors indicate the membership classification (Lupus or UCL) that results from our k-NN analysis. }
\end{center}
\end{figure*}

Figure~\ref{fig5} compares the performance of our refurbished weighted k-NN method with its classic version for various values of $k$. We conclude that our new strategy indeed yields better results and exhibits an accuracy of $\sim80\%$. The accuracy is defined by the fraction of test stars in our simulated test samples that are correctly classified by the k-NN algorithm. We conclude that any value of $k$ between 7 and 15 can be used in this work to segregate potential Lupus and UCL stars. We adopt $k=15$ under the assumption that more neighbors will help better define both groups. Doing so, we run the weighted k-NN method in our original test sample of 217 stars and identify 183~stars as potential Lupus members and 34~stars as potential UCL members. The distribution of Lupus and UCL stars is illustrated in Fig.~\ref{fig6} and shows that the majority of Lupus stars is located in the vicinity of the molecular clouds (as expected). 

%FIGURE 5

%----------------------------------------------------------------------------------------------------------------
\subsection{Convergent point analysis for the Lupus moving group} 

The procedure described in the previous section made it possible to separate some potential UCL members included in our initial sample of Lupus PMS population. However, in order to search for association members, proper motions must be considered at this stage. Our final analysis consists in running the CP search method on the sample of 183 stars classified by the k-NN method as Lupus stars (see Sect.~5.2). We identify a moving group with 109 members and CP located at 
\begin{center}
$(\alpha_{cp},\delta_{cp})=(112.2^{\circ},-48.6^{\circ})\pm(4.6^{\circ},3.6^{\circ})$
\end{center}
with chi-squared statistics $\chi^{2}_{red}=1.1$ (i.e., $\chi^{2}/\nu=118.8/107$). We note that the rejection of some likely UCL members in our sample dramatically changed the CP position of the moving group as compared to our first solution in Sect.~5.1, which now appears to be a mixed CP solution of Lupus and UCL stars, and is therefore not valid for either moving group. Tables~\ref{tab5}, \ref{tab6}, and \ref{tab7} present the 109 Lupus moving group members selected by the CP search method, together with their parallaxes (to be discussed in Sect.~6). We note that six stars that were classified as UCL members by \citet{Preibisch(2008)} and previously regarded as Lupus stars \citep{Krautter(1997),Wichmann(1997a),Wichmann(1997b)} have been accepted as Lupus members in our CP analysis. These stars are marked with the symbol ``*". On the other hand, we found that 19~stars previously regarded as WTTSs following their discovery in X-ray have been rejected in this analysis, and they are instead probable UCL members, as discussed by \cite{Preibisch(2008)}.

%----------------------------------------------------------------------------------------------------------------
\newpage
\subsection{Result check via Monte Carlo simulations}

\begin{figure*}[!btp]
\begin{center}
\sidecaption
\includegraphics[width=0.64\textwidth]{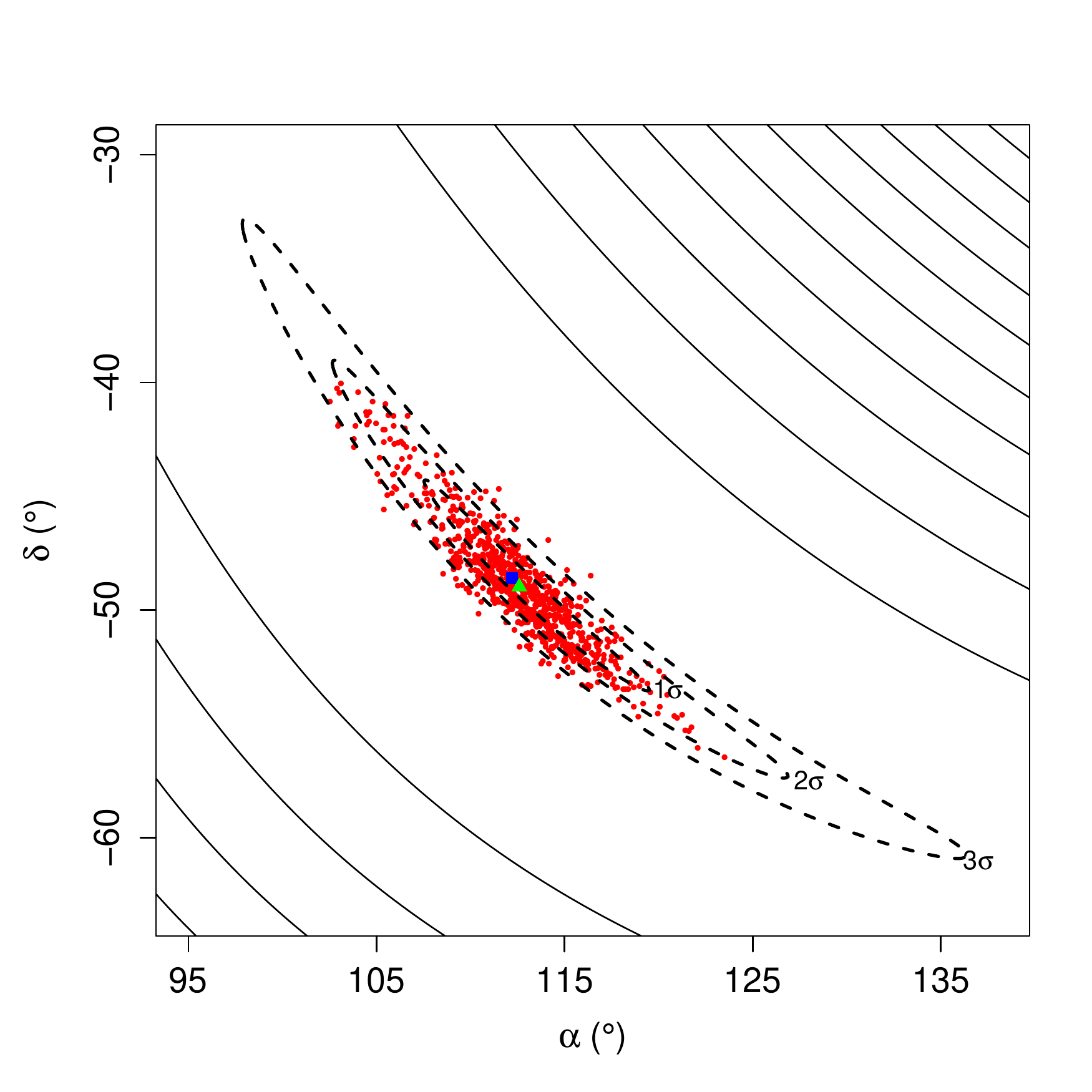}
\caption{
\label{fig7}
Realizations of CP coordinates for 1000 Monte Carlo simulations (red dots) overlaid on the $X^{2}$ contours (solid lines) for the CP solution derived in Sect.~5.3. The blue square denotes the CP coordinates for the Lupus moving group derived in that section, and the green triangle denotes the centroid distribution of simulated CPs (see Sect.~5.4). The dashed lines indicate the $1\sigma$, $2\sigma$, and $3\sigma$ contour levels of our CP solution.}
\end{center}
\end{figure*}

To assess the validity of the CP location presented above, we performed Monte Carlo simulations of the 109 Lupus moving group members. We constructed 1000 samples of moving groups by resampling the stellar proper motions from a Gaussian distribution where the mean and variance are equal to the individual stellar proper motions and its uncertainty (in both components). We ran the CP search method for each set of simulated stars and computed the CP of each moving group. The results of this study are presented in Fig.~\ref{fig7}. The CP solution derived in this paper is fully consistent with the centroid distribution of the Monte Carlo realizations, located at  
\begin{center}
$(\alpha_{cp},\delta_{cp})=(112.6^{\circ},-49.0^{\circ})\pm(3.2^{\circ},2.5^{\circ})$\, , 
\end{center}
\noindent and we therefore conclude that it is representative of the Lupus moving group.

%FIGURE 6

%FIGURE 7

%----------------------------------------------------------------------------------------------------------------									6. KINEMATIC PARALLAXES
%----------------------------------------------------------------------------------------------------------------
\section{Kinematic parallaxes}

\subsection{Parallaxes and space velocities of group members with known radial velocity}

Once the moving group is defined, it is possible to derive individual kinematic parallaxes $\pi_{ind}$ for group members if their RVs $V_{r}$ are known. Individual parallaxes are given by
\begin{equation}\label{eq_plx_ind}
\pi_{ind}=\frac{A\,\mu_{\parallel}}{V_{r}\tan\lambda},
\end{equation}
where $A=4.74047$~km~yr/s is the ratio of one astronomical unit in km to the number of seconds in one Julian year, $\lambda$ is the angular distance from the CP position to a given star in the moving group, and $\mu_{\parallel}$ the stellar proper motion component that points towards the CP  \citep[see][for more details]{Galli(2012)}. The parallax uncertainty is derived by error propagation of this equation and takes the error budget of proper motions, RVs, and the CP into account (see Appendix~A for more details).

We found RVs for only 60 stars in the sample of 109 moving group members. We reject (resolved) binaries since it will not be possible to derive their parallaxes from a single RV measurement. Stars that exhibit poor RVs because of their errors are also excluded from this analysis. The observed RV for Lupus stars is expected to be low and a small variation accounts for a more significant shift in the parallax and space velocity (see Appendix~B for more details). To spot possible errors on parallaxes and velocities, we define lower and upper limits for the space velocity of Lupus stars. To do so, we use five stars\footnote{Sz~120 is not considered in this analysis because it is a binary HAeBe star \citep{Correia(2006)}. } with known trigonometric parallax in the \textsc{Hipparcos} catalog that have been selected as moving group members in our CP analysis. The space velocities derived using \textsc{Hipparcos} parallaxes range from $V_{lower}\simeq17$~km/s and $V_{upper}\simeq33$~km/s, which define the limits for the space velocity of Lupus stars. This leaves us with a sample of 19 stars (see Table~\ref{tab5}) that we define here as the Lupus \textit{core moving group}. The mean parallax is $\overline{\pi}=6.8\pm0.4$~mas, which agrees well with the estimated distance of 150~pc assumed in the CP analysis (see Sect.~5.1).

\begin{figure*}[!btp]
\begin{center}
\includegraphics[width=0.24\textwidth]{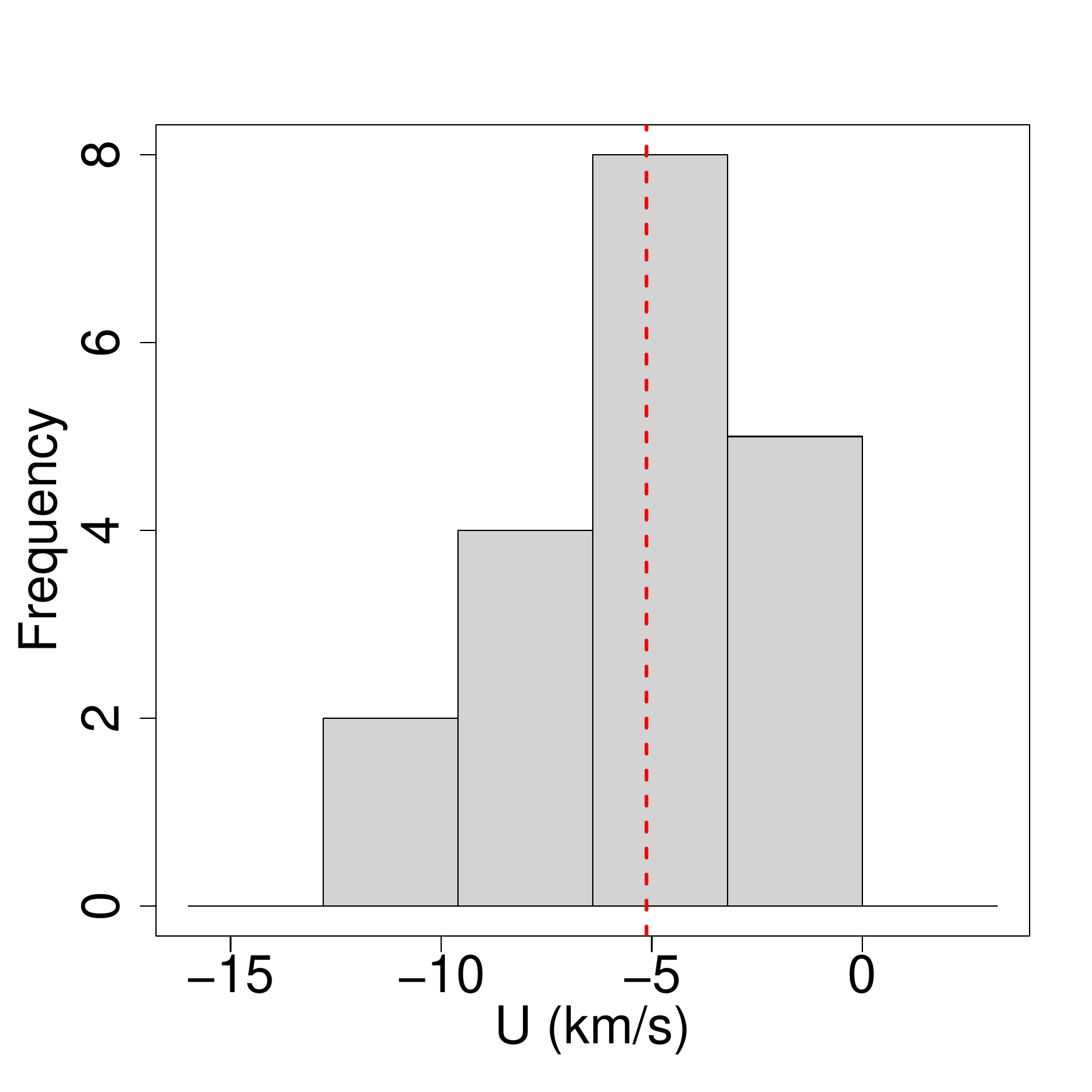}
\includegraphics[width=0.24\textwidth]{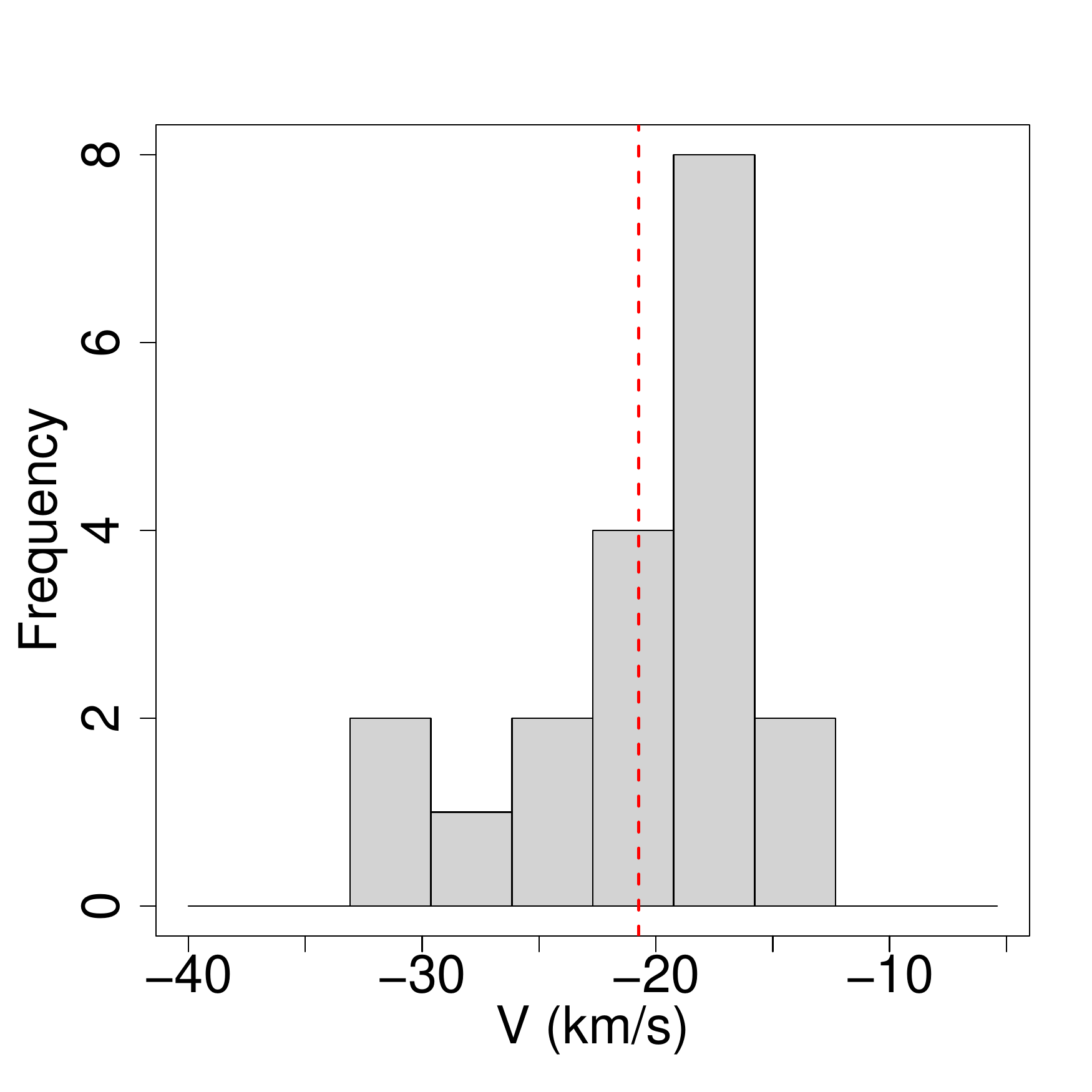}
\includegraphics[width=0.24\textwidth]{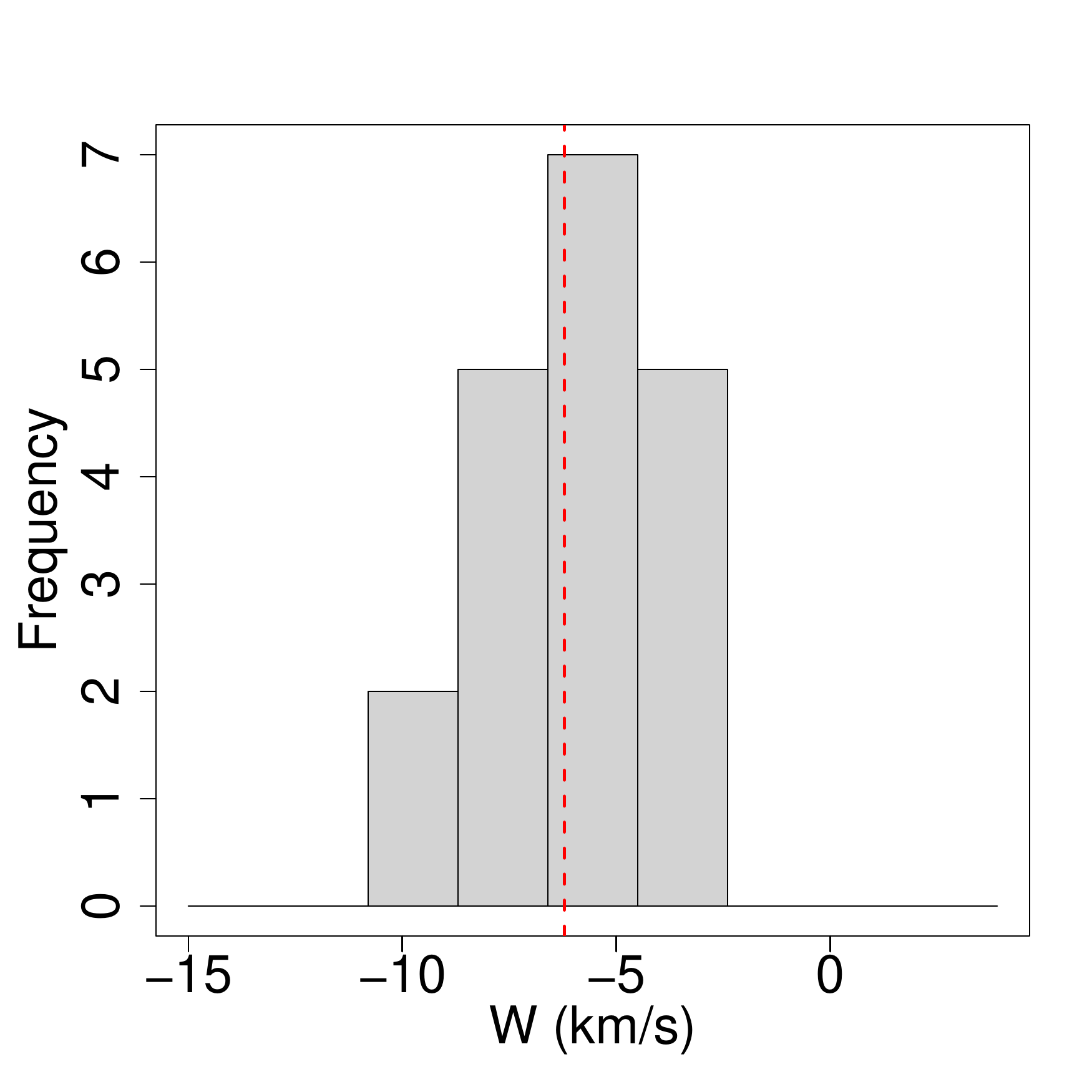}
\includegraphics[width=0.24\textwidth]{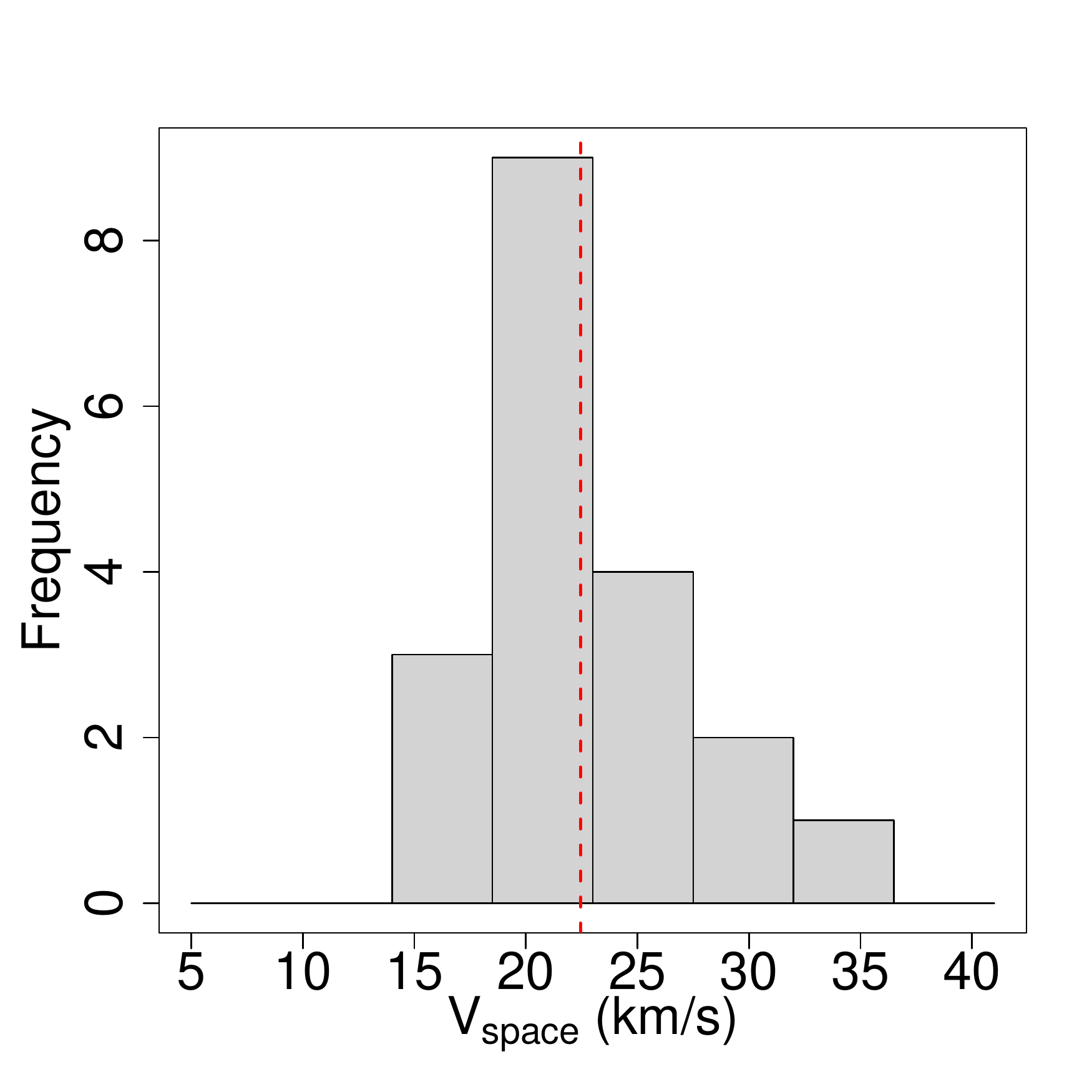}
\caption{Histograms of the Galactic velocity components for members of the Lupus moving group with known RVs. The red dashed line denotes the average values given in Sect.~6.1. 
\label{fig8} }
\end{center}
\end{figure*}

\begin{figure*}[!btp]
\begin{center}
\includegraphics[width=0.33\textwidth]{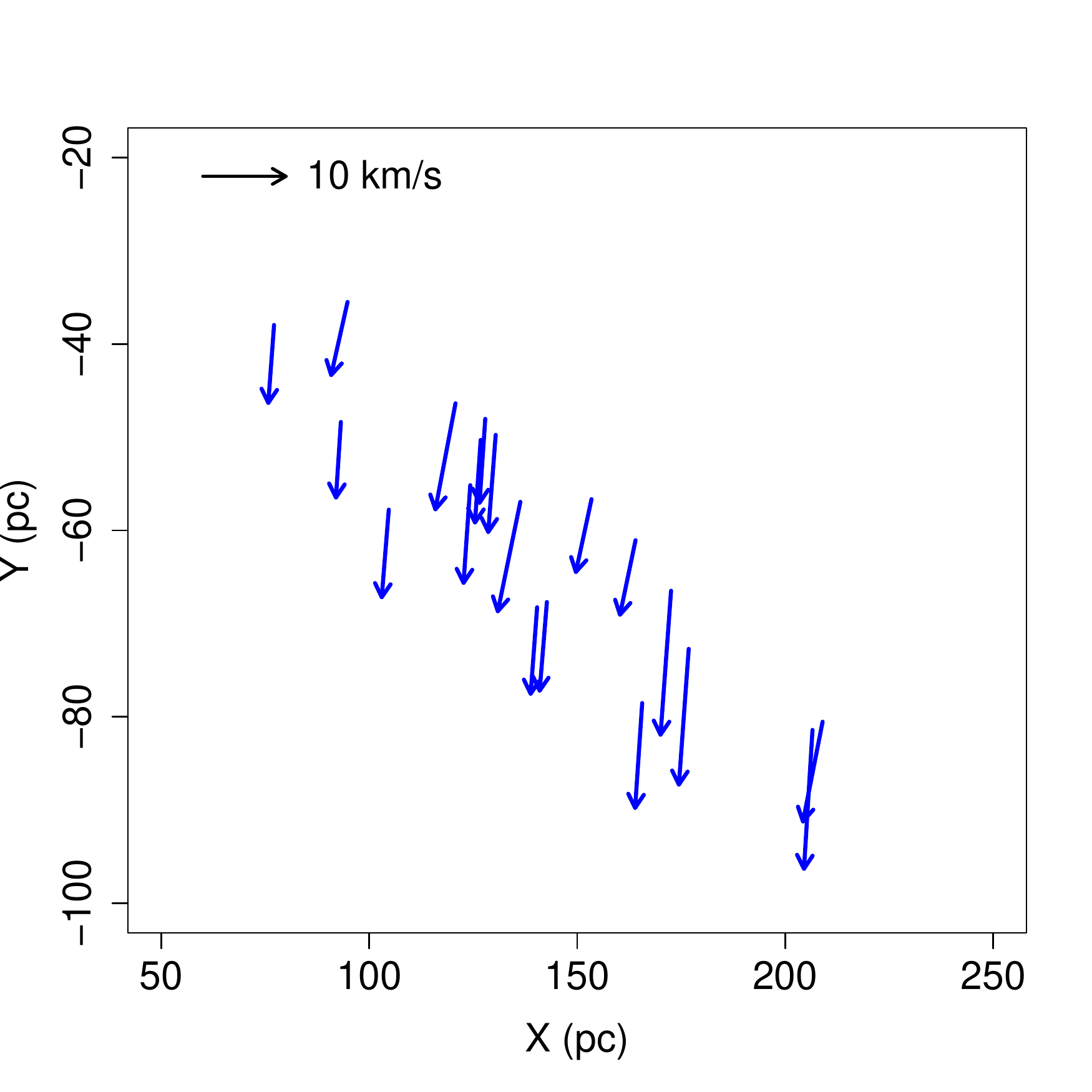}
\includegraphics[width=0.33\textwidth]{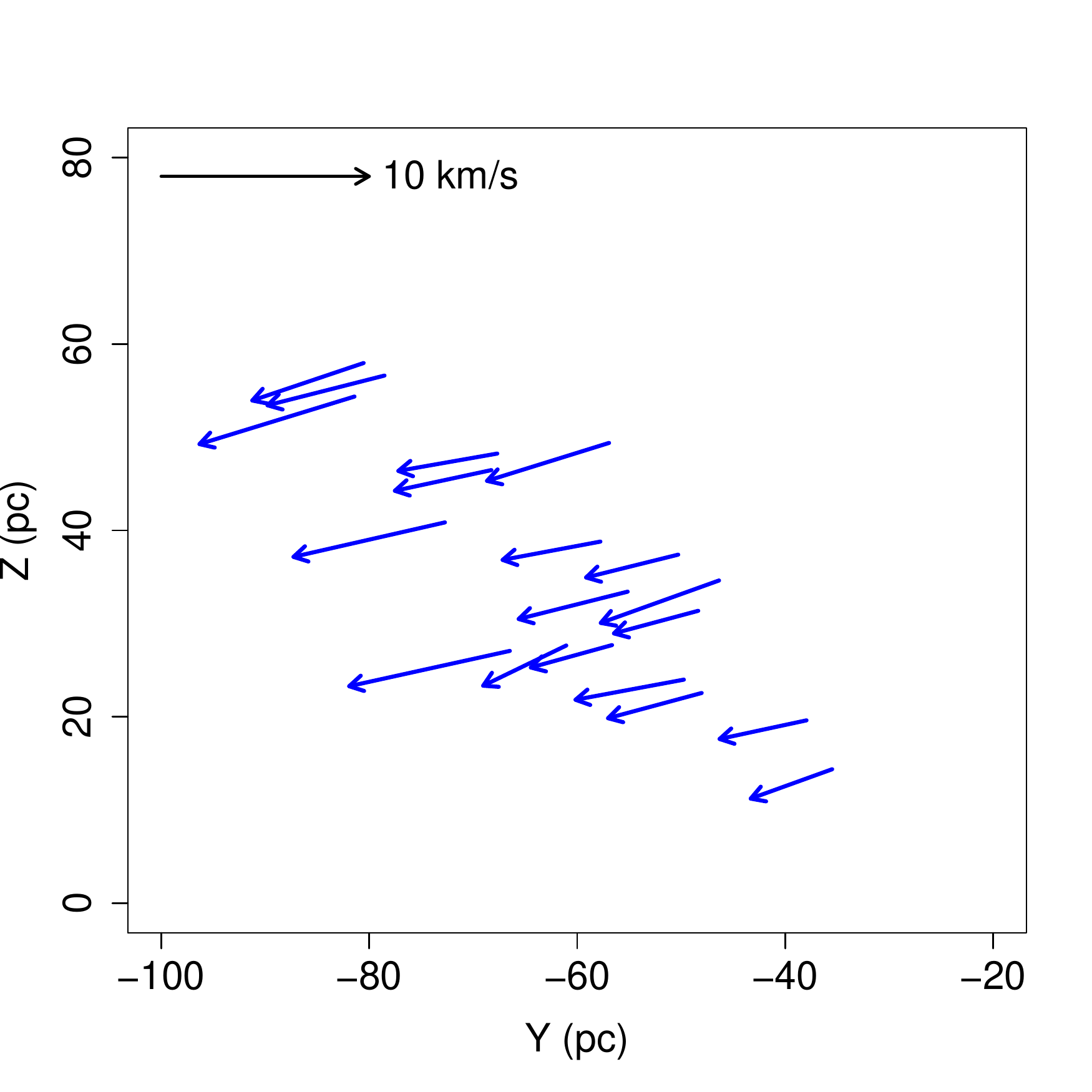}
\includegraphics[width=0.33\textwidth]{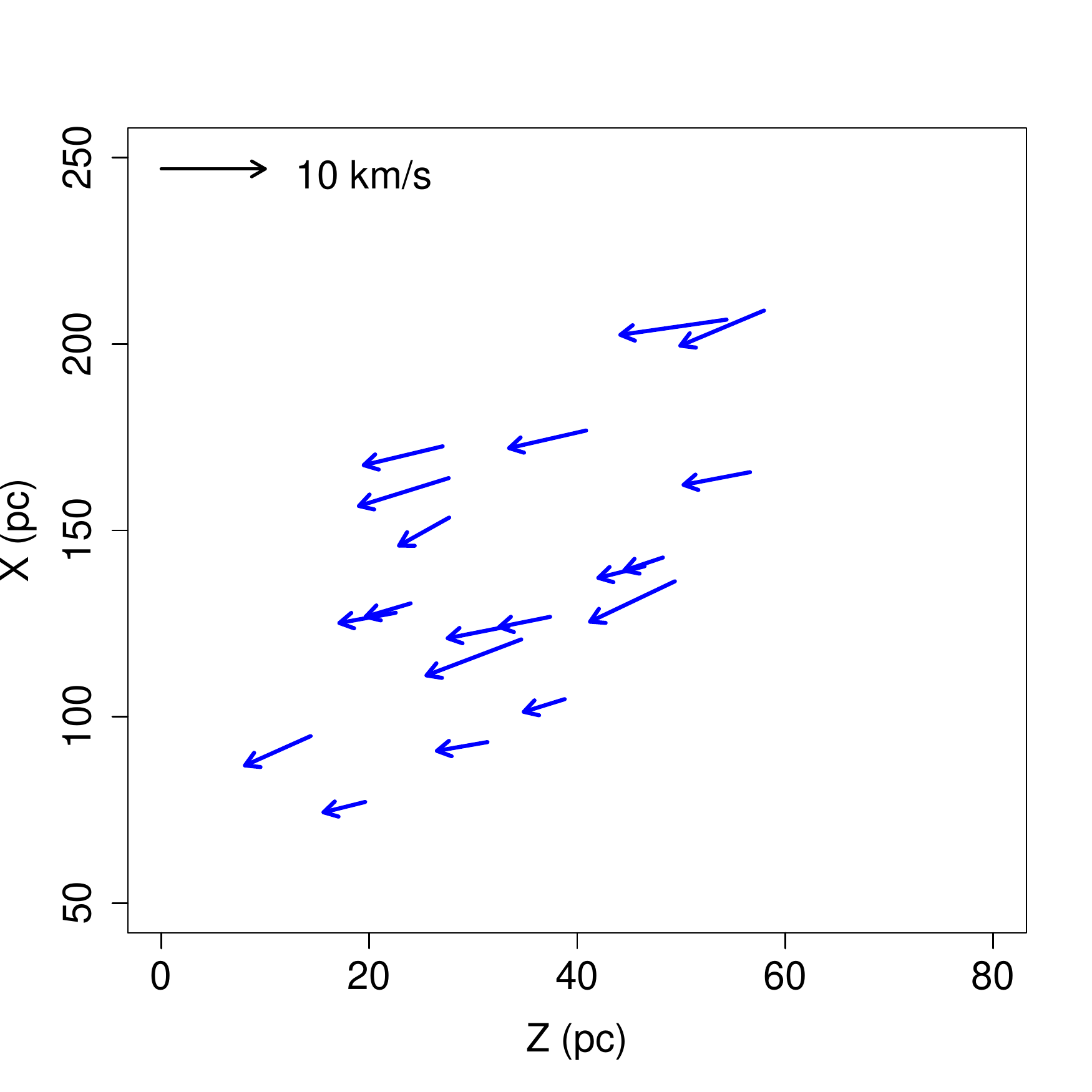}
\caption{Spatial velocities of the 19 group members of the Lupus moving group with known RVs projected on the $XY$, $YZ$ and $ZX$ planes. 
\label{fig9} }
\end{center}
\end{figure*}

We computed the Galactic velocities for each star using the procedure described in \citet{Soderblom(1987)}. Figure~\ref{fig8} displays the distribution of the $UVW$ Galactic velocity components while Fig.~\ref{fig9} shows the velocity vectors of the 19 stars with individual parallaxes in a $XYZ$ grid defined as follows. This reference system has its origin at the Sun where $X$ points to the Galactic center, $Y$ points in the direction of Galactic rotation, and $Z$ points to the Galactic north pole.

The average space velocity for the Lupus moving group derived in this paper is 
\begin{center}
$(U,V,W)=(-5.1,-20.7,-6.2)\pm(0.6,1.1,0.5)$~km/s\,, 
\end{center}
\begin{center}
$V_{space}=22.5\pm1.1$~km/s.
\end{center}
The space motion for the UCL subgroup was recently revised by \citet{Chen(2011)} to take into account the parallaxes from \textsc{Hipparcos} new reduction \citep{HIP07} and modern RV compilations. They report a space velocity of $(U,V,W)=(-5.1,-19.7,-4.6)\pm(0.6,0.4,0.3)$~km/s. The relative space motion between Lupus stars and the UCL subgroup is $(\Delta U,\Delta V,\Delta W )=(0.0,-1.0,-1.6)\pm(0.8,1.2,0.6)$~km/s. These results suggest that Lupus is moving at $1.9\pm1.6$~km/s with respect to UCL and that their velocities are statistically indistinguishable at the 1-2~km/s level which roughly corresponds to the velocity dispersion in each group. A similar conclusion was reached regarding Ophiuchus stars and the US subgroup of the Sco-Cen association \citep[see][]{Mamajek(2008)}.

One particular point of the Lupus association is that the observed RVs of young stars are expected to be low and exhibit both positive and negative values (see e.g. Table~4 and Fig.~\ref{fig4}). Among the moving group members with individual parallaxes presented in Table~\ref{tab5}, a total of six stars exhibit negative values for their RVs. In such cases we use the absolute value of the RVs to compute the individual parallaxes using Eq.~(\ref{eq_plx_ind}). In this context, \citet{Makarov(2007)} reports a significant mismatch between the observed spectroscopic radial velocities and the value inferred from his CP solution, implying a moderate degree of expansion. That the Lupus association of young stars is undergoing expansion with a velocity of $\simeq1$~km/s \citep[see][]{Makarov(2007)} and the RVs are also near zero possibly explains the existence of the two RV populations.  This situation contrasts with other SFRs, e.g. Taurus, where the observed RV is higher, $V_{r}\simeq+16$~km/s \citep[see][]{Luhman(2009)}, and the existence of group members with RVs changing signs (i.e., $V_{r}<0$) cannot be tolerated if assuming a one-dimensional velocity dispersion of only a few km/s. 

%FIGURE 8

%FIGURE 9

When we run the CP search method on the sample of stars with known $V_{r}<0$, we identify a moving group with  17~stars and  CP located at $(\alpha_{cp},\delta_{cp})=(99.9^{\circ},-36.1^{\circ})\pm(7.3^{\circ},9.6^{\circ})$. The large error bars in this CP solution arise from the small number of stars used for deriving the CP coordinates \citep[see][]{Galli(2012)}. As one should notice, the effect of linear expansion changes the CP position \citep[see also discussion in][]{Makarov(2007)}, but the CP mentioned above is still compatible with the one given in Sect.~5.3. In the following we seek to learn whether the use of either CP solutions lead to significant differences in the derived parallaxes. We computed the individual parallaxes for those six stars again with $V_{r}<0$ using the above mentioned CP and the procedure outlined in Appendix~A. The average difference between the recomputed parallaxes and those given in Table~\ref{tab6} is $\Delta\pi=0.3\pm0.1$~mas with rms of 1.1~mas, and they are statistically compatible within their error bars. The re-evaluated UVW velocity components for these stars has a negligible effect on the average space motion of the Lupus moving group, yielding instead
\begin{center}
$(U,V,W)=(-5.2,-20.9,-6.3)\pm(0.6,1.0,0.5)$~km/s\,, 
\end{center}
\begin{center}
$V_{space}=22.7\pm1.1$~km/s.
\end{center}
There is therefore no reason to reject those stars with $V_{r}<0$ from our analysis. Owing to the negligible impact to our results and for clarity of presentation we have decided not to present the re-evaluated parallaxes and space velocities.

%TABLE 5
\begin{table*}[!btp]
\centering
\caption{Proper motion and RV for the 19~stars that define the Lupus core moving group. 
\label{tab5}}
\resizebox{17cm}{!} {
\begin{tabular}{lcccccccc}
\hline
Star & $\alpha$& $\delta$& $\mu_{\alpha}\cos\delta$ & $\mu_{\delta}$ & Source &$V_{r}$&Ref.&2MASSJ\\ 
& (h:m:s) &($^{\circ}$ $^\prime$ $^\prime$$^\prime$) & (mas/yr) & (mas/yr)&&(km/s)&&\\
\hline		

RXJ1508.8-3715	&	15	08	53.8	&	-37	15	46	&$	-20.8	\pm	1.3	$&$	-24.1	\pm	1.2	$&	SPM4	&$	+4.5	\pm	0.3	$&	1	&	15085379-3715467	\\
RXJ1518.4-3738*	&	15	18	26.9	&	-37	38	02	&$	-18.1	\pm	1.3	$&$	-26.5	\pm	1.2	$&	SPM4	&$	+3.7	\pm	0.4	$&	2	&	15182692-3738021	\\
RXJ1524.5-3652	&	15	24	32.4	&	-36	52	02	&$	-14.8	\pm	2.7	$&$	-19.6	\pm	2.8	$&	SPM4	&$	+3.8	\pm	0.3	$&	1	&	15243236-3652027	\\
RXJ1525.0-3604	&	15	25	03.6	&	-36	04	45	&$	-14.4	\pm	2.3	$&$	-21.1	\pm	2.5	$&	SPM4	&$	+4.3	\pm	0.5	$&	3	&	15250358-3604455	\\
RXJ1525.5-3613	&	15	25	33.2	&	-36	13	46	&$	-15.7	\pm	2.1	$&$	-19.1	\pm	2.2	$&	SPM4	&$	+3.6	\pm	0.1	$&	4	&	15253316-3613467	\\
RXJ1531.3-3329	&	15	31	22.0	&	-33	29	39	&$	-21.7	\pm	1.7	$&$	-29.2	\pm	1.8	$&	SPM4	&$	-3.6	\pm	0.5	$&	5	&	15312193-3329394	\\
RXJ1534.6-4003K	&	15	34	38.2	&	-40	02	27	&$	-24.0	\pm	1.4	$&$	-32.8	\pm	1.4	$&	SPM4	&$	+3.8	\pm	0.9	$&	6	&	15343816-4002280	\\
RXJ1540.7-3756	&	15	40	41.2	&	-37	56	18	&$	-17.7	\pm	1.1	$&$	-27.4	\pm	1.1	$&	SPM4	&$	+3.9	\pm	1.0	$&	3	&	15404116-3756185	\\
RXJ1544.5-3521	&	15	44	35.3	&	-35	21	49	&$	-14.7	\pm	2.4	$&$	-23.2	\pm	2.4	$&	SPM4	&$	+2.5	\pm	0.2	$&	1	&	15443529-3521492	\\
Sz73	&	15	47	57.0	&	-35	14	35	&$	-21.0	\pm	13.2	$&$	-35.4	\pm	16.2	$&	SPM4	&$	-3.3	\pm	0.2	$&	7	&	15475693-3514346	\\
GQLup	&	15	49	12.1	&	-35	39	04	&$	-11.8	\pm	2.7	$&$	-19.0	\pm	2.5	$&	SPM4	&$	-3.2	\pm	0.7	$&	7	&	15491210-3539051	\\
RXJ1549.9-3629	&	15	49	59.2	&	-36	29	57	&$	-13.7	\pm	1.6	$&$	-25.6	\pm	1.6	$&	SPM4	&$	+4.4	\pm	1.0	$&	3	&	15495920-3629574	\\
RXJ1552.3-3819	&	15	52	19.5	&	-38	19	31	&$	-17.6	\pm	1.1	$&$	-27.2	\pm	1.2	$&	SPM4	&$	+4.9	\pm	1.0	$&	3	&	15521952-3819313	\\
RXJ1605.7-3905*	&	16	05	45.0	&	-39	06	06	&$	-17.7	\pm	2.2	$&$	-26.4	\pm	2.1	$&	SPM4	&$	+3.2	\pm	0.9	$&	6	&	16054499-3906065	\\
F304	&	16	08	11.0	&	-39	10	46	&$	-13.6	\pm	2.0	$&$	-25.2	\pm	2.0	$&	SPM4	&$	+2.8	\pm	0.1	$&	1	&	16081096-3910459	\\
RXJ1608.5-3847	&	16	08	31.6	&	-38	47	29	&$	-12.0	\pm	2.3	$&$	-19.2	\pm	2.4	$&	SPM4	&$	-2.5	\pm	1.0	$&	3	&	16083156-3847292	\\
RXJ1610.0-4016	&	16	10	04.8	&	-40	16	12	&$	-18.6	\pm	2.0	$&$	-30.7	\pm	2.0	$&	SPM4	&$	+5.1	\pm	1.0	$&	3	&	16100478-4016122	\\
Sz121	&	16	10	12.2	&	-39	21	18	&$	-8.3	\pm	2.4	$&$	-21.6	\pm	2.5	$&	SPM4	&$	-2.8	\pm	2.0	$&	8	&	16101219-3921181	\\
RXJ1613.0-4004	&	16	13	02.4	&	-40	04	33	&$	-17.2	\pm	1.5	$&$	-33.9	\pm	1.5	$&	SPM4	&$	-2.8	\pm	1.0	$&	3	&	16130240-4004329	\\

\hline

\end{tabular}
}

\tablefoot{The symbol ``*" indicates those stars whose membership status (Lupus or UCL) is doubtful in the literature. We provide for each star the most usual identifier, position (epoch 2000), proper motion, source of proper motion, RV, source of RV, and the 2MASS identifier. }

\tablebib{
Radial velocity sources:
(1)~This work;
(2)~\citet{James(2006)};
(3)~\citet{Wichmann(1999)};
(4)~\citet{Guenther(2007)};
(5)~\citet{White(2007)};
(6)~\citet{Torres(2006)}; 
(7)~\citet{Melo(2003)};
(8)~\citet{Dubath(1996)}.
}
\vspace{1cm}
\end{table*}

%TABLE 6
\begin{table*}[!btp]
\centering
\caption{Individual parallax and velocity components for Lupus stars with known RVs. 
\label{tab6}}
\resizebox{15cm}{!} {
\begin{tabular}{lcccccc}
\hline
Star&$\pi$&$U$&$V$&$W$&$V_{space}$\\
&(mas)&(km/s)&(km/s)&(km/s)&(km/s)\\
\hline

RXJ1508.8-3715	&$	8.0	\pm	1.4	$&$	-3.4	\pm	1.3	$&$	-18.8	\pm	2.1	$&$	-3.9	\pm	2.4	$&$	19.5	\pm	2.1	$\\
RXJ1518.4-3738	&$	9.1	\pm	1.8	$&$	-2.4	\pm	1.1	$&$	-16.2	\pm	2.2	$&$	-4.9	\pm	2.5	$&$	17.1	\pm	2.2	$\\
RXJ1524.5-3652	&$	6.1	\pm	1.4	$&$	-3.1	\pm	1.7	$&$	-18.6	\pm	3.4	$&$	-4.4	\pm	3.7	$&$	19.3	\pm	3.4	$\\
RXJ1525.0-3604	&$	5.2	\pm	1.3	$&$	-3.4	\pm	1.9	$&$	-22.5	\pm	4.1	$&$	-6.4	\pm	4.6	$&$	23.6	\pm	4.1	$\\
RXJ1525.5-3613	&$	6.1	\pm	1.3	$&$	-3.5	\pm	1.5	$&$	-19.0	\pm	3.1	$&$	-3.7	\pm	3.3	$&$	19.7	\pm	3.0	$\\
RXJ1531.3-3329	&$	6.4	\pm	2.0	$&$	-10.8	\pm	2.4	$&$	-23.5	\pm	5.5	$&$	-8.2	\pm	5.9	$&$	27.1	\pm	5.2	$\\
RXJ1534.6-4003K	&$	11.3	\pm	3.3	$&$	-2.8	\pm	1.6	$&$	-16.7	\pm	3.2	$&$	-4.0	\pm	3.6	$&$	17.4	\pm	3.2	$\\
RXJ1540.7-3756	&$	7.1	\pm	2.3	$&$	-3.2	\pm	2.0	$&$	-21.0	\pm	4.6	$&$	-5.9	\pm	5.1	$&$	22.0	\pm	4.6	$\\
RXJ1544.5-3521	&$	7.1	\pm	2.1	$&$	-2.7	\pm	1.5	$&$	-17.7	\pm	3.9	$&$	-4.9	\pm	4.2	$&$	18.6	\pm	3.9	$\\
Sz73	&$	7.5	\pm	3.4	$&$	-9.7	\pm	4.4	$&$	-22.7	\pm	11.8	$&$	-9.1	\pm	12.5	$&$	26.3	\pm	11.2	$\\
GQLup	&$	4.3	\pm	1.6	$&$	-9.4	\pm	2.5	$&$	-21.4	\pm	6.6	$&$	-8.0	\pm	7.0	$&$	24.7	\pm	6.2	$\\
RXJ1549.9-3629	&$	4.4	\pm	1.5	$&$	-4.1	\pm	2.5	$&$	-29.8	\pm	7.4	$&$	-10.2	\pm	8.0	$&$	31.8	\pm	7.4	$\\
RXJ1552.3-3819	&$	5.1	\pm	1.5	$&$	-4.7	\pm	2.4	$&$	-29.1	\pm	6.1	$&$	-7.4	\pm	6.6	$&$	30.4	\pm	6.0	$\\
RXJ1605.7-3905	&$	7.1	\pm	2.7	$&$	-3.6	\pm	2.1	$&$	-20.8	\pm	5.6	$&$	-4.3	\pm	5.9	$&$	21.6	\pm	5.5	$\\
F304	&$	7.2	\pm	1.8	$&$	-2.7	\pm	1.1	$&$	-18.0	\pm	3.5	$&$	-5.4	\pm	3.6	$&$	19.0	\pm	3.5	$\\
RXJ1608.5-3847	&$	6.0	\pm	2.9	$&$	-7.5	\pm	2.2	$&$	-15.6	\pm	6.1	$&$	-4.8	\pm	6.4	$&$	18.0	\pm	5.6	$\\
RXJ1610.0-4016	&$	5.4	\pm	1.6	$&$	-5.1	\pm	2.4	$&$	-30.9	\pm	6.7	$&$	-7.6	\pm	7.0	$&$	32.2	\pm	6.6	$\\
Sz121	&$	5.6	\pm	4.3	$&$	-7.5	\pm	3.3	$&$	-16.0	\pm	10.3	$&$	-8.6	\pm	10.7	$&$	19.6	\pm	9.7	$\\
RXJ1613.0-4004	&$	9.8	\pm	4.2	$&$	-7.9	\pm	1.9	$&$	-15.7	\pm	5.4	$&$	-6.3	\pm	5.6	$&$	18.6	\pm	5.0	$\\

\hline
\end{tabular}

}

\tablefoot{We provide the most usual identifier, individual parallax, and velocity components for each star.}

\end{table*}

%----------------------------------------------------------------------------------------------------------------
\subsection{Approximate parallaxes for other moving group members}

\begin{figure}[!h]
\begin{center}
%\sidecaption
\includegraphics[width=0.50\textwidth]{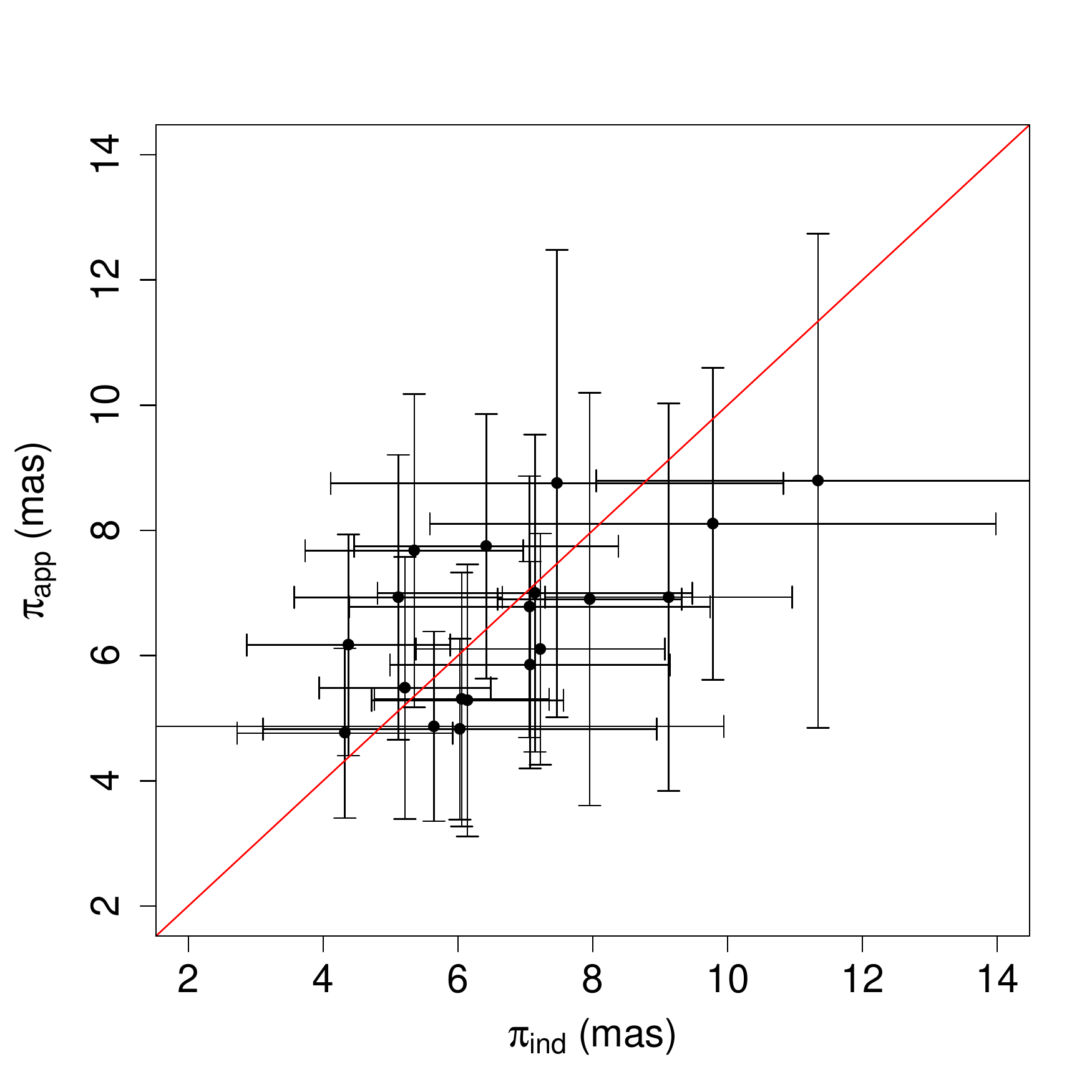}
\caption{
\label{fig10}
Comparison of parallaxes computed with RVs $(\pi_{ind})$ and the spatial velocity $(\pi_{app})$ for the Lupus core moving group of 19 stars. The red solid line indicates perfect correlation. The mean difference between parallaxes computed with both strategies is 0.2~mas, and the rms is 1.4~mas. }
\end{center}
\end{figure}

The hypothesis that all members of a moving group share the same space motion allows us to compute tentative parallaxes for group members with unknown RVs. First we derive the average spatial velocity $V_{space}$ from the Galactic velocity of the stars with known RVs using the 19~stars that define the Lupus core moving group (see Sect.~6.1). Then we compute an approximate parallax $\pi_{app}$ as
\begin{equation}\label{eq_pi_app}
\pi_{app}=\frac{A\mu_{\parallel}}{V_{space}\sin\lambda}.
\end{equation}
The uncertainty on this tentative parallax is again derived by error propagation and considers in this case the error budget of proper motions, space velocity, and the CP errors. We present in Fig.~\ref{fig10} a comparison between individual parallaxes (see Table~5) and approximate parallaxes for the Lupus core moving group. Both procedures return similar results within the admittedly large error bars, which tends to justify the assumption of a common space motion. 

%FIGURE 10

The velocity dispersion of the cluster prevents all stars from having exactly the same space velocity. However, the procedure described above for deriving approximate parallaxes considers a single value of the space velocity for all stars in the group. We performed Monte Carlo simulations by resampling the spatial velocity from a Gaussian distribution where the mean and variance correspond to the values given in Sect.~6.1 for the average space velocity of the Lupus core moving group. We constructed a total of 1000 realizations. In each run we assigned a different value of space velocity and derived the approximate parallax for each star by using Eq.~(\ref{eq_pi_app}). The average value of the computed parallaxes gives our final result, and the standard error of the mean is propagated into the parallax uncertainty. This strategy not only allows us to reproduce the effect of velocity dispersion, but also makes our parallax results less dependent on a single value used for the spatial velocity of the group. We present in Table~\ref{tab7} the approximate parallaxes derived for the remaining 90 Lupus stars with unknown RVs.

We do stress, however, that the individual approximate parallaxes derived in this way should only be seen as tentative values that will be useful for preliminary statistical analyses of the Lupus association. They should be superseded by more precise values when RV measurements become available for these association members.

%----------------------------------------------------------------------------------------------------------------
\subsection{Comparison with Hipparcos parallaxes}

\begin{figure}[!htp]
\begin{center}
\includegraphics[width=0.44\textwidth]{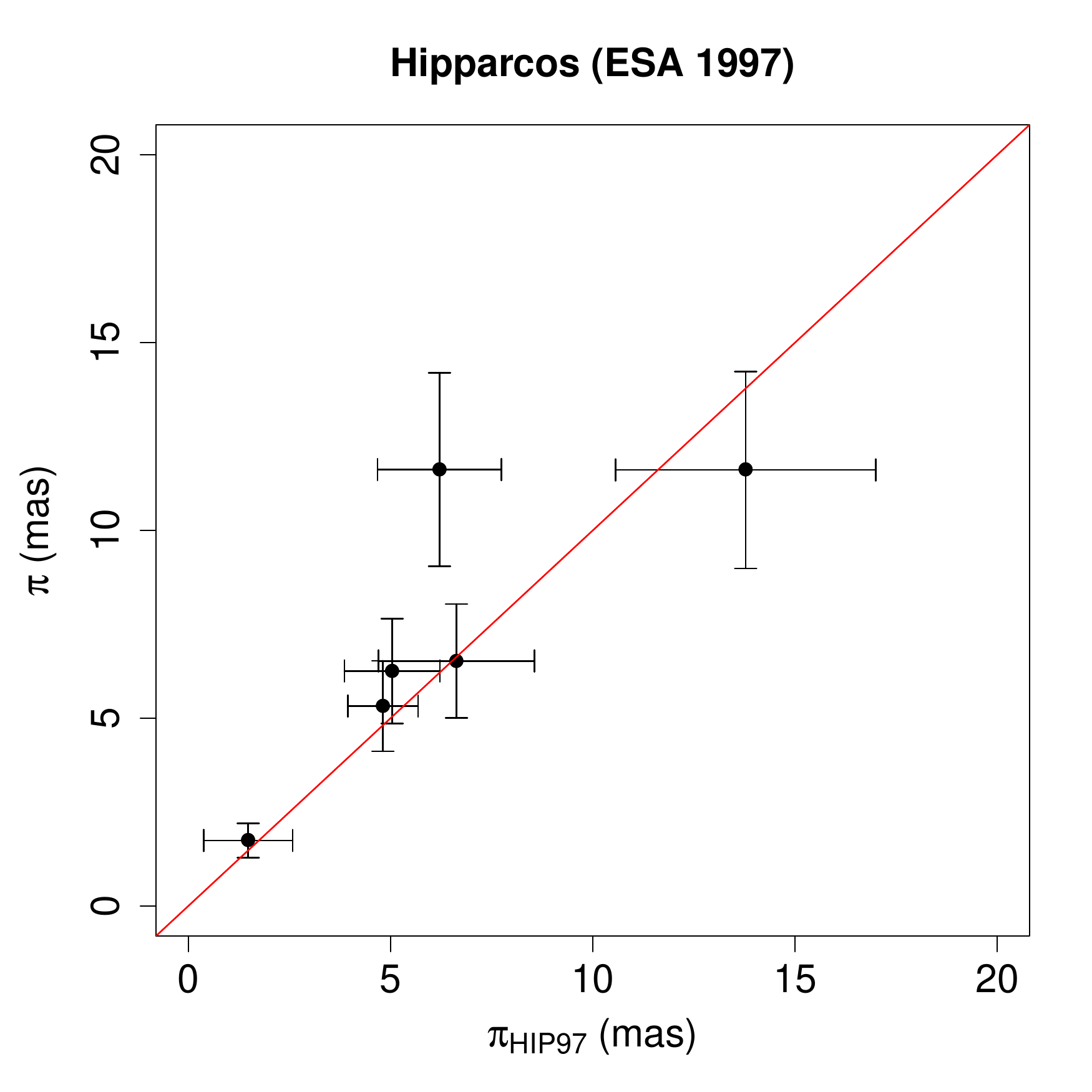}
\includegraphics[width=0.44\textwidth]{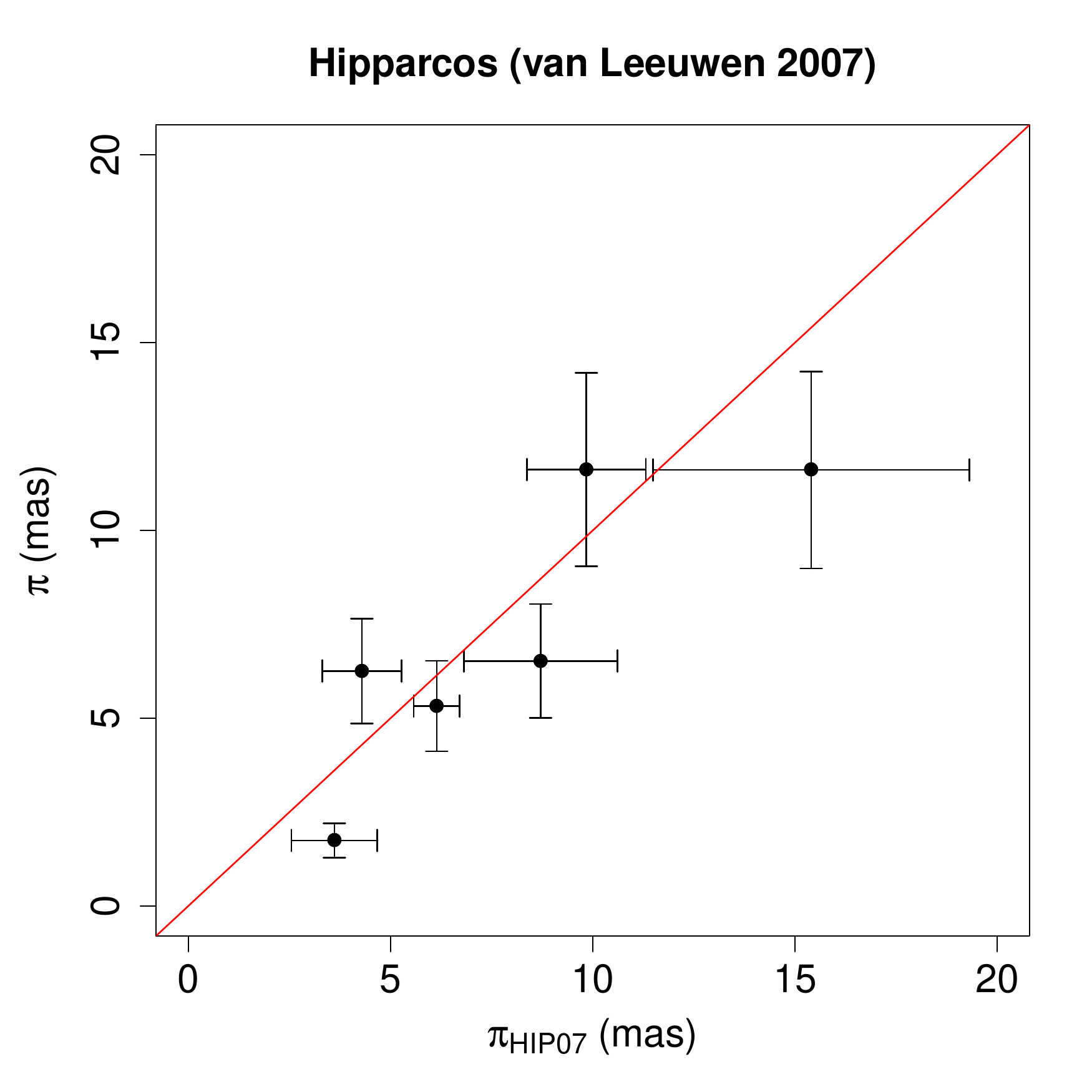}
\caption{
\label{fig11}
Parallaxes derived in this paper compared with the trigonometric parallaxes given in HIP97 (\textit{upper panel}) and HIP07 (\textit{lower panel}). The red solid line represents the expected distribution for equal results. }
\end{center}
\end{figure}

As a final check of our results we compare the parallaxes derived in this work with \textsc{Hipparcos} trigonometric parallaxes. We consider both versions of the catalog: the original catalog \citep[][hereafter HIP97]{HIP97}, and the new reduction of \textsc{Hipparcos} data \citep[][hereafter HIP07]{HIP07}. We found only six \textsc{Hipparcos} stars among the 109 moving group members and used the group spatial velocity to compute approximate parallaxes. The results of this comparison are presented in Fig.~\ref{fig11}. The rms with respect to HIP97 is 2.7~mas and 2.5~mas for HIP07. The mean difference between the parallaxes derived in this work and the ones in HIP97 and HIP07 are -0.9~mas and +0.8~mas, respectively. We conclude that our results are in good agreement with the trigonometric parallaxes given in \textsc{Hipparcos}.

%FIGURE 11

%----------------------------------------------------------------------------------------------------------------
\subsection{Comparison with \citet{Makarov(2007)} results}

\citet{Makarov(2007)} estimated kinematic distances for Lupus stars in a similar manner to the procedure described in Sect.~6.2. To do so, he assumed a mean spatial velocity of 22~km/s and used the derived distances to infer the depth of the Lupus association. In the following we discuss these findings in light of our own results derived in this study. 

\begin{figure}[!htp]
\begin{center}
\includegraphics[width=0.46\textwidth]{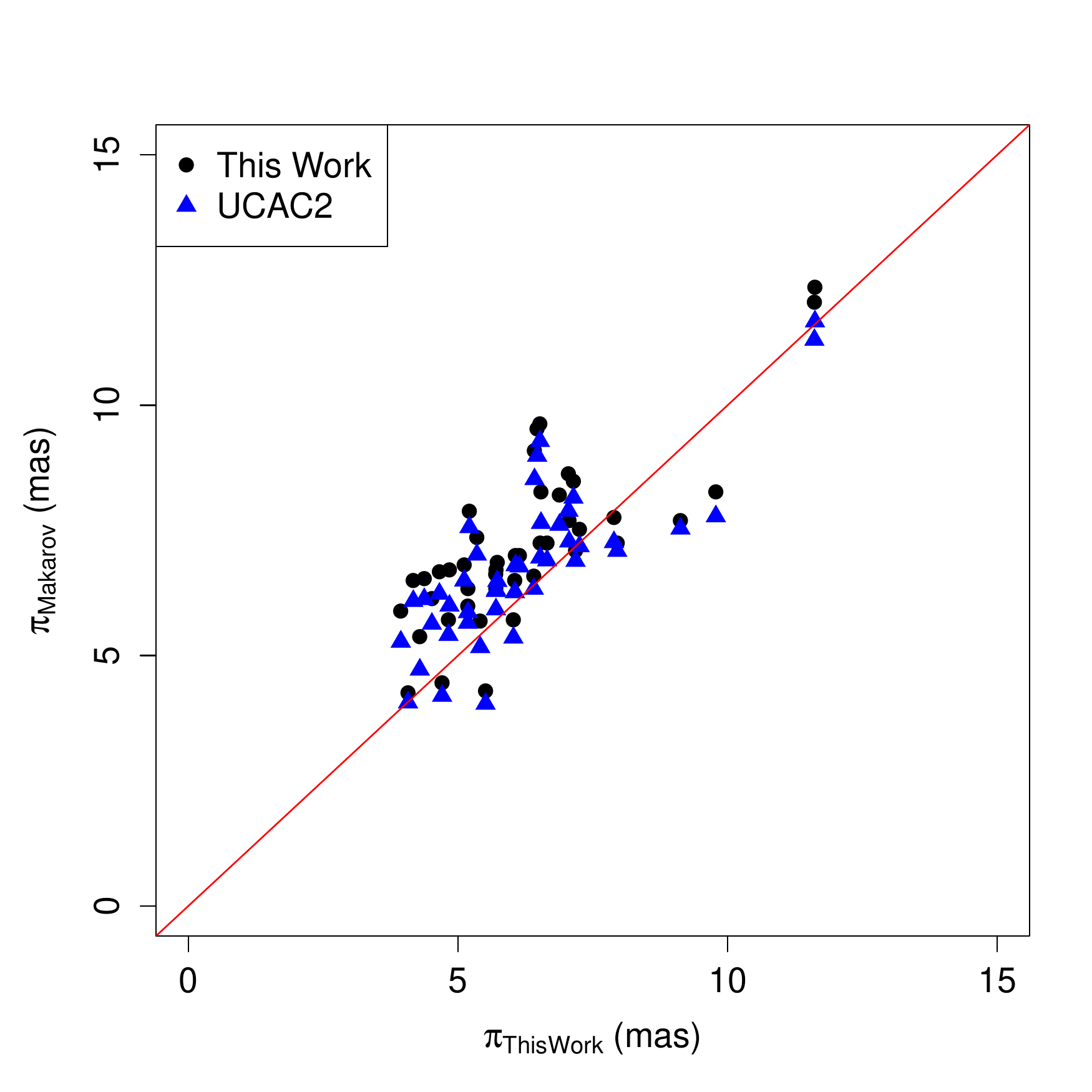}
\caption{
\label{fig12}
Comparison between the parallaxes derived in this work and by \citet{Makarov(2007)}. Parallaxes are calculated using the proper motion catalogs included in this work (see Sect.~3.1) and UCAC2  following \citet{Makarov(2007)}. The red solid line indicates perfect correlation. }
\end{center}
\end{figure}

The sample of 109 moving group members defined in our CP analysis is cross-correlated with the 93 stars used by \citet{Makarov(2007)} resulting in 44 stars. When comparing the parallaxes derived in both studies we find a mean difference of $\Delta\pi=-0.9\pm0.2$~mas (in the sense ``this work'' minus ``\citealt{Makarov(2007)}'')  with rms of 1.5~mas (see Fig.\ref{fig12}). In Sect.~6.1 we derived a mean space velocity of $V_{space}=22.5\pm1.1$~km/s that confirms the value of 22~km/s used by \citet{Makarov(2007)} in his analysis. Since the difference between both values amounts to only 2\%, it seems unlikely that the reported systematic effect between the two parallax sets can only be explained by the adopted value of the group spatial velocity. As discussed in Sect.~5.3 the CP position derived in both studies differs significantly by $\Delta\lambda=25.4^{\circ}\pm8.3^{\circ}$. 

It is also important to note that the proper motions used in each work play an important role not only for determining the CP position, but also for deriving the parallaxes of group members. When we use proper motions from the UCAC2 catalog as done by \citet{Makarov(2007)} with the CP derived in this paper, the mean difference between the parallaxes presented in both studies drops to $\Delta\pi=-0.5\pm0.2$~mas. This result is illustrated in Fig.~\ref{fig12}. However, given the more recent and precise astrometric catalogs available now, such as the ones used in this paper (see Sect.~3.1), there is no good reason to use the UCAC2 catalog where the proper motions for some stars are poorly defined with only two observational data points. That Table~2 of \citet{Makarov(2007)} does not include the uncertainty of the computed distances makes it difficult to compare the quality of both parallax results. The average error of the proper motions used  in this paper and by \citet{Makarov(2007)} is, respectively, 1.9~mas/yr and 3.3~mas/yr in each component. Given the proper motion and CP errors (see Sects.~5.1 and 5.3) presented in both studies we conclude that the parallaxes derived in this paper are more precise and accurate, because our results take the various sources of errors included in the parallax computation into account.

\begin{figure}[!htp]
\begin{center}
\includegraphics[width=0.46\textwidth]{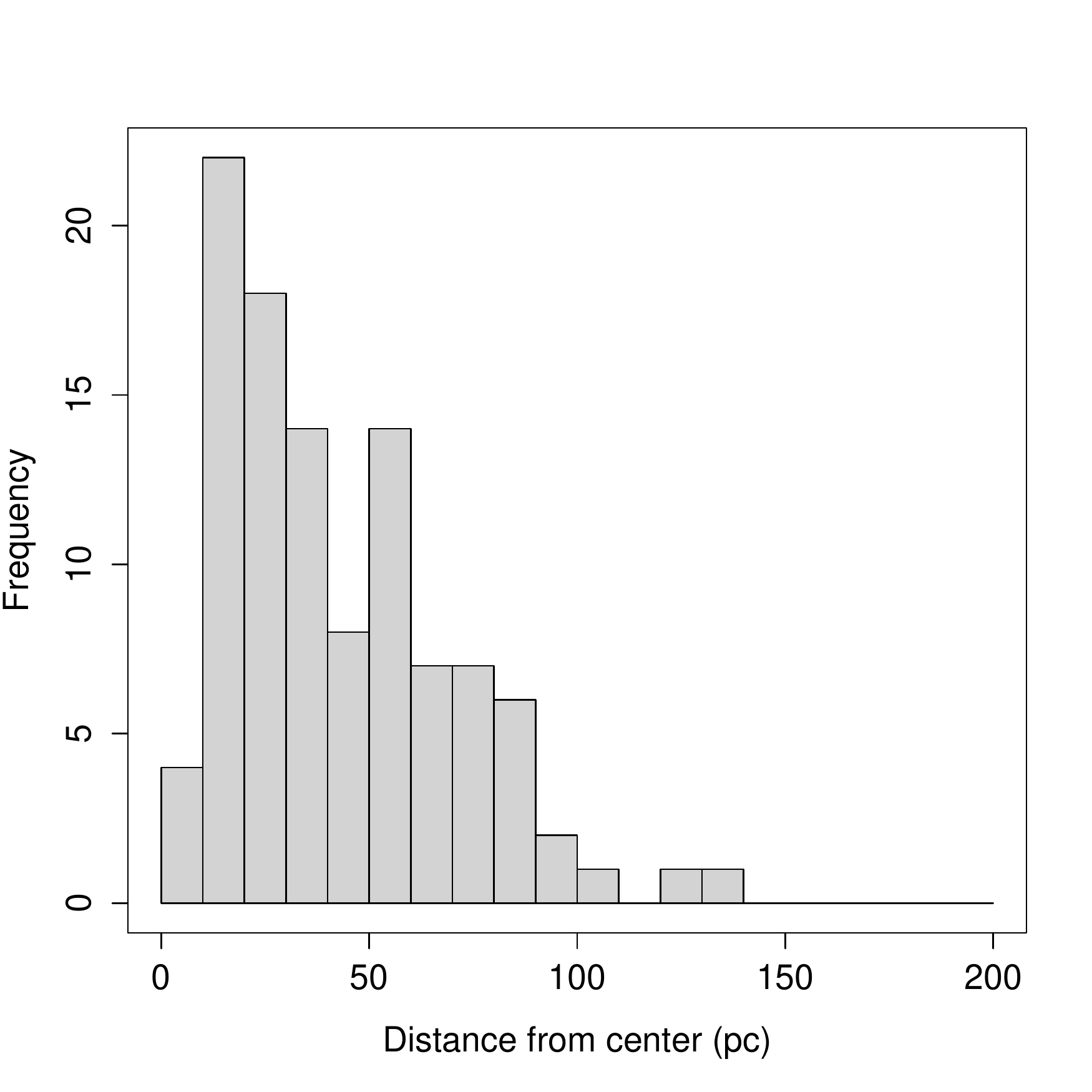}
\caption{
\label{fig13}
Histogram of the individual distances of the 109 Lupus moving group members to the center of the association. }
\end{center}
\end{figure}

To investigate the depth of the Lupus moving group we first determine the so-assumed center of the association, $(X_{C},Y_{C},Z_{C})=(142,-61,38)$~pc, as defined by  the 19~stars in Table~\ref{tab5}, and then compute the distance of all 109 group members to the center of the association. Figure~\ref{fig13} displays the histogram of the computed distances. We conclude that the Lupus association shows a large depth  of at least 100~pc, exceeding the value of 80~pc reported by \citet{Makarov(2007)}. However, we stress that a more detailed study about the spatial distribution of Lupus stars and the size of the association is necessary when more RVs become available in the future.   

%FIGURE 12

%FIGURE 13

%TABLE 7
\begin{longtab}
\small{
\begin{longtable}{lcccccccc}

\caption{Approximate parallax derived in this paper for Lupus stars with unknown RVs.
\label{tab7}  }\\
\hline
Star & $\alpha$& $\delta$  & $\mu_{\alpha}\cos\delta$ & $\mu_{\delta}$ & Source &$\pi$&2MASSJ\\ 
& (h:m:s) &($^{\circ}$ $^\prime$ $^\prime$$^\prime$)  & (mas/yr) & (mas/yr)&&(mas)&\\
\hline
\endfirsthead
\caption{continued.}\\
\hline
Star & $\alpha$& $\delta$  & $\mu_{\alpha}\cos\delta$ & $\mu_{\delta}$ & Source &$\pi$&2MASSJ\\ 
& (h:m:s) &($^{\circ}$ $^\prime$ $^\prime$$^\prime$)  & (mas/yr) & (mas/yr)&&(mas)&\\
\hline
\endhead
\hline
\endfoot

RXJ1511.0-3252AB	&	15	11	04.6	&	-32	51	30	&$	-14.0	\pm	3.0	$&$	-24.0	\pm	3.0	$&	D05	&$	5.9	\pm	1.5	$&	15110450-3251304	\\
RXJ1511.6-3550	&	15	11	37.0	&	-35	50	42	&$	-17.9	\pm	1.6	$&$	-21.7	\pm	1.6	$&	SPM4	&$	6.1	\pm	1.4	$&	15113696-3550417	\\
RXJ1512.6-3417	&	15	12	39.8	&	-34	16	59	&$	-15.6	\pm	1.8	$&$	-18.3	\pm	1.7	$&	SPM4	&$	5.2	\pm	1.2	$&	15123981-3416591	\\
HD135127	&	15	14	39.6	&	-34	45	41	&$	-17.3	\pm	1.3	$&$	-24.3	\pm	1.3	$&	TYCHO2	&$	6.4	\pm	1.4	$&	15143959-3445412	\\
RXJ1515.7-3332K	&	15	15	45.4	&	-33	31	59	&$	-23.1	\pm	1.6	$&$	-26.4	\pm	1.6	$&	SPM4	&$	7.5	\pm	1.7	$&	15154537-3331597	\\
GSC6770-655	&	15	19	53.0	&	-28	02	26	&$	-36.0	\pm	3.0	$&$	-45.0	\pm	3.0	$&	D05	&$	12.2	\pm	2.8	$&	15195295-2802266	\\
RXJ1525.6-3537	&	15	25	36.7	&	-35	37	32	&$	-21.0	\pm	2.2	$&$	-23.1	\pm	2.3	$&	SPM4	&$	6.7	\pm	1.5	$&	15253666-3537319	\\
RXJ1527.3-3603	&	15	27	22.9	&	-36	04	09	&$	-23.6	\pm	2.3	$&$	-34.4	\pm	2.4	$&	SPM4	&$	8.9	\pm	2.0	$&	15272286-3604087	\\
RXJ1529.3-3737	&	15	29	19.0	&	-37	37	20	&$	-19.2	\pm	0.9	$&$	-23.5	\pm	0.9	$&	SPM4	&$	6.5	\pm	1.4	$&	15291901-3737205	\\
RXJ1529.7-3628	&	15	29	47.3	&	-36	28	37	&$	-16.6	\pm	1.6	$&$	-20.8	\pm	1.7	$&	SPM4	&$	5.7	\pm	1.3	$&	15294727-3628374	\\
Sz65	&	15	39	27.8	&	-34	46	17	&$	-9.9	\pm	2.9	$&$	-21.5	\pm	2.8	$&	SPM4	&$	5.0	\pm	1.3	$&	15392776-3446171	\\
RXJ1539.7-3450	&	15	39	46.4	&	-34	51	02	&$	-14.9	\pm	2.0	$&$	-19.3	\pm	2.1	$&	SPM4	&$	5.2	\pm	1.2	$&	15394637-3451027	\\
SSTc2dJ154013.7-340142	&	15	40	13.7	&	-34	01	43	&$	-4.5	\pm	3.3	$&$	-14.0	\pm	3.1	$&	SPM4	&$	3.0	\pm	1.0	$&	15401371-3401429	\\
RXJ1540.3-3426A	&	15	40	18.5	&	-34	26	15	&$	-14.7	\pm	3.7	$&$	-19.5	\pm	3.6	$&	SPM4	&$	5.2	\pm	1.4	$&	15401850-3426146	\\
SSTc2dJ154148.3-350145	&	15	41	48.3	&	-35	01	46	&$	-13.0	\pm	4.1	$&$	-17.2	\pm	4.0	$&	SPM4	&$	4.6	\pm	1.3	$&	15414827-3501458	\\
RXJ1542.0-3601	&	15	42	05.2	&	-36	01	32	&$	-19.5	\pm	2.3	$&$	-23.4	\pm	2.3	$&	SPM4	&$	6.5	\pm	1.5	$&	15420518-3601317	\\
RXJ1544.0-3311*	&	15	44	03.8	&	-33	11	11	&$	-18.1	\pm	1.3	$&$	-26.9	\pm	1.4	$&	SPM4	&$	6.9	\pm	1.5	$&	15440376-3311110	\\
RXJ1546.6-3618	&	15	46	41.2	&	-36	18	47	&$	-12.7	\pm	1.9	$&$	-23.8	\pm	2.0	$&	SPM4	&$	5.7	\pm	1.3	$&	15464121-3618472	\\
RXJ1546.7-3459	&	15	46	45.1	&	-34	59	47	&$	-17.4	\pm	3.6	$&$	-20.6	\pm	3.4	$&	SPM4	&$	5.7	\pm	1.5	$&	15464506-3459473	\\
RXJ1547.1-3540	&	15	47	08.4	&	-35	40	19	&$	-13.3	\pm	2.2	$&$	-25.7	\pm	2.2	$&	SPM4	&$	6.1	\pm	1.4	$&	15470841-3540195	\\
RXJ1547.6-4018	&	15	47	41.8	&	-40	18	26	&$	-18.7	\pm	1.1	$&$	-27.6	\pm	1.1	$&	SPM4	&$	7.2	\pm	1.6	$&	15474176-4018267	\\
HMLup	&	15	47	50.6	&	-35	28	35	&$	-9.7	\pm	4.1	$&$	-22.0	\pm	3.9	$&	SPM4	&$	5.0	\pm	1.4	$&	15475062-3528353	\\
HNLup	&	15	48	05.2	&	-35	15	53	&$	-10.3	\pm	7.2	$&$	-19.6	\pm	7.1	$&	SPM4	&$	4.7	\pm	1.8	$&	15480523-3515526	\\
RXJ1548.1-3452	&	15	48	08.9	&	-34	52	53	&$	-17.9	\pm	2.2	$&$	-22.1	\pm	2.1	$&	SPM4	&$	6.0	\pm	1.4	$&	15480893-3452531	\\
RXJ1548.7-3520	&	15	48	42.5	&	-35	20	07	&$	-11.2	\pm	2.9	$&$	-16.1	\pm	2.8	$&	SPM4	&$	4.2	\pm	1.1	$&	15484253-3520066	\\
RXJ1548.9-3513	&	15	48	54.1	&	-35	13	18	&$	-17.4	\pm	2.1	$&$	-27.7	\pm	2.1	$&	SPM4	&$	7.0	\pm	1.6	$&	15485411-3513186	\\
Sz76	&	15	49	30.7	&	-35	49	51	&$	-16.1	\pm	2.8	$&$	-20.8	\pm	2.7	$&	SPM4	&$	5.6	\pm	1.4	$&	15493074-3549514	\\
HD141277*	&	15	49	45.0	&	-39	25	09	&$	-18.2	\pm	2.3	$&$	-24.4	\pm	2.2	$&	TYCHO2	&$	6.5	\pm	1.5	$&	15494499-3925089	\\
RXJ1550.7-3828	&	15	50	46.7	&	-38	29	27	&$	-10.3	\pm	1.3	$&$	-16.0	\pm	1.3	$&	SPM4	&$	4.1	\pm	0.9	$&	15504672-3829267	\\
Sz77	&	15	51	47.0	&	-35	56	43	&$	-12.5	\pm	2.2	$&$	-20.5	\pm	2.1	$&	SPM4	&$	5.1	\pm	1.2	$&	15514695-3556440	\\
RXJ1555.4-3338	&	15	55	26.3	&	-33	38	22	&$	-17.8	\pm	1.8	$&$	-28.1	\pm	1.8	$&	SPM4	&$	7.0	\pm	1.6	$&	15552621-3338232	\\
Sz81	&	15	55	50.3	&	-38	01	33	&$	-15.8	\pm	1.4	$&$	-22.4	\pm	1.5	$&	SPM4	&$	5.8	\pm	1.3	$&	15555030-3801329	\\
RXJ1556.0-3655	&	15	56	02.1	&	-36	55	28	&$	-9.3	\pm	2.5	$&$	-19.2	\pm	2.4	$&	SPM4	&$	4.5	\pm	1.1	$&	15560210-3655282	\\
Sz82	&	15	56	09.2	&	-37	56	06	&$	-12.7	\pm	3.9	$&$	-21.5	\pm	4.0	$&	SPM4	&$	5.3	\pm	1.4	$&	15560921-3756057	\\
Hip78092	&	15	56	41.9	&	-42	19	23	&$	-13.9	\pm	1.0	$&$	-25.5	\pm	1.0	$&	TYCHO2	&$	6.3	\pm	1.4	$&	15564188-4219232	\\
Sz126	&	15	57	24.0	&	-42	40	04	&$	-13.1	\pm	1.9	$&$	-22.3	\pm	2.1	$&	SPM4	&$	5.6	\pm	1.3	$&	15572401-4240044	\\
Sz127	&	15	57	30.4	&	-42	10	32	&$	-10.0	\pm	1.7	$&$	-12.3	\pm	1.8	$&	SPM4	&$	3.4	\pm	0.8	$&	15573035-4210324	\\
Sz128	&	15	58	07.3	&	-41	51	48	&$	-12.9	\pm	3.3	$&$	-18.0	\pm	3.3	$&	SPM4	&$	4.8	\pm	1.3	$&	15580732-4151479	\\
RXJ1558.9-3646	&	15	58	59.8	&	-36	46	20	&$	-11.6	\pm	3.0	$&$	-23.1	\pm	2.9	$&	SPM4	&$	5.5	\pm	1.4	$&	15585980-3646206	\\
CD-3610569	&	15	59	49.5	&	-36	28	28	&$	-29.4	\pm	2.7	$&$	-46.0	\pm	2.8	$&	TYCHO2	&$	11.6	\pm	2.6	$&	15594951-3628279	\\
RXJ1559.9-3750	&	15	59	54.2	&	-37	50	47	&$	-11.0	\pm	1.3	$&$	-19.8	\pm	1.3	$&	SPM4	&$	4.8	\pm	1.1	$&	15595416-3750469	\\
SSTc2dJ160000.6-422158	&	16	00	00.6	&	-42	21	57	&$	-10.4	\pm	2.3	$&$	-15.8	\pm	2.5	$&	SPM4	&$	4.1	\pm	1.0	$&	16000060-4221567	\\
Sz131	&	16	00	49.4	&	-41	30	04	&$	-8.3	\pm	5.4	$&$	-21.2	\pm	5.3	$&	SPM4	&$	4.8	\pm	1.6	$&	16004943-4130038	\\
RXJ1601.9-3613	&	16	01	59.2	&	-36	12	55	&$	-18.6	\pm	2.6	$&$	-24.7	\pm	2.6	$&	SPM4	&$	6.5	\pm	1.5	$&	16015918-3612555	\\
EXLup	&	16	03	05.5	&	-40	18	25	&$	-9.8	\pm	2.3	$&$	-18.2	\pm	2.4	$&	SPM4	&$	4.4	\pm	1.1	$&	16030548-4018254	\\
RXJ1603.8-3938*	&	16	03	52.5	&	-39	39	01	&$	-17.1	\pm	2.1	$&$	-29.3	\pm	2.1	$&	SPM4	&$	7.3	\pm	1.7	$&	16035250-3939013	\\
HD143978	&	16	04	57.1	&	-38	57	15	&$	-27.8	\pm	1.2	$&$	-46.8	\pm	1.6	$&	TYCHO2	&$	11.6	\pm	2.6	$&	16045707-3857157	\\
RXJ1605.5-3837	&	16	05	33.3	&	-38	37	45	&$	-12.4	\pm	2.5	$&$	-22.7	\pm	2.5	$&	SPM4	&$	5.5	\pm	1.3	$&	16053329-3837451	\\
HOLup	&	16	07	00.6	&	-39	02	19	&$	-10.1	\pm	2.5	$&$	-18.1	\pm	2.5	$&	SPM4	&$	4.4	\pm	1.1	$&	16070061-3902194	\\
Sz90	&	16	07	10.1	&	-39	11	03	&$	-6.9	\pm	3.3	$&$	-21.4	\pm	3.3	$&	SPM4	&$	4.7	\pm	1.2	$&	16071007-3911033	\\
Sz91	&	16	07	11.6	&	-39	03	47	&$	-14.5	\pm	2.6	$&$	-17.6	\pm	2.6	$&	SPM4	&$	4.8	\pm	1.2	$&	16071159-3903475	\\
RXJ1607.2-3839	&	16	07	13.7	&	-38	39	24	&$	-12.9	\pm	2.1	$&$	-18.0	\pm	2.2	$&	SPM4	&$	4.7	\pm	1.1	$&	16071370-3839238	\\
Sz95	&	16	07	52.3	&	-38	58	06	&$	-8.2	\pm	2.6	$&$	-21.7	\pm	2.6	$&	SPM4	&$	4.9	\pm	1.2	$&	16075230-3858059	\\
RXJ1608.0-3857	&	16	08	00.0	&	-38	57	51	&$	-13.4	\pm	2.1	$&$	-20.3	\pm	2.1	$&	SPM4	&$	5.2	\pm	1.2	$&	16075996-3857510	\\
Sz96	&	16	08	12.6	&	-39	08	33	&$	-8.7	\pm	2.2	$&$	-20.1	\pm	2.3	$&	SPM4	&$	4.6	\pm	1.1	$&	16081263-3908334	\\
RXJ1608.3-3843	&	16	08	18.3	&	-38	44	05	&$	-20.8	\pm	2.8	$&$	-30.7	\pm	2.8	$&	SPM4	&$	7.9	\pm	1.8	$&	16081824-3844052	\\
Sz97	&	16	08	21.8	&	-39	04	21	&$	-10.3	\pm	2.7	$&$	-19.8	\pm	2.7	$&	SPM4	&$	4.8	\pm	1.2	$&	16082180-3904214	\\
Sz99	&	16	08	24.0	&	-39	05	49	&$	-14.6	\pm	4.5	$&$	-25.1	\pm	4.5	$&	SPM4	&$	6.2	\pm	1.7	$&	16082404-3905494	\\
RXJ1608.4-3840	&	16	08	25.2	&	-38	40	56	&$	-13.2	\pm	2.5	$&$	-16.6	\pm	2.5	$&	SPM4	&$	4.5	\pm	1.1	$&	16082519-3840558	\\
Sz102	&	16	08	29.7	&	-39	03	11	&$	-12.7	\pm	4.3	$&$	-19.7	\pm	4.4	$&	SPM4	&$	5.0	\pm	1.4	$&	16082972-3903110	\\
Sz104	&	16	08	30.8	&	-39	05	49	&$	-21.3	\pm	6.6	$&$	-22.5	\pm	6.6	$&	SPM4	&$	6.5	\pm	2.0	$&	16083081-3905488	\\
V856Sco	&	16	08	34.3	&	-39	06	18	&$	-12.5	\pm	1.2	$&$	-21.6	\pm	1.6	$&	TYCHO2	&$	5.3	\pm	1.2	$&	16083427-3906181	\\
RXJ1608.6-3922	&	16	08	36.2	&	-39	23	02	&$	-10.6	\pm	2.2	$&$	-23.2	\pm	2.2	$&	SPM4	&$	5.4	\pm	1.3	$&	16083617-3923024	\\
SSTc2dJ160839.8-392922	&	16	08	39.7	&	-39	29	23	&$	-18.0	\pm	3.9	$&$	-27.5	\pm	3.9	$&	SPM4	&$	7.0	\pm	1.8	$&	16083974-3929228	\\
Sz107	&	16	08	41.8	&	-39	01	37	&$	-6.1	\pm	3.7	$&$	-18.6	\pm	3.7	$&	SPM4	&$	4.1	\pm	1.2	$&	16084179-3901370	\\
SSTc2dJ160853.2-391440	&	16	08	53.2	&	-39	14	40	&$	-19.9	\pm	5.0	$&$	-23.3	\pm	4.9	$&	SPM4	&$	6.4	\pm	1.8	$&	16085324-3914401	\\
RXJ1608.9-3905	&	16	08	54.3	&	-39	06	06	&$	-8.4	\pm	2.5	$&$	-20.4	\pm	2.5	$&	SPM4	&$	4.7	\pm	1.2	$&	16085427-3906057	\\
RXJ1608.9-3945	&	16	08	54.3	&	-39	46	05	&$	-8.2	\pm	2.4	$&$	-21.5	\pm	2.5	$&	SPM4	&$	4.8	\pm	1.2	$&	16085429-3946046	\\
Sz111	&	16	08	54.7	&	-39	37	43	&$	-8.1	\pm	2.3	$&$	-18.9	\pm	2.3	$&	SPM4	&$	4.4	\pm	1.1	$&	16085468-3937431	\\
Sz112	&	16	08	55.5	&	-39	02	34	&$	-9.6	\pm	2.8	$&$	-18.4	\pm	2.9	$&	SPM4	&$	4.4	\pm	1.1	$&	16085553-3902339	\\
Sz113	&	16	08	57.8	&	-39	02	23	&$	-12.5	\pm	3.5	$&$	-19.8	\pm	3.5	$&	SPM4	&$	5.0	\pm	1.3	$&	16085780-3902227	\\
V908Sco	&	16	09	01.9	&	-39	05	12	&$	-7.7	\pm	2.1	$&$	-18.4	\pm	2.2	$&	SPM4	&$	4.2	\pm	1.0	$&	16090185-3905124	\\
SSTc2dJ160904.6-392112	&	16	09	04.5	&	-39	21	13	&$	-10.6	\pm	2.6	$&$	-16.4	\pm	2.7	$&	SPM4	&$	4.2	\pm	1.1	$&	16090452-3921125	\\
Sz115	&	16	09	06.2	&	-39	08	52	&$	-12.9	\pm	3.4	$&$	-18.4	\pm	3.4	$&	SPM4	&$	4.8	\pm	1.3	$&	16090621-3908518	\\
Sz134	&	16	09	12.3	&	-41	40	25	&$	-13.3	\pm	2.1	$&$	-20.4	\pm	2.2	$&	SPM4	&$	5.2	\pm	1.2	$&	16091226-4140249	\\
RXJ1609.4-3850	&	16	09	27.4	&	-38	50	19	&$	-10.3	\pm	2.0	$&$	-17.3	\pm	2.0	$&	SPM4	&$	4.3	\pm	1.0	$&	16092739-3850186	\\
Sz116	&	16	09	42.6	&	-39	19	41	&$	-16.4	\pm	2.1	$&$	-26.1	\pm	2.2	$&	SPM4	&$	6.6	\pm	1.5	$&	16094258-3919407	\\
Sz117	&	16	09	44.4	&	-39	13	30	&$	-12.7	\pm	2.5	$&$	-20.1	\pm	2.5	$&	SPM4	&$	5.1	\pm	1.2	$&	16094434-3913301	\\
Sz118	&	16	09	48.6	&	-39	11	17	&$	-13.3	\pm	9.0	$&$	-13.4	\pm	8.7	$&	SPM4	&$	3.9	\pm	2.1	$&	16094864-3911169	\\
RXJ1609.9-3923	&	16	09	54.0	&	-39	23	27	&$	-7.6	\pm	2.1	$&$	-14.9	\pm	2.2	$&	SPM4	&$	3.6	\pm	0.9	$&	16095399-3923275	\\
Sz119	&	16	09	57.1	&	-38	59	48	&$	-11.9	\pm	2.3	$&$	-24.6	\pm	2.4	$&	SPM4	&$	5.8	\pm	1.4	$&	16095707-3859479	\\
Sz120	&	16	10	10.6	&	-40	07	44	&$	-5.2	\pm	1.1	$&$	-6.4	\pm	1.2	$&	TYCHO2	&$	1.7	\pm	0.5	$&	16101054-4007437	\\
Sz122	&	16	10	16.4	&	-39	08	05	&$	-15.9	\pm	2.4	$&$	-21.9	\pm	2.4	$&	SPM4	&$	5.7	\pm	1.4	$&	16101642-3908050	\\
Sz123	&	16	10	51.6	&	-38	53	14	&$	-7.1	\pm	2.5	$&$	-16.9	\pm	2.6	$&	SPM4	&$	3.9	\pm	1.0	$&	16105158-3853137	\\
RXJ1612.0-3840	&	16	12	01.4	&	-38	40	27	&$	-9.8	\pm	1.0	$&$	-15.7	\pm	1.1	$&	SPM4	&$	3.9	\pm	0.9	$&	16120140-3840276	\\
SSTc2dJ161207.6-381324	&	16	12	07.6	&	-38	13	24	&$	-6.9	\pm	1.3	$&$	-13.6	\pm	1.3	$&	SPM4	&$	3.2	\pm	0.8	$&	16120761-3813242	\\
RXJ1612.3-4012	&	16	12	22.1	&	-40	12	52	&$	-22.8	\pm	1.6	$&$	-42.6	\pm	1.7	$&	SPM4	&$	10.3	\pm	2.3	$&	16122204-4012522	\\
SSTc2dJ161243.8-381503	&	16	12	43.7	&	-38	15	03	&$	-6.8	\pm	1.4	$&$	-12.4	\pm	1.4	$&	SPM4	&$	3.0	\pm	0.7	$&	16124373-3815031	\\
RXJ1614.4-3808	&	16	14	26.4	&	-38	08	00	&$	-15.2	\pm	0.7	$&$	-22.2	\pm	0.7	$&	SPM4	&$	5.7	\pm	1.3	$&	16142637-3807597	\\
HD147402*	&	16	23	29.6	&	-39	58	00	&$	-11.6	\pm	1.0	$&$	-24.2	\pm	1.2	$&	SPM4	&$	5.7	\pm	1.3	$&	16232955-3958008	\\

\end{longtable}
}
\tablefoot{The symbol ``*'' indicates those stars whose membership status (Lupus or UCL) is doubtful in the literature. For each star we provide the most usual identifier, position (epoch 2000), proper motion, source of proper motion, parallax, and the 2MASS identifier.}
\end{longtab}

%----------------------------------------------------------------------------------------------------------------
%						7. DISCUSSION
%----------------------------------------------------------------------------------------------------------------
\newpage
\section{Discussion} 

\subsection{Positions and parallaxes for pre-main sequence subclasses}

\begin{figure*}[!btp]
\begin{center}
\includegraphics[width=0.59\textwidth,angle=-90]{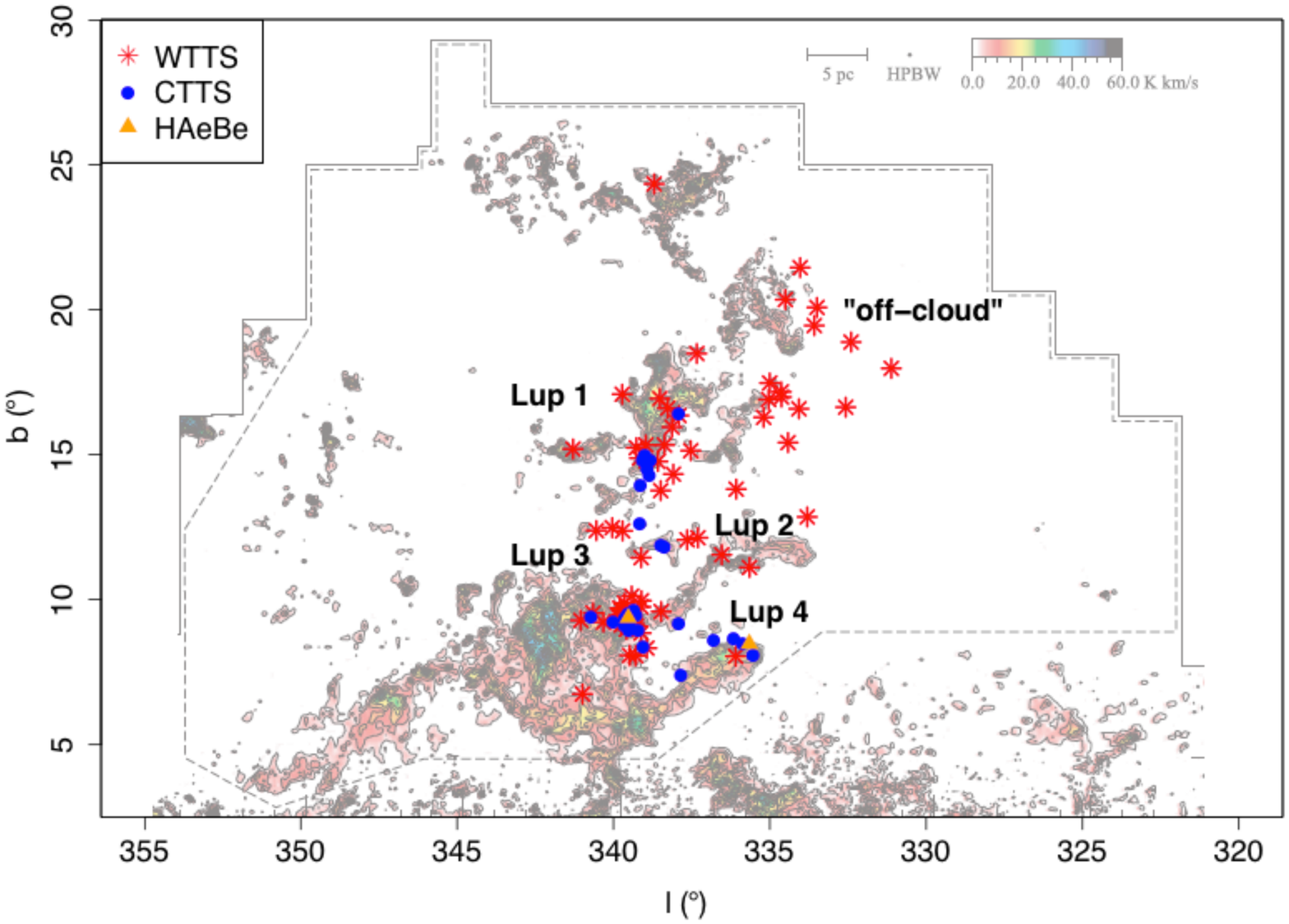}
\caption{Location of the moving group members overlaid on the $^{12}$CO intensity map from \citet{Tachihara(2001)}. Different symbols and colors mark the various YSO subclasses.
\label{fig14} }
\end{center}
\end{figure*}

The Lupus moving group identified in this paper contains 39~CTTSs, 68~WTTSs, and 2~HAeBes. Among the WTTSs we have 49~on-cloud and 19~off-cloud stars. Figure~\ref{fig14} displays the location of the various PMS subclasses. We note that the Lupus~4 population consists mainly of CTTSs, while in the remaining clouds we find both CTTSs and WTTSs, which may represent a selection effect of our input list of Lupus candidate stars (see Sect.~3). As already predicted, the off-cloud population contains only WTTSs while the CTTSs are located in the immediate vicinity of the molecular clouds. 

An interesting point arises when we compare the distances of these various TTS populations. Indeed, we find that the off-cloud WTTSs tend to be closer to us and the CTTSs more distant. We present in Table~\ref{tab8} the average parallaxes (distances) for each subgroup in the Lupus SFR. 

%TABLE 8
\begin{table}[h]
\centering
\caption{Properties of the TTSs in Lupus. 
\vspace{0.5cm}
\label{tab8}}
\begin{tabular}{lccc}
\hline
Sample&Stars&$\pi$&$d$\\
&&(mas)&(pc)\\
\hline
CTTS&39&$4.8\pm0.2$&$208^{+9}_{-8}$\\
WTTS \footnotesize{(on-cloud)}&49&$6.0\pm0.3$&$167_{-8}^{+9}$\\
WTTS \footnotesize{(off-cloud)}&19&$7.2\pm0.5$&$139_{-9}^{+10}$\\
\hline
\end{tabular}
\tablefoot{For each TTS subclass we provide the number of stars, average parallax and average distance with the corresponding uncertainties. }
\end{table}

%FIGURE 14

\subsection{Notes on the parallaxes of Lupus subgroups}

\begin{figure*}[!btp]
\begin{center}
\includegraphics[width=0.59\textwidth,angle=-90]{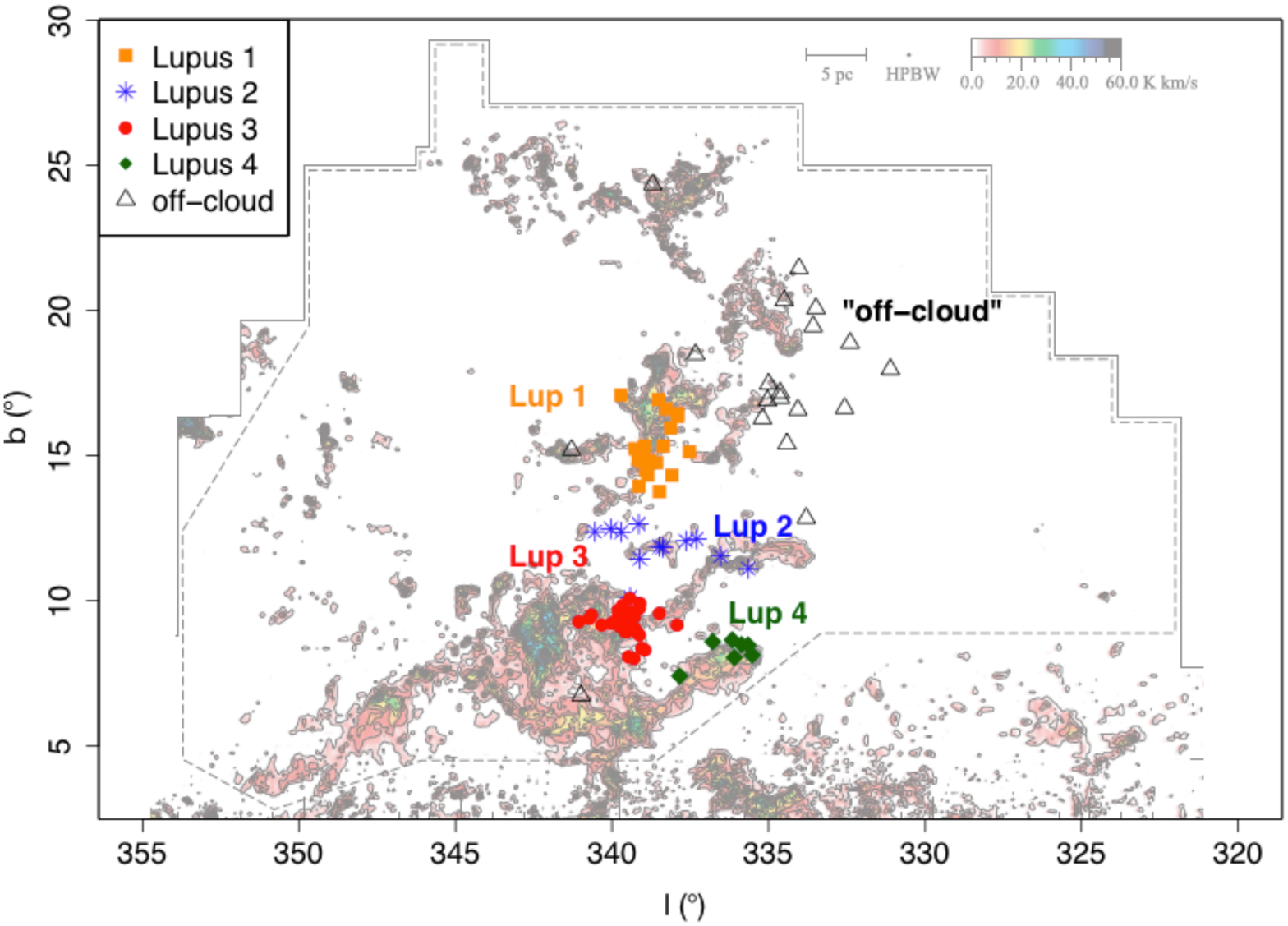}
\caption{Location of the moving group members overlaid on the $^{12}$CO intensity map from \citet{Tachihara(2001)}. Different symbols and colors mark the various clouds of the Lupus complex.
\label{fig15} }
\end{center}
\end{figure*}

\begin{figure}[h]
\begin{center}
\includegraphics[width=0.23\textwidth]{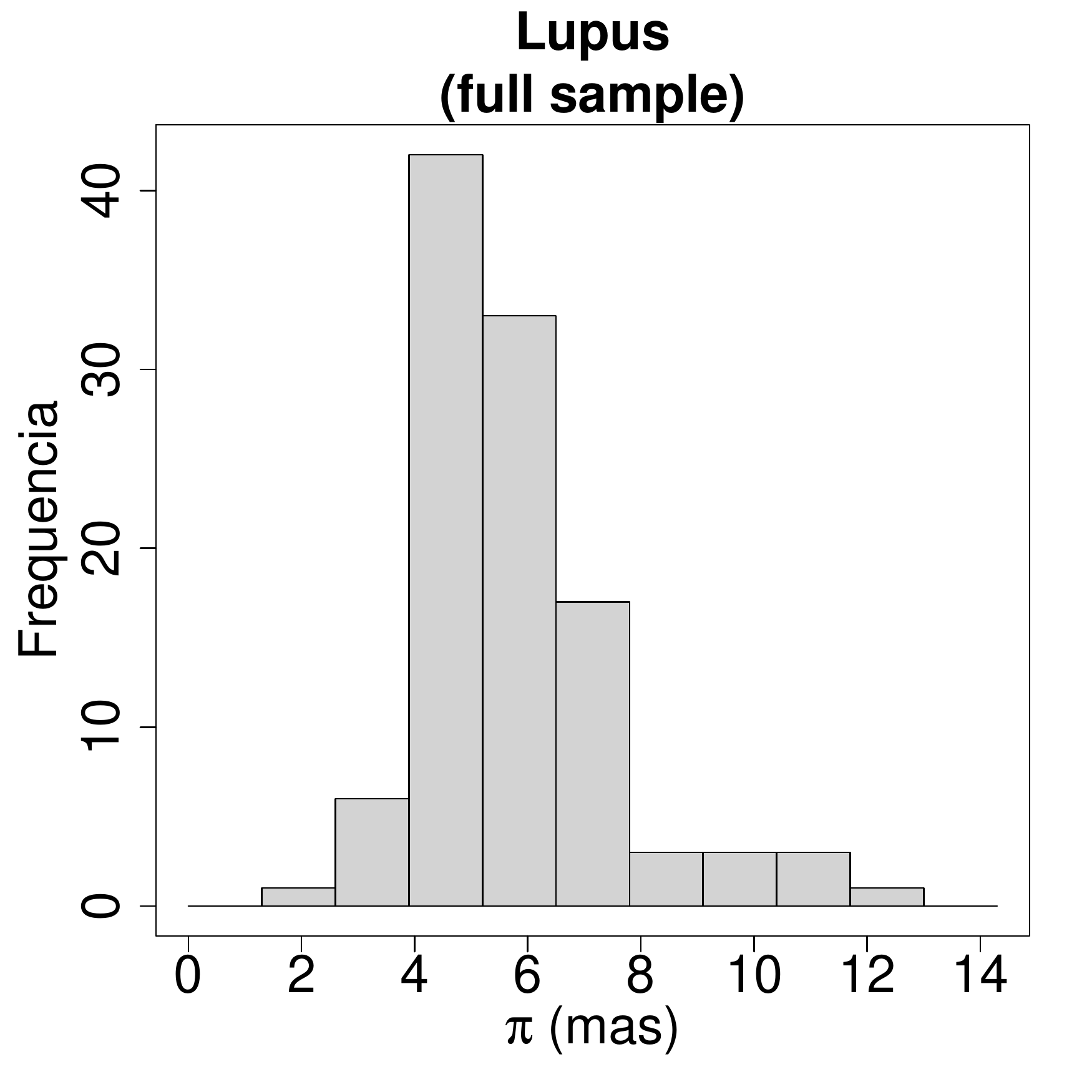}
\includegraphics[width=0.23\textwidth]{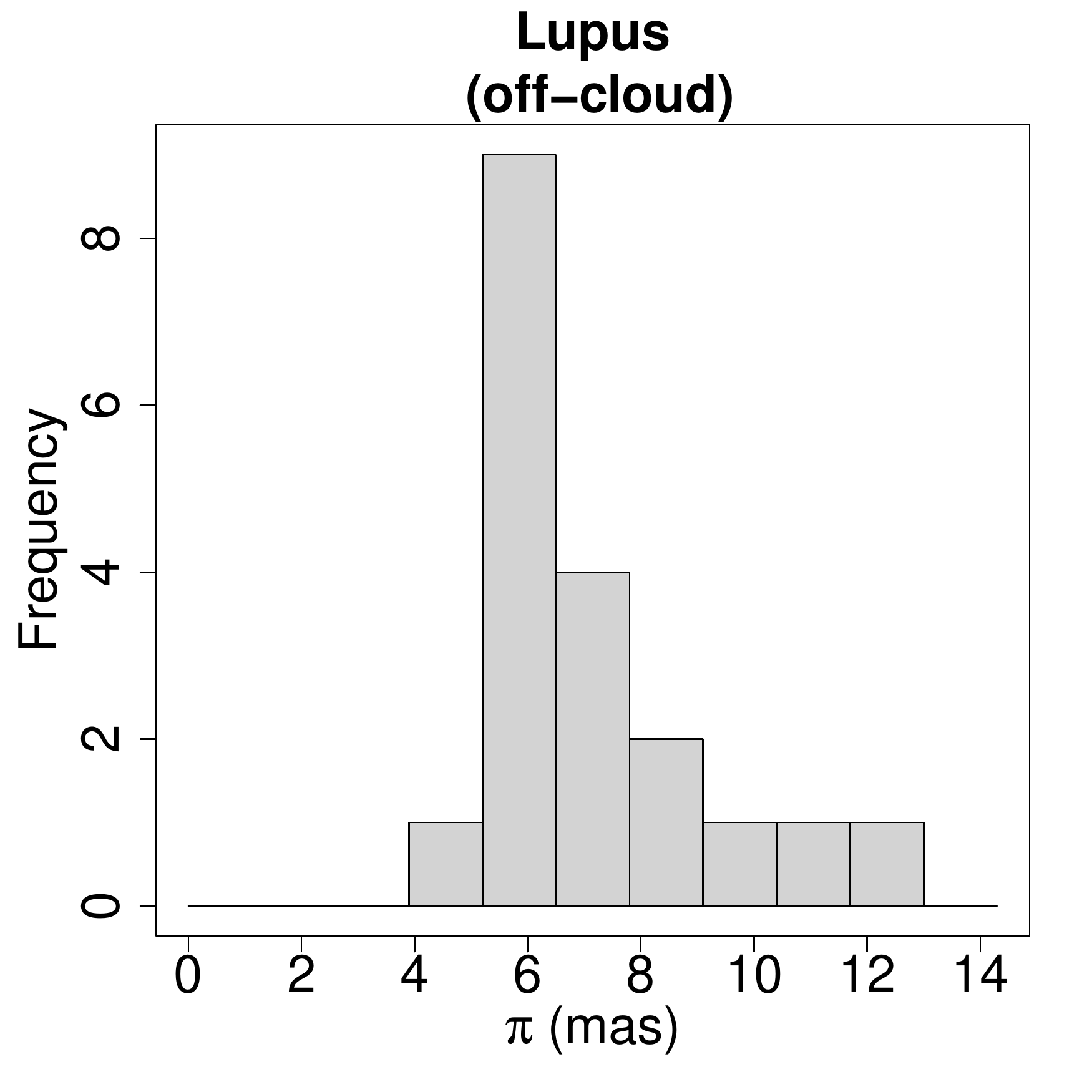}
\includegraphics[width=0.23\textwidth]{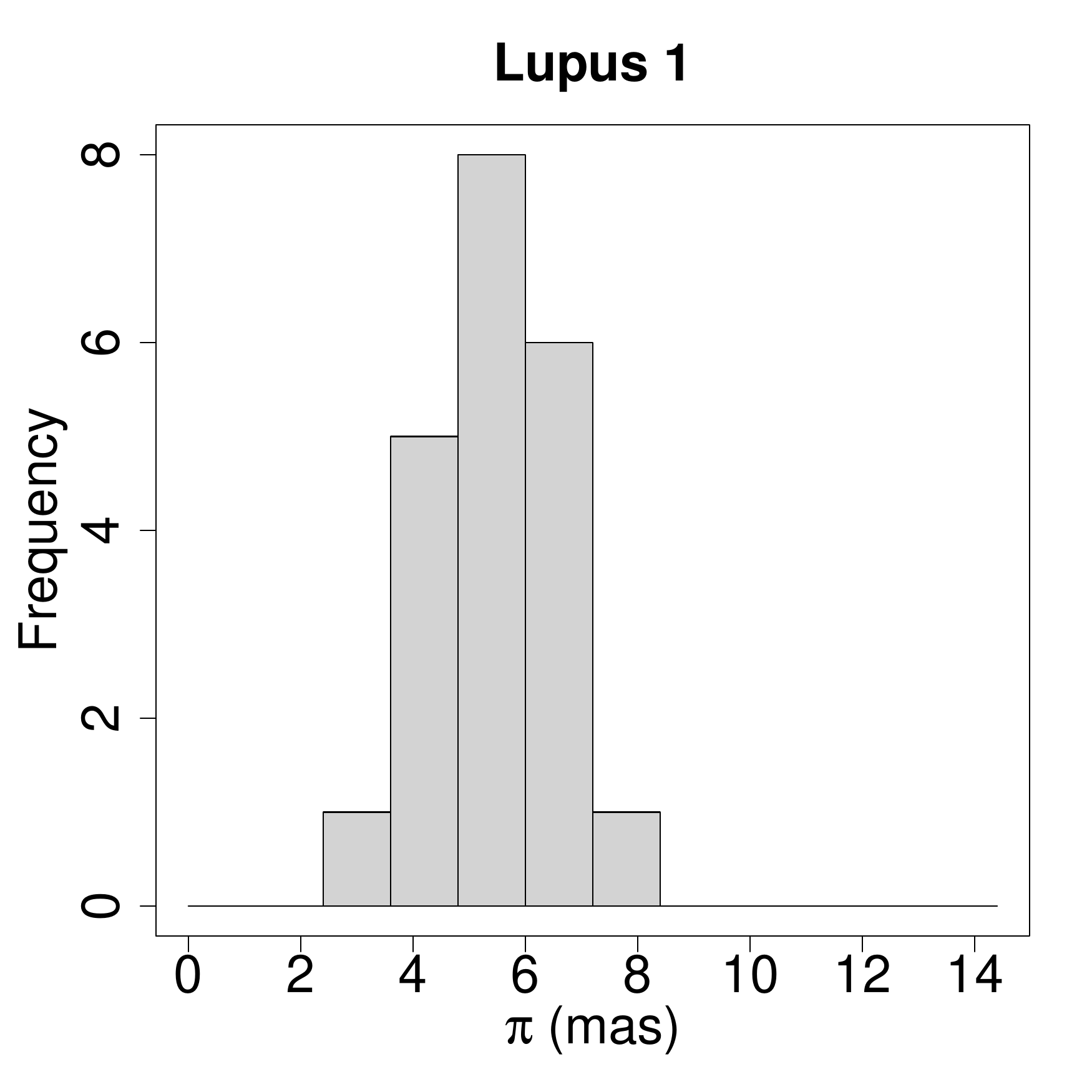}
\includegraphics[width=0.23\textwidth]{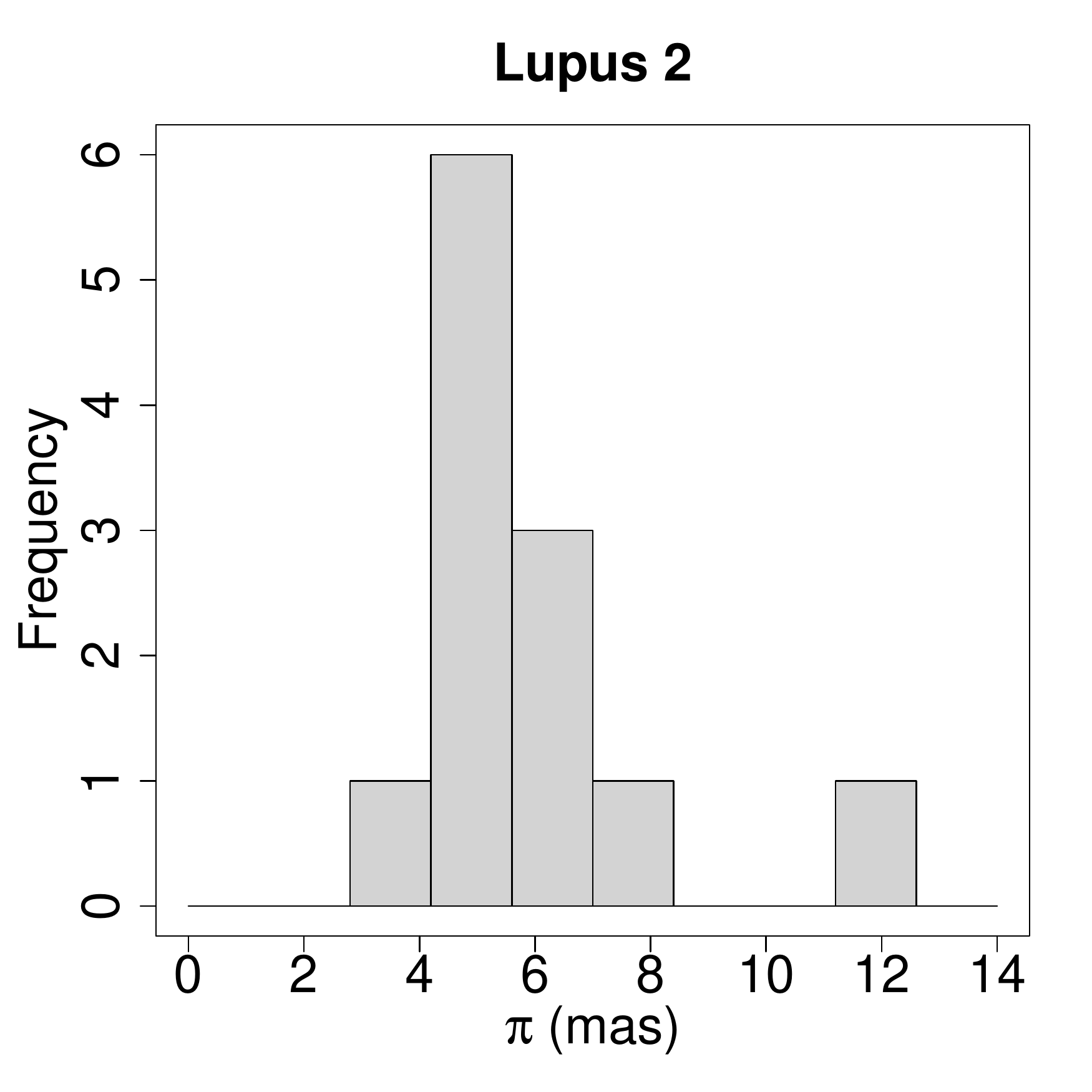}
\includegraphics[width=0.23\textwidth]{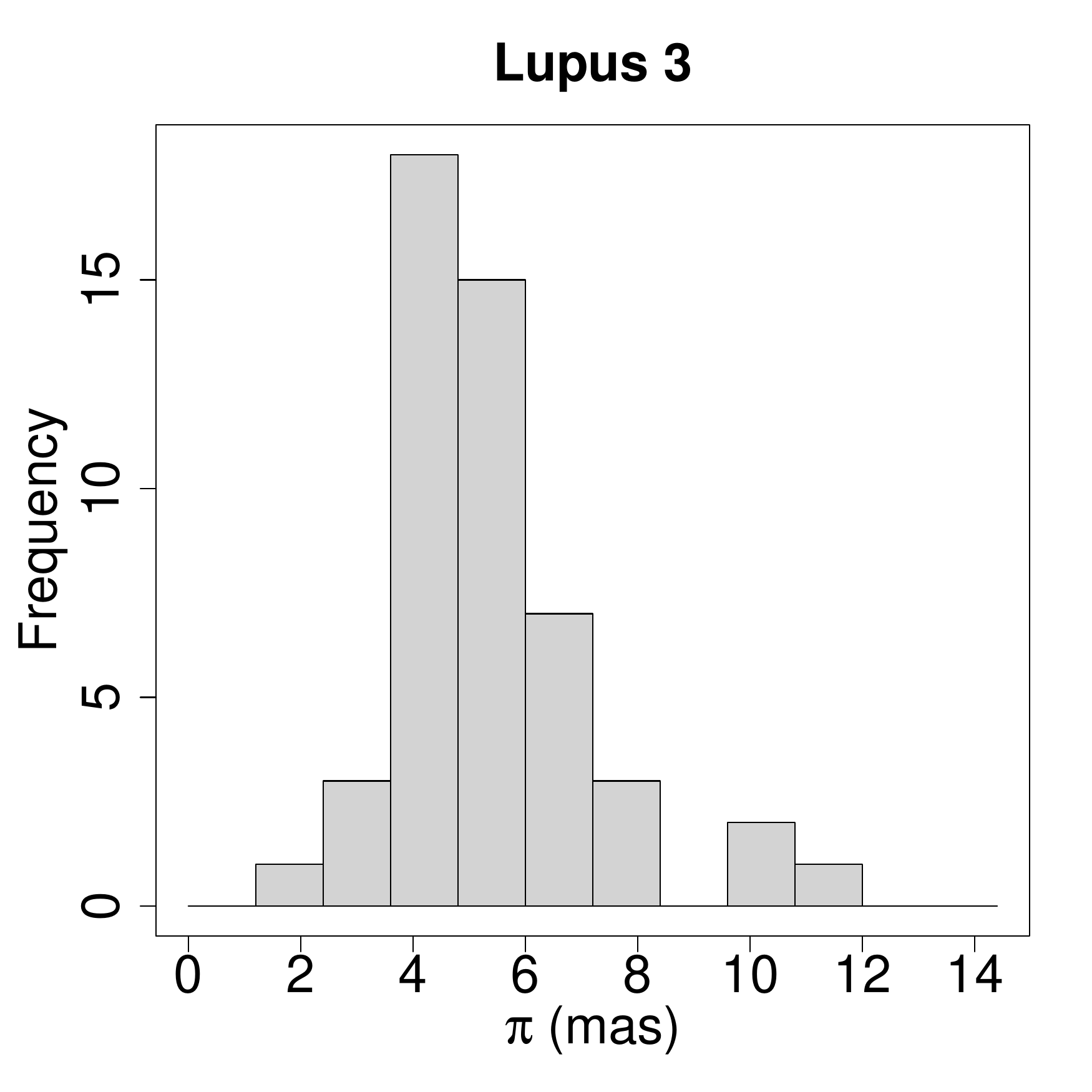}
\includegraphics[width=0.23\textwidth]{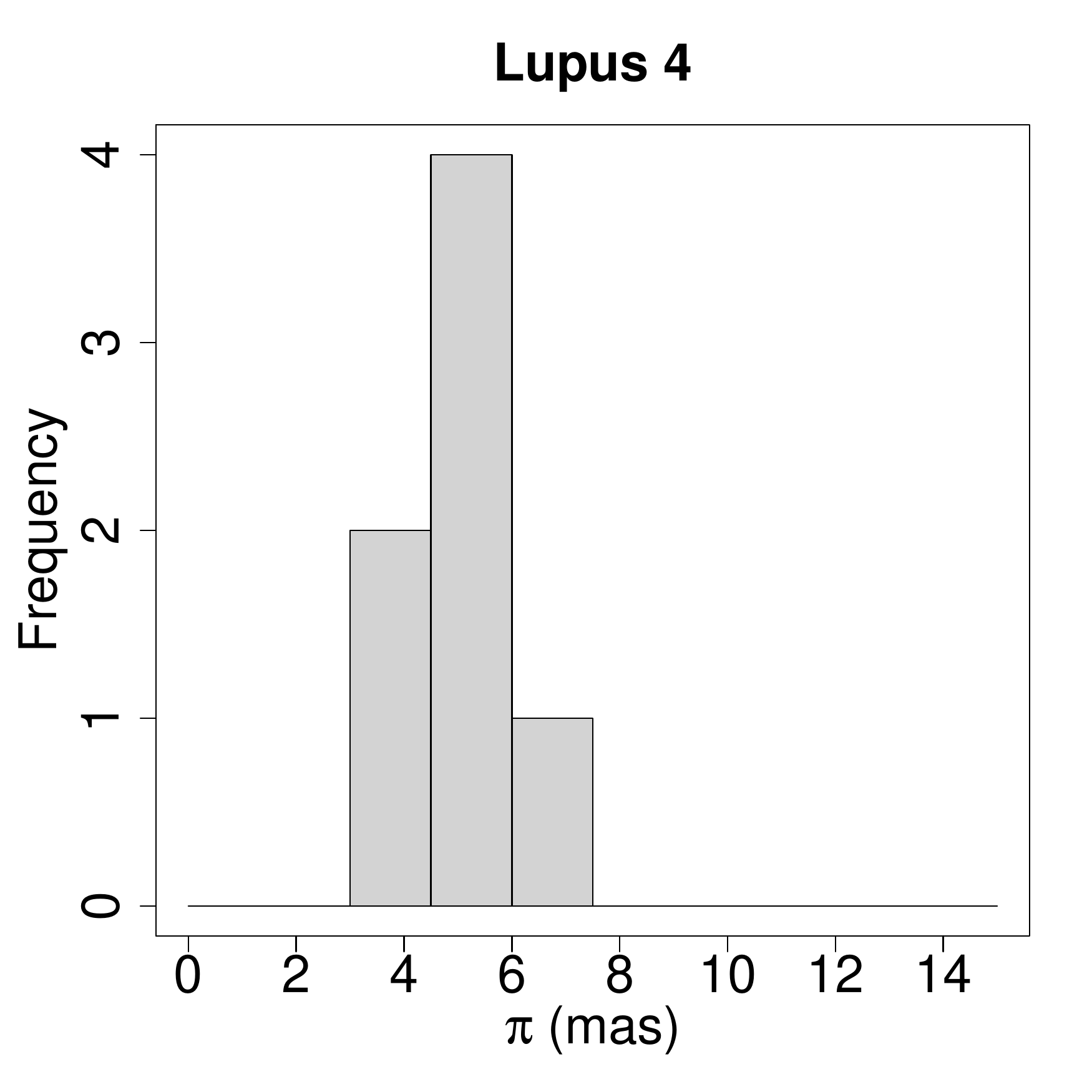}
\caption{Histogram of parallaxes for the various subgroups of the Lupus complex.
\label{fig16} }
\end{center}
\end{figure}

The parallaxes derived in this paper allow us to investigate the properties of the various subgroups in this cloud complex. Figure~\ref{fig15} displays the location of Lupus members and corresponding clouds. The histogram of parallaxes is presented in Fig.~\ref{fig16}. In the following we discuss the parallaxes derived for these subgroups based on the results presented in Table~\ref{tab9}. Using the information provided in Table~\ref{tab2} we calculate the mean UVW velocities for each subgroup and confirm that they exhibit coherent motions with a one-dimensional velocity dispersion of $\sim$1~km/s among the various subgroups (see Table~\ref{tab9}). 

%TABLE 9
\begin{table}[!h]
\centering
\caption{Properties of the various subgroups in Lupus. 
\label{tab9} }
\begin{tabular}{lcccc}

\hline
Sample&Stars&$\pi$&$d$&$(U,V,W)$\\
&&(mas)&(pc)&(km/s)\\
\hline
Lupus~1&21&$5.5\pm0.2$&$182_{-6}^{+7}$&(-4,-21,-5)\\
Lupus~2&12&$6.0\pm0.6$&$167_{-15}^{+19}$&(-5,-21,-5)\\
Lupus~3&50&$5.4\pm0.3$&$185_{-10}^{+11}$&(-5,-19,-6)\\
Lupus~4&7&$4.9\pm0.4$&$204_{-15}^{+18}$&(-7,-20,-9)\\
Lupus \footnotesize{(off-cloud)}&19&$7.2\pm0.5$&$139_{-9}^{+10}$&(-4,-19,-4)\\
\hline
Lupus \footnotesize{(full sample)}&109&$5.8\pm0.2$&$172_{-6}^{+6}$&(-5,-20,-6)\\
\hline

\end{tabular}
\tablefoot{ We provide for each subgroup the number of stars, average parallax, and average distance with the corresponding uncertainties and the mean UVW velocity components. Uncertainties in the mean velocities are 1-2~km/s. Those stars in the on-cloud region (see definition in Sect.~3.2) located beyond the limits of the main molecular clouds (Lupus 1-4) have been assigned to the off-cloud population (see Fig.~\ref{fig15}). 
}
\end{table}

%FIGURE 15

\paragraph{\textit{Full sample:}} The average error on parallaxes is $\overline{\sigma}_{\pi}=1.5$~mas, yielding an average precision of about 25\% in our distance results derived in this paper. We note the existence of background and foreground populations (as anticipated in Sect.~3),  confirming that the Lupus complex occupies a large volume in space.

\paragraph{\textit{Off-cloud population:}} The WTTSs that form this population are scattered not only in angular extent but also in depth (see Fig.~\ref{fig16}). Different scenarios have been suggested to explain the dispersed population of WTTSs: the stars may have been formed in the vicinity of the clouds and have reached their present location because of the velocity dispersion in the group \citep{Wichmann(1997a)}; they might have been ejected with high velocities by dynamical interactions in multiple systems \citep{Sterzik(1995)}; star formation could have occurred in small cloudlets that have now dispersed \citep{Feigelson(1996)}. One important parameter in this discussion is the stellar age that can now be accurately determined for those stars with individual parallaxes. Age determination goes beyond the scope of the present work and will be discussed in a forthcoming paper.

\paragraph{\textit{Lupus~1, 2, and 4:}} All together they contain 37\% of moving group members. While Lupus~1 and 2 are somewhat closer, Lupus~4 is more distant (see Table~\ref{tab9}). We derived individual parallaxes for four stars in Lupus~1 and one star in Lupus~2. Our results for Lupus~4 are less accurate, because we only computed tentative parallaxes, and our sample is biased towards CTTSs as mentioned before. However, Hip~78092 is an HAeBe star projected in the direction of Lupus~4, and \textsc{Hipparcos} results for this star, $\pi_{HIP97}=5.04\pm1.18$~mas and $\pi_{HIP07}=4.29\pm0.98$~mas, confirm the approximate distance to the cloud derived in this paper.

\paragraph{\textit{Lupus~3:}}  We find significant depth effects in Lupus~3, while the stars in the remaining clouds are less dispersed along the line of sight (see, e.g., Fig.~\ref{fig16}). One possibility for explaining this result is the existence of various components along the line of sight that lie at different distances. We computed the average parallax of the CTTSs (26 stars) and WTTSs (23 stars) in Lupus~3 and find
\begin{center}
$\overline{\pi}_{CTTS}=4.6\pm0.2$~mas\,,
\end{center}
\begin{center}
$\overline{\pi}_{WTTS}=6.1\pm0.4$~mas\,.
\end{center}
The depth of Lupus~3 derived from the closest and remotest parts of this cloud is in good agreement with \citet[][see Sect.~2]{Lombardi(2008)}. This preliminary result must be confirmed by further investigations since we derived accurate parallaxes for only a few stars.

%FIGURE 16

%----------------------------------------------------------------------------------------------------
\subsection{Expansion of the Lupus association}

In Sect.~6.1 we mentioned that the Lupus association is undergoing expansion \citep[see also][]{Makarov(2007)}. From the proper motions alone, it is impossible to distinguish between  the state of linear expansion and parallel space motions, so RVs are needed to distinguish between them. In the following we use the Blaauw's expansion model \citep{Blaauw(1964)}, combined with the radial velocities and individual parallaxes given in Tables~\ref{tab5} and \ref{tab6}, to investigate whether the expansion rate of the association can be detected.

\begin{figure}[h]
\begin{center}
\includegraphics[width=0.49\textwidth]{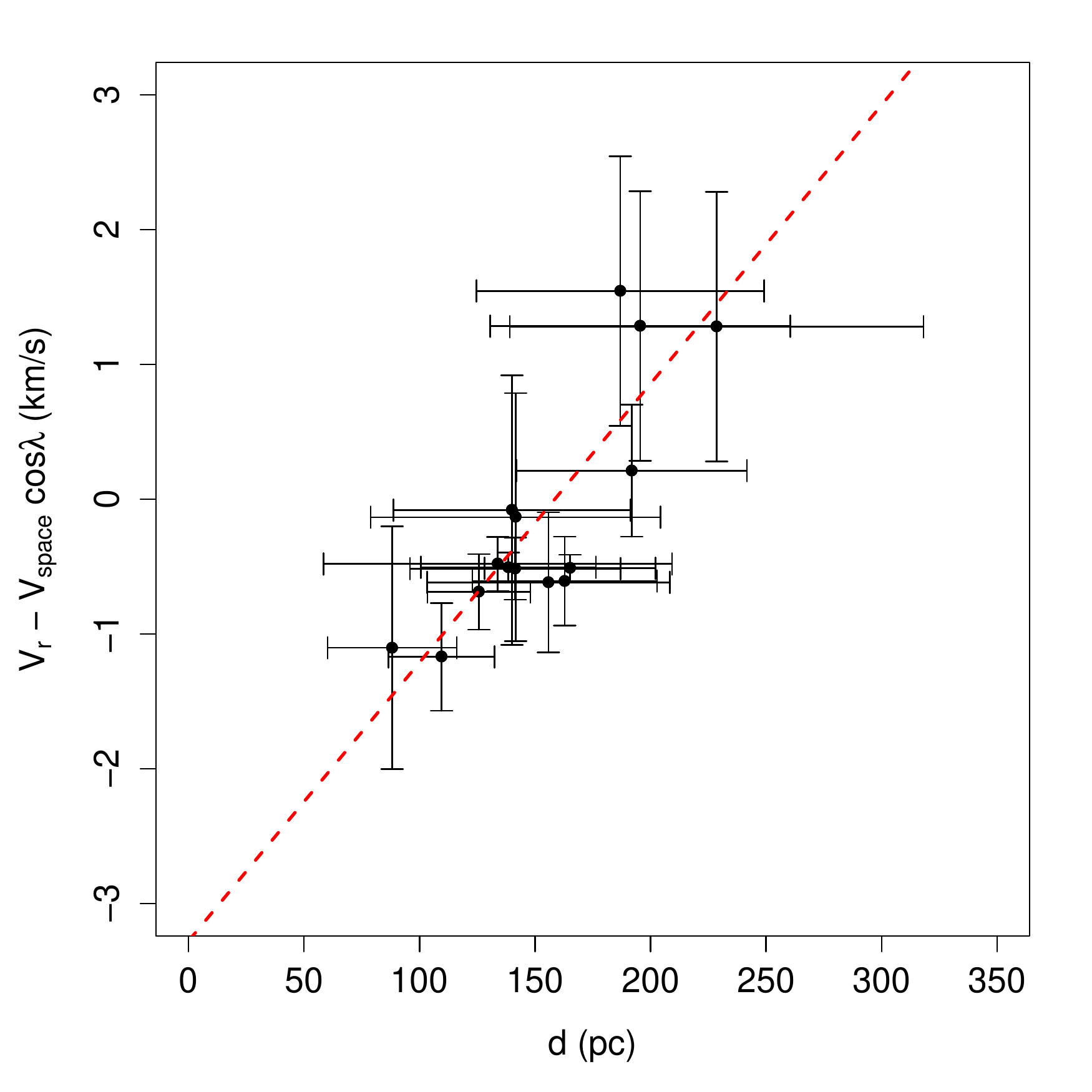}
\caption{Blaauw's linear expansion model applied to the Lupus association of young stars. The red dashed line indicates the weighted least-squares fit to the data. }
\label{fig17} 
\end{center}
\end{figure}

In the Blaauw model for linear expansion the stellar RVs are given by 

\begin{equation}\label{eq.6}
V_{r}= V_{space}\cos\lambda + \kappa\, d + K\, ,
\end{equation}
where $\kappa$ is the expansion term, $K$ a systematic error in the radial velocities, and $d$ the individual distance of the star. A linear expansion exists when $\kappa>0$ in this equation. To derive $\kappa$, we plot the difference between the observed spectroscopic RVs and the predicted RVs ($V_{r}^{pred}=V_{space}\cos\lambda$) inferred from the CP coordinates as a function of the individual distances. Then, we solve for the slope $\kappa$ in Eq.~\ref{eq.6} (see Figure~\ref{fig17}). The best fit to the data yields $\kappa=0.021\pm0.004$~km\,s$^{-1}$\,pc$^{-1}$. We reject GQ~Lup, RXJ1608.5-3847, and Sz121 from this analysis, because of the large errors in distance (see Table~\ref{tab6}), and RXJ1613.0-4004 because of its large deviation, presumably due to a poor RV. Although the derived expansion term $\kappa$ is small when compared to, say, the TW Hydrae association \citep[see][]{Mamajek(2005)}, it is positive within 3$\sigma$ of the computed uncertainty. This result confirms that the linear expansion in the Lupus association of young stars is real but small.

%FIGURE 17

%----------------------------------------------------------------------------------------------------------------			
%							8. CONCLUSIONS
%----------------------------------------------------------------------------------------------------------------
\newpage
\section{Conclusions} 

We have identified a moving group of 109 stars in the Lupus SFR by applying our new CP search method combined with the k-NN algorithm, which made it possible to perform a membership analysis and to distinguish between Lupus and UCL stars. We applied the CP search method to different subsets of association members and confirmed our solution with Monte Carlo simulations. Because of the close proximity to the UCL subgroup and the difficulties encountered in separating these groups, we claim that we have detected a \textit{minimum} moving group in our CP analysis that may not contain all stars kinematically associated to Lupus. We derived accurate parallaxes for members with known RVs that we define as the Lupus core moving group and used the group spatial velocity to tentatively calculate approximate parallaxes for the remaining stars.

Determination of individual parallaxes is restricted to moving group members with known RVs. The RVs of many stars in the group either have never been measured or are of low quality so cannot be used to derive accurate parallaxes. We presented new RV measurements for 52 PMS stars of our starting sample of Lupus candidate stars based on spectroscopic observations performed with FEROS. We encourage observers to employ the available precise spectrographs to perform spectroscopic surveys of this complex SFR. Additional observations will also allow one to detect new binaries and confirm the youth of many PMS candidates.

We show that the CTTSs are located close to or within the molecular clouds, while the WTTSs are dispersed not only in angular extent but also in depth. We find evidence of depth effects in Lupus~3 that must be confirmed by further investigations. Based on the individual distances and radial velocities derived in this paper, we also confirm that the Lupus association is undergoing expansion. These results represent a first step toward better understanding of the structure of the Lupus cloud complex. The individual distances derived in this paper will be used in a forthcoming paper to investigate the physical properties of Lupus stars.

%----------------------------------------------------------------------------------------------------------------
%						ACKNOWLEDGEMENTS
%----------------------------------------------------------------------------------------------------------------
\begin{acknowledgements}
It is a pleasure to thank Johny Setiawan and Maren Mohler for their assistance with the FEROS Data Reduction Software and cross-correlation techniques used for radial velocity computations. We thank Kengo Tachihara for providing the image for Figures 3, 6, 14, and 15, and the PASJ editorial office for granting permission to reproduce this material. We also acknowledge the referee for the careful report. This research was funded by a doctoral/post-doctoral fellowship from FAPESP and made use of the SIMBAD database operated at the CDS, Strasbourg, France.  
\end{acknowledgements}

%----------------------------------------------------------------------------------------------------------------
%							BIBLIOGRAPHY
%----------------------------------------------------------------------------------------------------------------
\bibliographystyle{aa}
\bibliography{references.bib}

%----------------------------------------------------------------------------------------------------------------
%							APPENDIX A
%----------------------------------------------------------------------------------------------------------------
\appendix
\onecolumn
\section{Error propagation of parallaxes}

The parallax uncertainty for those stars with known RVs is obtained by error propagation of Eq.~(\ref{eq_plx_ind}) and it is given by
\begin{equation}\label{eq.A1}
\sigma_{\pi}^{2}=\left(\frac{A}{V_{r}\tan\lambda}\right)^{2}\sigma_{\mu_{\parallel}}^{2}+\left(\frac{A\,\mu_{\parallel}}{V_{r}^{2}\tan\lambda}\right)^{2}\sigma_{V_{r}}^{2}+\left(\frac{A\, \mu_{\parallel}}{V_{r}\sin^{2}\lambda}\right)^{2}\sigma_{\lambda}^{2}\, .
\end{equation}
While $\sigma_{\mu_{\parallel}}$ and $\sigma_{V_{r}}$ refer to errors on observational quantities (i.e., proper motions and RVs), the use of $\sigma_{\lambda}$ in eq.~(\ref{eq.A1}) is not straightforward. \citet{Galli(2012)} performed extensive simulations that convincingly demonstrate that the precision of the CP position is influenced by several parameters. The increasing velocity dispersion of the cluster and the existence of possible interlopers in the moving group also play an important role when evaluating the precision and accuracy of the CP position. These sources of errors can be roughly divided into two parts: (i) observational errors that arise mainly from proper motion errors (since errors on stellar position can be neglected to a first-order approximation), and (ii) geometric effects (e.g., angular distance from the CP to the moving group, cluster concentration, distance, and number of group members). To correctly introduce the CP errors in the parallax uncertainty, one needs to separate the contribution of these effects, otherwise the error budget due to proper motions will be considered twice in Eq.~(\ref{eq.A1}). To do so, we set the proper motion errors to zero (i.e, $\sigma_{\mu_{\alpha,\delta}}=0$). Then, we use Eq.~(19) of \citet{deBruijne(1999b)} to estimate the velocity dispersion of the Lupus moving group that results from the peculiar motion of the stars given by the $\mu_{\perp}$ statistics. This procedure yields $\sigma_{v,\perp}=0.7^{+0.4}_{-0.2}$~km/s and confirms the adopted value of 1~km/s as the one-dimensional velocity dispersion of the moving group used in our CP analysis (see Sect.~5.1). By adopting $\sigma_{v}=0.7$~km/s and $\sigma_{\mu_{\alpha,\delta}}=0$, the uncertainties on the CP position decrease to $(\sigma_{\alpha_{cp}},\sigma_{\delta_{cp}})=(1.6^{\circ},1.2^{\circ})$ yielding $\sigma_{\lambda}\simeq2.0^{\circ}$. This result represents a more realistic estimate of the CP error budget due to geometric effects, not including proper motions errors, to be used in the error propagation of parallaxes.

%----------------------------------------------------------------------------------------------------------------
%							APPENDIX B
%----------------------------------------------------------------------------------------------------------------
\section{Notes on radial velocities}

One important point is that the observed RVs for Lupus stars are expected to be lower than to other SFRs, such as Taurus-Auriga and Chamaeleon \citep[see][]{Bertout(2006), James(2006)}. The average value and standard deviation of the RVs presented in Table~\ref{tab4} for Lupus stars is $\overline{V}_{r}=2.7\pm1.9$~km/s, and the average error is $\sigma_{V_{r}}=0.5$~km/s. Here we investigate the errors on parallaxes and space velocities caused by uncertainties in RVs. A small variation $\Delta V_{r}$ in RVs accounts for the variation $\Delta \pi$ in parallaxes that can be approximated by

\begin{equation}
\Delta\pi\simeq\frac{A\, \mu_{\parallel}}{\tan\lambda}\left(\frac{\Delta V_{r}}{V_{r}^2}\right).
\end{equation}
This shifts the space velocity of the star by $\Delta V_{space}$ that is given as
\begin{equation}
\Delta V_{space}\simeq\frac{A\,\mu_{\parallel}}{\sin\lambda}\left(\frac{\Delta\pi}{\pi^{2}}\right).
\end{equation} 
Using the values of position, proper motion, and radial velocities given in Table~\ref{tab2} for the on-cloud population, and the CP solution derived in Sect.~5.3, we note that a small shift $\Delta V_{r}=0.5$~km/s in the measured RVs accounts for $\Delta\pi\simeq$ 1.4~mas, and consequently, $\Delta V_{space}\simeq$ 4~km/s. We thus emphasize that high-precision RVs are needed to derive reliable kinematic parallaxes of Lupus stars.

\end{document}